\pdfoutput=1
\documentclass[10pt, letterpaper]{article}
\usepackage{style}
\usepackage{shortcuts}
\usepackage[mathlines]{lineno}

\title{Connectivity in a New Bounded-Space Dynamic Model}
\author[1]{Bardiya Aryanfard}
\author[1]{Monika Henzinger}
\author[2]{John Kuszmaul}
\author[2]{Jakob Nogler}
\author[2]{Virginia~Vassilevska~Williams}
\affil[1]{Institute of Science and Technology Austria, Klosterneuburg, Austria}
\affil[2]{Massachusetts Institute of Technology, Cambridge, MA, USA}
\date{}

\begin{document}

\hypersetup{pageanchor=false}
\maketitle
\begin{abstract}

We introduce a new model for dynamic algorithms in which the algorithm
has read-only access to a dynamically changing input and random access to its own bounded
workspace for computation. The model is a dynamic analogue of the
time--space tradeoffs for static inputs studied in complexity theory, in
which the input is stored separately and
 space is only measured through the algorithm's own writable
memory usage. We study lower and upper bounds in our new model
for dynamic undirected $s$--$t$ connectivity: after every edge update to an $n$-vertex graph, an
algorithm with $S$ words of workspace must report whether two fixed
vertices $s$ and $t$ are in the same connected component. The algorithm has oracle access to
the dynamically changing graph, either through adjacency list queries (give the degree of $v$ or its $j$th neighbor) or through adjacency matrix queries (are $u$ and $v$ adjacent?).

Our algorithms are randomized, and our lower bounds apply to randomized
algorithms with up to $cn$ bits of workspace, for a small constant $c>0$
(up to $cn/\log n$ bits for the worst-case bound with an adjacency matrix).
We prove:
\begin{itemize}[itemsep=5pt]
\item With adjacency lists, worst-case time $\widetilde\Theta(n^2/S^2)$
per insertion is necessary and sufficient for incremental algorithms.
\item With an adjacency matrix, worst-case time $\widetilde\Omega(n^2/S)$
per insertion is necessary for incremental algorithms, and it is
sufficient for $S\ge\sqrt n$.
\item Amortization helps:
With adjacency lists, we obtain incremental algorithms
with amortized time $\widetilde O(n/S)$ per insertion. With an adjacency
matrix, we get $\widetilde O(n^2/S)$ amortized time for every $S$. 
We complement these with lower bounds, showing that every incremental algorithm needs amortized time
$\widetilde\Omega(\sqrt{n/S})$ with adjacency lists and
$\widetilde\Omega(n\sqrt{n/S})$ with an adjacency matrix.

\item We obtain strong unconditional lower bounds for fully dynamic algorithms even when the graph is always a forest:
Amortized time $\widetilde\Omega(n/S)$
per update is necessary with adjacency lists and $\widetilde\Omega(n^2/S)$ is necessary with an
adjacency matrix.
These lower bounds are tight for forests: we give a deterministic fully dynamic
algorithm for forests with worst-case time $O(n^2/S)$ per update with an
adjacency matrix, and $\widetilde O(n/S)$ with adjacency lists when the
maximum degree is bounded. 
\end{itemize}
In contrast to conventional dynamic connectivity, which can be maintained 
for all pairs of vertices using $\Omega(n)$ words of space with polylogarithmic update time [Henzinger-King J. ACM 1999]
and with inverse-Ackermann amortized time under insertions
only [Tarjan J. ACM 1975],  our lower bounds
show that polynomial update time is necessary whenever
$S\le n^{1-\varepsilon}$ for a constant $\varepsilon>0$, even when only a
single pair of vertices must be maintained, and even for incremental
algorithms with amortized guarantees.
\end{abstract}
\nolinenumbers

\thispagestyle{empty}
\clearpage
\setcounter{page}{1}



\providecommand{\softO}{\widetilde{O}}

\section{Introduction}\label{sec:intro}

Undirected $s$--$t$ connectivity, the problem of deciding whether two given
vertices of a graph lie in the same connected component, is one of the most
basic problems in both algorithms and complexity theory. In complexity
theory, it has long been the main problem in the study of deterministic
logarithmic space. Aleliunas et al.~\cite{AKLLR79}  use random walks to solve it in
randomized logarithmic space, and Lewis and Papadimitriou \cite{LP82}
showed that it is complete for symmetric logarithmic space ($\mathsf{SL}$). A long sequence
of deterministic algorithms using ever less space
\cite{Savitch70,Nisan94,NSW92,ATWZ00,Trifonov08} culminated in Reingold's
celebrated theorem  \cite{Reingold08} that undirected $s$--$t$ connectivity can be solved in
deterministic logarithmic space, and 
$\mathsf{SL}=\mathsf{L}$.

In algorithms design, undirected $s$--$t$ connectivity has been a central testbed
for trading off time and space, for more than four decades. On an $n$-vertex,
$m$-edge graph, one can use
breadth-first search (BFS) to solve the problem in $O(m+n)$ time, but BFS
stores a search frontier and a visited set and uses $O(n)$ words\footnote{A \emph{word} consists of $\Theta(\log n)$ bits, enough to store a vertex or an integer bounded by a polynomial in $n$. We write $\softO(\cdot)$, $\widetilde\Omega(\cdot)$, and $\widetilde\Theta(\cdot)$ to suppress factors polynomial in $\log n$; throughout, we do not attempt to optimize such factors.}. At the
other extreme, the random walk of Aleliunas et al.\ uses only $O(\log n)$
bits of space and $O(mn)$ time, with constant error probability. A long
line of work on random walks and graph exploration has charted the space in
between \cite{BKRU94,BF96,Feige97,Kosowski13}: Feige \cite{Feige97}
achieved time--space product $\softO(mn)$ throughout the full space range, and
Kosowski \cite{Kosowski13}, assuming stronger access to the graph,
achieved $\softO(m+n^2/W)$ time with $\softO(W)$ words for each $W$. In all of these
results the graph is a read-only input, and one only pays for the algorithm's writable
workspace: the algorithms succeed with little memory precisely
because they can reread the graph instead of remembering it.

All of this work concerns a single, static graph. Dynamic connectivity, in
which the graph changes over time, has been studied from a very different
starting point: the goal is to maintain enough information about the graph
to answer connectivity queries quickly. For insertions alone, a disjoint-set
structure supports every operation in inverse-Ackermann amortized time with
$O(n)$ words \cite{Tarjan75}. Fully dynamic data structures support both
insertions and deletions in polylogarithmic time
\cite{HK99,HLT01,HHKPT23}, and very recent work achieves polylogarithmic
worst-case update time even against adaptive updates \cite{MG26,MGY26}.
Because these data structures answer queries about arbitrary pairs from
their own memory, without consulting the graph, they must in effect encode
the partition of the vertices into components, which takes
$\Omega(n\log n)$ bits, that is, $\Omega(n)$ words.

Neither line of work addresses the setting in which the two lines (static time-space trade-offs and dynamic updates) meet: the
graph changes over time, it is stored outside the algorithm's memory where
the algorithm can reread it, and the algorithm must keep track of whether
one fixed pair of vertices $s$ and $t$ is connected. In this setting the
$\Omega(n\log n)$ barrier from having to keep the connected components no longer applies. The answer is a single bit, and,
as in the static small-space algorithms above, information about the graph
need not be stored, since it can be recovered by rereading the graph. Can the answer
then be kept up to date with far fewer than $n$ words of memory, and at what
cost in time?

This combination arises when a graph is owned by one system and monitored
by another. The owner stores and maintains the graph as part of an
application's data, while a separate monitor with limited private memory is
notified of each update and can inspect the graph, but may not annotate it
or build an unrestricted auxiliary index: the graph is read-only \emph{to
the monitor}, even though its owner keeps modifying it. For example, a
network-management system maintains a topology of bidirectional links that
changes as links fail and recover, and a monitoring task with a small
memory allocation must track whether two designated sites are connected.
Similarly, a data-integration pipeline accepts links asserting that two
records describe the same entity, and an auditor must raise an alarm as soon
as a chain of accepted links joins two records known to describe different
entities; this problem is incremental unless links can later be withdrawn.
In both cases the relation is symmetric, so the graph is undirected, the
pair to monitor is fixed in advance, and the monitor is responsible for a
single bit, whereas conventional connectivity data structures maintain the
entire partition into components. Appendix~\ref{sec:motivation} discusses
such scenarios in more detail.

\begin{figure}[t]
\centering
\begin{tikzpicture}[>=stealth, font=\small,
  vtx/.style={circle, fill=black, inner sep=1.3pt},
  term/.style={circle, draw, fill=white, inner sep=1.6pt}]
\node[align=center] (upd) at (0.1,1.1) {updates, one at a time:\\
  insert or delete $\{u,v\}$};
\draw[rounded corners=3pt, thick] (2.3,0) rectangle (7.2,2.2);
\node at (4.75,1.85) {\textbf{Algorithm}};
\draw[rounded corners=2pt, dashed, fill=gray!15] (2.5,0.22) rectangle (7.0,1.53);
\node[align=center, font=\footnotesize] at (4.75,0.875)
  {writable workspace: $S$ words\\ persistent \emph{and} temporary\\
   (all charged)};
\draw[rounded corners=3pt, thick] (9.2,-0.25) rectangle (13.6,2.4);
\node at (11.4,2.05) {\textbf{Read-only oracle}};
\node[term, label={[font=\footnotesize]left:$s$}] (s) at (9.95,1.05) {};
\node[vtx] (a) at (10.7,1.55) {};
\node[vtx] (b) at (10.85,0.65) {};
\node[vtx] (c) at (11.65,1.2) {};
\node[vtx] (d) at (12.05,0.45) {};
\node[term, label={[font=\footnotesize]right:$t$}] (t) at (12.8,1.15) {};
\draw (s)--(a)--(c)--(t) (s)--(b) (b)--(d);
\node[font=\footnotesize] at (11.4,0.05) {current graph (not charged)};
\draw[->, thick] (upd.east) -- (2.3,1.1);
\draw[->] (7.2,1.35) -- node[above, font=\footnotesize]{probe} (9.2,1.35);
\draw[<-] (7.2,0.85) -- node[below, font=\footnotesize]{reply} (9.2,0.85);
\draw[->, thick] (4.75,0) -- (4.75,-0.55)
  node[below, align=center]{answer after each update:\\
  are $s$ and $t$ connected?};
\node[below, align=left, font=\footnotesize] at (11.4,-0.35)
  {list model: $\operatorname{Degree}(v)$,
   $\operatorname{Neighbor}(v,j)$\\
   matrix model: $\operatorname{Adjacent}(u,v)$};
\end{tikzpicture}%
\caption{The dynamic oracle model. Updates arrive one at a time. While
processing an update, the algorithm may probe the graph as it was before
the update through a read-only oracle, at unit cost per probe, and it must
then report whether $s$ and $t$ are connected. All writable memory counts
toward the space bound; the graph itself does not.}
\label{fig:model}
\end{figure}
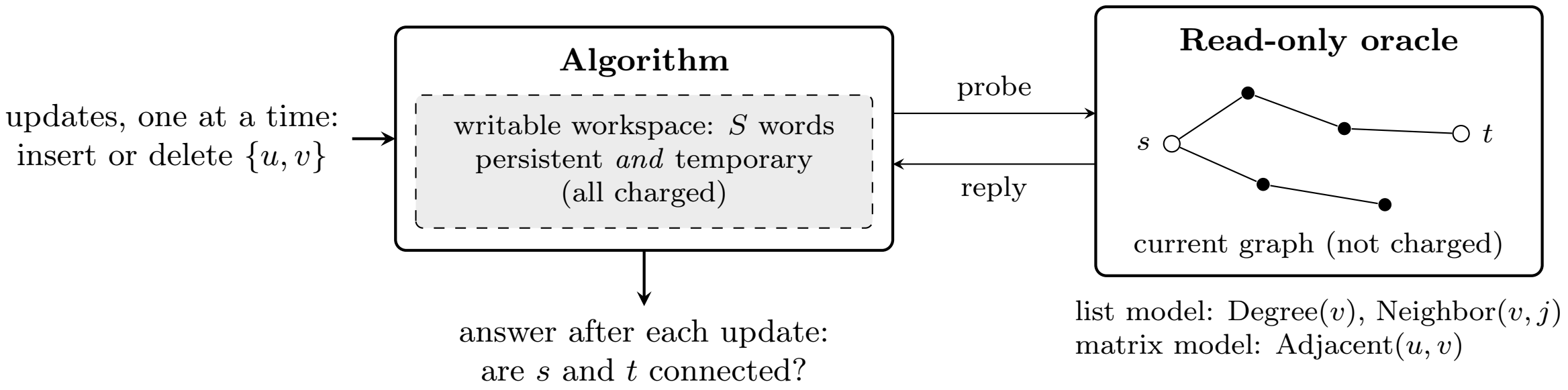

Our setting restricts the monitor's memory in two ways: it limits both the
information the monitor can retain between updates and the scratch space it
can use while processing a single update. 
All persistent summaries, random seeds, caches, and temporary computation count toward the budget; in particular, any information about past updates, such as component labels or a log of the updates, must be kept in this memory. What the monitor can do is reread
the graph, exactly as the static small-space algorithms do. The difference
is that a static algorithm answers a single query about a single graph,
whereas the monitor must answer after every update and cannot afford to
start from scratch each time. The tension between retaining and recovering
information is the time--space tradeoff we study:
\begin{center}
\begin{minipage}{0.86\linewidth}
\centering
\emph{How much time per update is needed to maintain $s$--$t$ connectivity
in a changing graph with $S\ll n$ words of writable memory, given read
access to the current graph?}
\end{minipage}
\end{center}
The answer depends on the interface through which the graph can be
inspected. We consider the two standard ones. In the
\emph{adjacency-matrix model} (the
\emph{matrix model} for short), the algorithm can ask whether two given
vertices are adjacent.
In the
\emph{adjacency-list model} (the \emph{list model}), it can instead ask for
the degree of a vertex and for its $j$th neighbor. A system that supports
only pairwise edge tests need not support neighbor enumeration, and comparing
the two models asks how trading pairwise tests for neighbor enumeration
changes the private memory needed for efficient dynamic algorithms in our model. Each such query is an
\emph{input probe} and costs unit time. Section~\ref{sec:model} defines the
model precisely, and Figure~\ref{fig:model} illustrates it.

Two baselines are immediate. With $\Theta(n)$ words, a disjoint-set
structure handles insertions in nearly constant amortized time. With $S$
words, one can rerun a static time--space tradeoff after every update;
with Feige's algorithm this takes $\softO(mn/S)$ time per update, and
$\softO(m^2n/S)$ time in total over a sequence of $m$ insertions.

\begin{figure}[t]
\centering
\definecolor{LandLower}{HTML}{C62828}
\definecolor{LandUpper}{HTML}{2265AC}
\definecolor{LandTight}{HTML}{7B3294}
\definecolor{LandGap}{HTML}{FCE7A0}
\begin{tikzpicture}[x=4.6cm, y=1.55cm, >=stealth, font=\small]
\begin{scope}
\fill[LandGap] (0,0.5) -- (0,1) -- (1,0) -- cycle;
\draw[->] (0,0) -- (1.1,0);
\draw[->] (0,0) -- (0,2.25);
\foreach \x/\l in {0/{$1$},0.5/{$\sqrt n$},1/{$n$}}
  \draw (\x,0) -- (\x,-0.05) node[below] {\l};
\foreach \y/\l in {0/{$1$},0.5/{$\sqrt n$},1/{$n$},2/{$n^2$}}
  \draw (0,\y) -- (-0.015,\y) node[left] {\l};
\draw[very thick, LandTight] (0,2) -- (1,0);
\draw[very thick, dashed, LandUpper] (0,1) -- (1,0);
\draw[very thick, densely dotted, LandLower] (0,0.5) -- (1,0);
\node[anchor=west, font=\footnotesize, LandTight] at (0.2,1.65) {$n^2/S^2$};
\node[anchor=west, font=\footnotesize, LandUpper] at (0.33,0.78) {$n/S$};
\node[anchor=north east, font=\footnotesize, LandLower] at (0.3,0.33) {$\sqrt{n/S}$};
\node at (0.55,2.25) {(a) list model};
\node at (0.55,-0.5) {workspace $S$ (words)};
\node[rotate=90] at (-0.2,1.1) {time per update};
\end{scope}
\begin{scope}[xshift=6.4cm]
\fill[LandGap] (0,1.5) -- (0,2) -- (1,1) -- cycle;
\draw[->] (0,0) -- (1.1,0);
\draw[->] (0,0) -- (0,2.25);
\foreach \x/\l in {0/{$1$},0.5/{$\sqrt n$},1/{$n$}}
  \draw (\x,0) -- (\x,-0.05) node[below] {\l};
\foreach \y/\l in {0/{$1$},1/{$n$},1.5/{$n^{3/2}$},2/{$n^2$}}
  \draw (0,\y) -- (-0.015,\y) node[left] {\l};
\draw[very thick, dashed, LandUpper] (0,2) -- (0.5,1.5);
\draw[very thick, LandTight] (0.5,1.5) -- (1,1);
\draw[very thick, densely dotted, LandLower] (0,1.5) -- (1,1);
\draw[gray, thin] (0.5,0) -- (0.5,1.5);
\node[anchor=south west, font=\footnotesize, LandTight] at (0.55,1.47) {$n^2/S$};
\node[anchor=north east, font=\footnotesize, LandLower] at (0.42,1.3) {$n\sqrt{n/S}$};
\node at (0.55,2.25) {(b) matrix model};
\node at (0.55,-0.5) {workspace $S$ (words)};
\end{scope}
\begin{scope}[yshift=-1.5cm, xshift=-0.5cm, font=\footnotesize]
\draw[very thick, LandTight] (0,0) -- (0.12,0);
\node[anchor=west] at (0.14,0) {tight worst-case bound};
\draw[very thick, densely dotted, LandLower] (0,-0.42) -- (0.12,-0.42);
\node[anchor=west] at (0.14,-0.42) {incremental amortized lower bound};
\draw[very thick, dashed, LandUpper] (1.32,0) -- (1.44,0);
\node[anchor=west] at (1.46,0)
  {incremental amortized upper $=$ fully dynamic lower bound};
\fill[LandGap] (1.32,-0.48) rectangle (1.44,-0.36);
\node[anchor=west] at (1.46,-0.42) {open gap for incremental amortized time};
\end{scope}
\end{tikzpicture}%
\caption{The time--space landscape for general graphs on a log--log
scale, ignoring polylogarithmic factors; lower bounds hold for workspace up
to $cn$ bits, and up to $cn/\log n$ bits for Theorem~\ref{thm:lb-matrix-wc}. (a)~In the list model, worst-case time
$\widetilde\Theta(n^2/S^2)$ is tight (Theorems~\ref{thm:metropolis}
and~\ref{thm:lb-list-wc}), and amortized time $\softO(n/S)$ suffices
(Theorem~\ref{thm:smallspace}). (b)~In the matrix model, $n^2/S$ is the
amortized upper bound for incremental connectivity, the amortized lower
bound for fully dynamic connectivity (Theorem~\ref{cor:fd-lower}), and the
worst-case lower bound for incremental connectivity
(Theorem~\ref{thm:lb-matrix-wc}); the latter is tight for $S\ge\sqrt n$,
within the range of the lower bounds (Theorem~\ref{thm:largespace}). The
upper bounds shown hold throughout the update sequence;
Theorem~\ref{thm:upper} gives smaller bounds for early updates.}
\label{fig:landscape}
\end{figure}
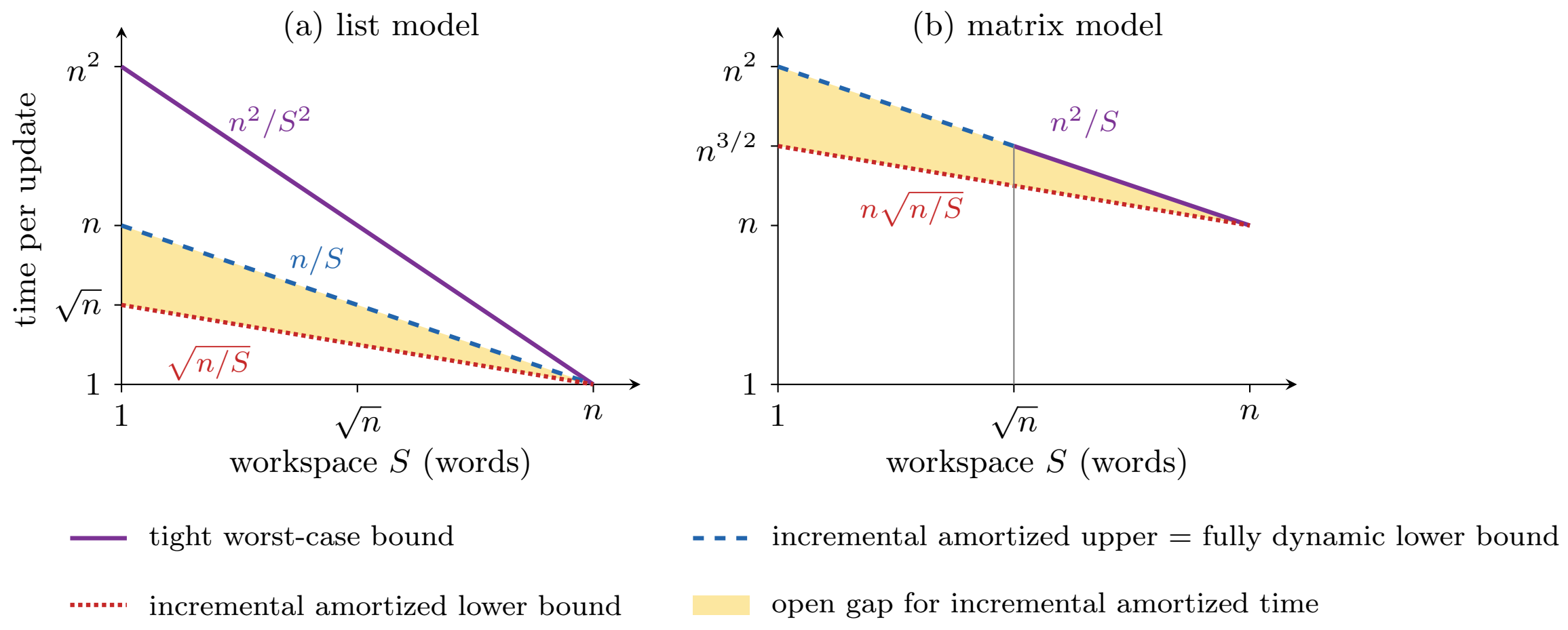

In this paper we formalize this question and study it mainly in the
incremental setting. The answer turns out to depend sharply on whether one
asks for worst-case or amortized guarantees. Our main findings are the
following, and Figure~\ref{fig:landscape} plots them against the
workspace~$S$. All of our lower bounds hold for workspace of up to $cn$
bits, that is, up to $\Theta(n/\log n)$ words, for a small constant $c>0$,
except for Theorem~\ref{thm:lb-matrix-wc}, which holds for up to $cn/\log n$ bits.
\begin{itemize}
\item \emph{Tight worst-case bounds.} Within this range, with $S$ words of
workspace, worst-case time $\widetilde\Theta(n^2/S^2)$ per insertion is necessary and
sufficient in the list model. In the matrix model, worst-case time
$\widetilde\Omega(n^2/S)$ is necessary, and it is sufficient for
$S\ge\sqrt n$. Worst-case hardness appears only after many insertions:
with $\softO(S)$ words, the $i$th insertion takes worst-case time
$\softO(1+i^2/S^2)$ in the list model, for every $S$, and, for
$S\ge\sqrt n$, worst-case time $\softO(1+i/S)$ in the list model and
$\softO(n(1+i/S))$ in the matrix model.
\item \emph{Amortization helps.} In the list model, amortized time
$\softO(n/S)$ per insertion suffices, a factor of $n/S$ below the
worst-case bound, which is polynomial whenever $S\le n^{1-\varepsilon}$.
Over the first $m$ insertions, the total time is
$\softO(m(1+\min\{n,m\}/S))$, which is at most $\softO(mn/S)$.
Perhaps surprisingly, keeping the answer up to date after \emph{all}
insertions thus costs, up to polylogarithmic factors, no more than Feige's
static time bound for a single query on the final graph.
\item \emph{Lower bounds for amortized and fully dynamic algorithms.}
Every incremental algorithm needs amortized time
$\widetilde\Omega(\sqrt{n/S})$ per insertion in the list model and
$\widetilde\Omega(n\sqrt{n/S})$ in the matrix model. Every fully dynamic
algorithm needs amortized time $\widetilde\Omega(n/S)$ per update in the
list model and $\widetilde\Omega(n^2/S)$ in the matrix model, and both bounds hold
even when the graph is at all times a forest. The
amortized time of our incremental algorithms matches these fully dynamic
lower bounds, up to polylogarithmic factors.
\end{itemize}
Table~\ref{tab:results} in Section~\ref{sec:results} lists these bounds
together with the theorems that establish them.

\subsection{The Dynamic Oracle Model}\label{sec:model}

Our model is the dynamic counterpart of the standard model of computation
on read-only inputs (see, e.g., \cite{EHK15}). The graph is stored
externally and is available only through a read-only oracle; the algorithm
has a separate read--write working memory. All writable memory counts
toward the space bound, including both the information retained between
updates and the temporary storage used while processing an update. The
storage and maintenance of the oracle are not charged to the algorithm.
Thus the resource we measure is exactly the memory needed to compute with
the evolving input, beyond the input representation itself. A word
contains $\Theta(\log n)$ bits.

The input is a simple undirected graph on the fixed vertex set $[n]$,
initially with no edges, together with two fixed terminals $s,t\in[n]$.
At initialization, the algorithm is given $n$, $s$, $t$, and $S$. Each update is announced before it takes effect: while
processing an insertion or deletion, the algorithm can inspect the graph as
it was before the update, and it must report whether $s$ and $t$ are
connected after the update; the oracle then advances to the updated graph.
In the \emph{incremental} setting all updates are insertions, while the
\emph{fully dynamic} setting permits both insertions and deletions.

Formally, we consider the following input representations. In the
\emph{list model}, $\operatorname{Degree}(v)$ returns the degree of $v$,
and $\operatorname{Neighbor}(v,j)$ returns the $j$th neighbor of $v$ in an
arbitrary ordering; the oracle does not support reverse lookup of a
neighbor's position. In the \emph{matrix model},
$\operatorname{Adjacent}(u,v)$ tests whether $\{u,v\}$ is an edge. Each
list-model query can be simulated by $O(n)$ matrix-model queries, by
scanning a row of the adjacency matrix.

We measure three quantities of an algorithm: its \emph{space}, the number
$S$ of words of its working memory; its \emph{queries}, the number of
oracle calls it makes while processing an update; and its \emph{time}, the
number of word operations, including oracle calls, it performs while
processing an update. In particular, the number of queries never exceeds the time. We prove our lower bounds for the number of queries and our upper bounds for the time, so each of them holds for both measures. A time bound $T$ is \emph{worst-case} if every
update costs at most $T$, and \emph{amortized} if, for every $k\ge1$, the
first $k$ updates cost at most $kT$ in total. An expected amortized bound
requires $\mathbb E[\operatorname{cost}(\sigma)]\le|\sigma|T$ for every
fixed update sequence $\sigma$. Since the graph is initially empty, we
allow no initialization time: any such work could equally be done during
the first update. A randomized algorithm may additionally read
independent, uniformly random words. 

Our randomized upper bounds hold against output-adaptive adversaries,
which may choose updates and adjacency-list orderings based on previous
connectivity answers, but do not observe the algorithm's internal
state, random choices, oracle queries, or running times.
As explained in \cref{sec:upper-bound}, it suffices to establish the
stated guarantees for every fixed update sequence.
Unless stated otherwise, all answers and
resource bounds hold simultaneously with high probability. %

\paragraph{Modeling choices.}
The model measures the auxiliary space of the algorithm (the monitor of
our examples) and the cost of maintaining its answer, not the total storage
used by the graph's owner. Our formal interface serializes updates and
gives the algorithm a consistent view of the old graph while it processes
an announced edge change. An implementation would need suitable
synchronization or versioned access, since an asynchronous notification
feed alone does not guarantee this snapshot and ordering. Unit-cost oracle
operations abstract away network latency, batching, and block transfers.
Finally, the empty initial graph matters: construction is part of the
update sequence, and amortized guarantees cover every prefix, whereas
monitoring an already populated graph would add a preprocessing phase over
its existing edges, which our update bounds do not cover. These choices
isolate the computational tradeoff, leaving system-specific storage and
concurrency costs outside the model.

\paragraph{Other interfaces.}
Other oracle interfaces are possible, and some of them move state from the
algorithm to the oracle. For example, an adjacency list might be accessible
only through an iterator, as lists are in many programming languages: an
operation $\operatorname{NextNeighbor}(v)$ returns the neighbor of $v$
that follows the one it returned last, and the oracle stores the current
position in each list. Such an interface restricts access, since a
uniformly random neighbor, which our random-walk algorithms obtain with a
single $\operatorname{Neighbor}$ probe, can then take time proportional
to the degree. At the same time, the positions of the iterators are state
that the algorithm modifies, and might use to store information, but that
is not charged to its workspace. A model with iterators would therefore
have to specify whether these positions count toward the space bound and
how they behave under updates. Our list model provides the indexed access
to adjacency arrays that is commonly assumed for read-only graphs (see,
e.g., \cite{EHK15}), and we leave other interfaces to future work.

\subsection{Our Results}\label{sec:results}

Our incremental algorithms achieve the following bounds.

\begin{maintheorem}[Upper bounds]
\label{thm:upper}
Let $c>0$ be a constant, and let $n\ge2$ and $1\le S\le n$ be integers.
The following hold for \STConn in the incremental setting with worst-case per update guarantees:
\begin{enumerate}[label=(\alph*),ref=\themaintheorem(\alph*),nosep,topsep=3pt]
    \item \label[maintheorem]{thm:largespace}
    There is an algorithm that, during its $i$-th update, uses space $\widetilde O(S+\min\{n,i\}/S)$ and takes time per update of $\widetilde O(n(1+\min\{n,i\}/S))$ in the matrix model and $\widetilde O(\min\{n^2/S^2,1+i/S\})$ in the list model.
    \item \label[maintheorem]{thm:metropolis}
    There is an algorithm that, during its $i$-th update, uses space $\widetilde O(S)$ and takes time per update of $\widetilde O(1+\min\{n,i\}^2/S^2)$ in the list model.
\end{enumerate}
And the following in the incremental setting with amortized guarantees:
\begin{enumerate}[label=(\alph*),ref=\themaintheorem(\alph*),nosep,topsep=3pt, resume]
    \item \label[maintheorem]{thm:smallspace}
    There is an algorithm that, during its first $k$ updates, for every $k\ge1$, uses space $\widetilde O(S)$ and takes total time $\widetilde O(nk(1+\min\{n,k\}/S))$ in the matrix model and $\widetilde O(k(1+\min\{n,k\}/S))$ in the list model.
\end{enumerate}
All algorithms are randomized and, for every update sequence, with probability at least $1-n^{-c}$, all of their answers are correct.
\end{maintheorem}

Observe that
in the matrix model, \cref{thm:largespace,thm:smallspace} have the same time bound $\softO(n(1+\min\{n,i\}/S))$ per update, in the worst case and amortized, respectively, but only \cref{thm:smallspace} uses space $\softO(S)$ for all $S$, whereas \cref{thm:largespace} does so only if $S\ge\sqrt n$ or $i\le S^2$.
In the list model, for $S\ge\sqrt n$, \cref{thm:largespace} uses space $\softO(S)$ and is never slower than \cref{thm:metropolis}, so \cref{thm:metropolis} matters for $S<\sqrt n$.
Its worst-case time $\softO(1+\min\{n,i\}^2/S^2)$ exceeds the amortized time $\softO(1+\min\{n,k\}/S)$ of \cref{thm:smallspace} by a factor of up to $n/S$, and this gap is inherent: by \cref{thm:lb-list-wc}, every algorithm with $\Theta(S\log n)$ bits of memory needs worst-case time $\widetilde\Omega(n^2/S^2)$ per update in the list model, so no worst-case algorithm can match \cref{thm:smallspace}.

All of our algorithms follow a common framework. Like the static algorithm
of Broder et al.~\cite{BKRU94}, they maintain a small set of
\emph{landmark} vertices, together with a disjoint-set structure that
records which landmarks are connected. To process an insertion, they
locate a landmark in the component of each endpoint, or certify that the
component contains none, and merge the corresponding classes. The
algorithms differ in how they search for landmarks with little memory
(see also Section~\ref{sec:overview}):
\begin{itemize}
  \item The algorithm of Theorem~\ref{thm:largespace} searches by breadth-first
search and stores the discovered vertices in $\softO(\min\{n,i\}/S)$
additional words, so its space is $\softO(S)$ whenever $S\ge\sqrt n$ or
$i\le S^2$.
  \item The algorithm of Theorem~\ref{thm:metropolis} searches with Metropolis walks, which remember only their current vertex, avoiding the need to store the entire search frontier.
  \item Finally, the algorithm of Theorem~\ref{thm:smallspace}, our main
algorithmic result, searches with random walks and combines several
ideas, which allow it to add landmarks in regions where they are sparse;
its analysis requires several more sophisticated arguments based on
effective resistances. Its total
time over $m$ insertions is $\softO(mn/S)$, which Feige's static
tradeoff~\cite{Feige97} needs for a \emph{single} query on the final
graph. As $S$ grows, this bound interpolates between two extremes: with
polylogarithmic workspace, it is $\softO(mn)$, the time of a single random
walk~\cite{AKLLR79} on the final graph, and with $\softO(n)$ words, it is
$\softO(m)$, nearly matching the disjoint-set structure.
\end{itemize}

\paragraph{Worst-case lower bounds.}
Our lower bounds hold for working memory of $B$ bits, where
$B=\Theta(S\log n)$ for a budget of $S$ words, as long as $B\le cn$ for a
small constant $c>0$, and $B\le cn/\log n$ for Theorem~\ref{thm:lb-matrix-wc}. Our lower bounds assume that the time bounds hold on every execution. Over all updates, the worst-case time of
Theorem~\ref{thm:largespace} is $\softO(n^2/S)$ in the matrix model, and
this is optimal.

\begin{restatable}[Worst-case lower bound, matrix model]{maintheorem}{matrixwclb}
\label{thm:lb-matrix-wc}
There is a constant $c>0$ such that every randomized algorithm for
\STConn in the incremental setting and the matrix model with
$B=\Theta(S\log n)\le cn/\log n$ bits of working memory such that, for every
update sequence, with probability at least $2/3$, all of its answers are
correct, needs worst-case time $\widetilde\Omega(n^2/S)$ per update.
\end{restatable}

As the upper and lower bounds match (up to
polylogarithmic factors)  for $\sqrt n\le S=O(n/\log^2 n)$, Theorem~\ref{thm:largespace} is optimal in the matrix model up to
polylogarithmic factors and the lower bound is tight.
The situation is similar for Theorem~\ref{thm:metropolis} and the list model: 
The next theorem shows that the bound of
Theorem~\ref{thm:metropolis} for the list model is optimal up to polylogarithmic factors
for all $S=O(n/\log n)$.

\begin{restatable}[Worst-case lower bound, list model]{maintheorem}{listwclb}
\label{thm:lb-list-wc}
There is a constant $\eta>0$ such that every randomized algorithm for
\STConn in the incremental setting and the list model with
$B=\Theta(S\log n)\le \eta n$ bits of working memory such that, for every
update sequence, with probability at least $3/4$, all of its answers are
correct, needs worst-case time $\widetilde\Omega(n^2/B^2)$ per update, on
sequences of $\Theta(n^2/B)$ insertions.
\end{restatable}

Note that the hard instances consist of $\Theta(B)$ cliques of size
$\Theta(n/B)$, built with $\Theta(n^2/B)$ insertions. Thus, the lower bound
applies only after many insertions, and our upper bounds show that this is
necessary, as earlier insertions can be processed faster; once the lower
bound applies, it is tight by Theorem~\ref{thm:metropolis}. In detail, by
Theorem~\ref{thm:metropolis}, the $i$th insertion takes time
$\softO(1+i^2/S^2)$, which stays below $n^2/S^2$ until
$i=\widetilde\Omega(n)$, so every hard sequence has length
$\widetilde\Omega(n)$. Furthermore, for $S\ge\sqrt n$, the algorithm of
Theorem~\ref{thm:largespace} uses space $\softO(S)$ and takes time
$\softO(1+i/S)$ for the $i$th insertion, which stays
below $n^2/S^2$ until $i=\widetilde\Omega(n^2/S)$. Thus, for $S\ge\sqrt n$,
the length of our hard sequences is optimal up to polylogarithmic factors.
In this sense, the lower bound is also optimal with respect to the number of
insertions, which, as the graph is initially empty, equals the number of
edges of the graph, i.e., its sparsity.

\paragraph{Amortized lower bound for fully dynamic algorithms.}
The amortized bounds of Theorem~\ref{thm:smallspace}, in either model, are
optimal for algorithms that also support deletions. This is our main lower
bound, which we state precisely below. Probabilities and
expectations are over the algorithm's randomness.

\begin{restatable}[Fully dynamic amortized lower bound]{maintheorem}{fdlower}\label{cor:fd-lower}
    There exist constants $\eta,c>0$ such that for any randomized algorithm $\calA$ that uses at most $B$ bits of memory and at most $T$ adjacency probes per update (amortized) for \STConn in the fully dynamic setting, such that $B\leq \eta n$ and $BT\leq cn^2$, there exists an input sequence $\sigma$ of length $\Theta(n)$ such that the algorithm answers all queries correctly with probability less than $2/3$ and the graph is a forest throughout the entire stream.
    For the adjacency-list model, the same lower bound holds for $BT\leq cn$.
\end{restatable}

Translating a budget of $S$ words into $B=\Theta(S\log n)$ bits,
Theorem~\ref{cor:fd-lower} gives amortized time $\Omega(n^2/(S\log n))$ in the
matrix model whenever $\Theta(S\log n)\le\eta n$. Since each list query can
be simulated by $O(n)$ matrix queries with $O(1)$ additional words, it also
gives amortized time $\widetilde\Omega(n/S)$ in the list model. Thus no
fully dynamic algorithm can beat the amortized bounds of
Theorem~\ref{thm:smallspace}, in either model.

The lower bound is polynomial throughout its space range: with polylogarithmic
workspace, every fully dynamic algorithm must make $n^{2-o(1)}$ amortized
probes per update, and even with $\eta n$ bits it must make $\Omega(n)$.
This is in sharp contrast with the conventional setting, in which fully
dynamic connectivity admits polylogarithmic update and query times
\cite{HLT01,HHKPT23} albeit with $\Omega(m)$ space and the cell-probe lower bounds for the time per operation of
P{\u a}tra{\c s}cu and Demaine, which are logarithmic, even on paths \cite{PD06}.
More generally, the strongest unconditional cell-probe lower bound known
for any dynamic problem is only $\Omega((\log n/\log\log n)^2)$
\cite{Lar12}, a bound recently attained also for problems with Boolean
answers \cite{KoBool26} and for incremental reachability in directed
acyclic graphs \cite{Ko26}. Due to the workspace constraint in our models, the models differ (see
Appendix~\ref{sec:related}), but the comparison is telling: in our matrix
model, restricting the workspace to at most $\eta n$ bits makes fully
dynamic $s$--$t$ connectivity polynomially hard, even on forests.

\paragraph{Fully dynamic forests.}
Theorem~\ref{cor:fd-lower} is tight on the class of its hard instances: for update sequences under which the graph stays a forest, we obtain a matching upper bound.

\begin{restatable}[Fully dynamic forests]{maintheorem}{forestsUB}\label{thm:forests}
Let $n\ge2$ and $1\le S\le n$ be integers.
There is a deterministic algorithm for \STConn on forests in the fully dynamic setting that uses $O(S)$ words of space and takes worst-case time $O(n^2/S)$ per update in the matrix model, and $O((n/S)(\Delta+\log S))$ in the list model if the maximum degree of the forest never exceeds $\Delta\ge1$.
\end{restatable}

For $B=\Theta(S\log n)\le\eta n$ bits, this matches Theorem~\ref{cor:fd-lower} up to a logarithmic factor in the matrix model, and up to polylogarithmic factors and a factor of $\Delta$ in the list model, where $\Delta\le3$ on the hard instances of Theorem~\ref{cor:fd-lower}, even deterministically and in the worst case; we regard it as a further result, showing the tightness of Theorem~\ref{cor:fd-lower}, and prove it in Section~\ref{sec:forests}.
The algorithm sparsifies Euler-tour trees~\cite{HK99,Tarjan97}: it stores only $O(S)$ oriented edges, spaced $\Theta(n/S)$ apart along the Euler tours of the trees, and recovers the others by navigating the forest using the oracle.
In contrast, dynamic trees~\cite{ST83}, Euler-tour trees, and top trees~\cite{AHLT05} support links, cuts, and connectivity queries in logarithmic time but use $\Theta(n)$ words, and succinct dynamic trees~\cite{FM11,NS14,GW21} still encode the whole forest in $2n+o(n)$ bits.
The closest precedent is the macro--micro decomposition of Gagie and Wild's succinct Euler-tour trees~\cite{GW21}, which stores the micro trees, whereas we do not store the parts of the Euler tours between landmarks at all.

\paragraph{Amortized lower bounds for incremental algorithms.}
For incremental algorithms, we show the following amortized lower bounds.

\begin{maintheorem}[Incremental amortized lower bound, informal version of \cref{thm:io-amortized-lb}]
\label{thm:lb-amortized}
There is a constant $c>0$ such that every randomized algorithm for
\STConn in the incremental setting with $B=\Theta(S\log n)\le cn$ bits of
working memory such that, for every update sequence, with probability at
least $2/3$, all of its answers are correct, needs amortized time
$\widetilde\Omega(n\sqrt{n/S})$ per
update in the matrix model, on sequences of $\Theta(n)$ insertions, and
hence amortized time $\widetilde\Omega(\sqrt{n/S})$ in the list model.
\end{maintheorem}

\paragraph{Comparing the bounds.}
Table~\ref{tab:results} and Figure~\ref{fig:landscape} collect our bounds.
Within the range of our lower bounds, the worst-case picture is complete
up to polylogarithmic factors, except for the matrix model with
$S<\sqrt n$. In both models, our amortized upper bounds for \emph{incremental
connectivity} match the amortized lower bounds for \emph{fully dynamic
connectivity,} up to polylogarithmic factors. Any faster incremental
algorithm must therefore exploit the absence of deletions, whereas any
fully dynamic algorithm that matches our incremental bounds would be
optimal.

\definecolor{BoundUpper}{HTML}{2265AC}
\definecolor{BoundLower}{HTML}{18835C}
\definecolor{BoundConjecture}{HTML}{FFEF9A}
\providecommand{\UpperBoundLabel}{\textcolor{BoundUpper}{\textbf{Upper}}}
\providecommand{\LowerBoundLabel}{\textcolor{BoundLower}{\textbf{Lower}}}
\providecommand{\ConjecturedBoundLabel}{\colorbox{BoundConjecture}{\strut Conjecture}}

\begin{table}[t]
\centering
\small
\caption{Our time bounds for general graphs with $S$ words of workspace,
up to polylogarithmic factors; lower bounds are green and upper bounds
blue, and numbers in parentheses refer to theorems. Lower bounds hold for
$B=\Theta(S\log n)\le cn$ bits, for a small constant $c>0$, and for
$B\le cn/\log n$ bits in the case of Theorem~\ref{thm:lb-matrix-wc}. The incremental
amortized lower bounds use sequences of $\Theta(n)$ insertions, and the
list-model worst-case lower bound uses $\Theta(n^2/B)$ insertions. 
A dash marks an entry not
summarized in this table; amortized lower bounds also bound
worst-case time, and lower bounds for incremental algorithms also apply to
fully dynamic ones. The upper bounds shown hold throughout the update
sequence;
Theorem~\ref{thm:upper} gives smaller bounds for early updates.
$^\dagger$With $\softO(S+\min\{n,i\}/S)$ words during the $i$th update,
that is, $\softO(S)$ words for $S\ge\sqrt n$; in the list model, the same
algorithm takes worst-case time $\softO(\min\{n^2/S^2,1+i/S\})$ for the
$i$th update.}
\label{tab:results}
\setlength{\tabcolsep}{5pt}
\def\introtabref#1{\,{\scriptsize(#1)}}
\begin{tabular}{@{}ll>{\color{BoundLower}}c>{\color{BoundUpper}}c>{\color{BoundLower}}c>{\color{BoundUpper}}c@{}}
\toprule
\multirow{2}{*}[-0.55ex]{\textbf{Updates}} & \multirow{2}{*}[-0.55ex]{\textbf{Model}}
 & \multicolumn{2}{c}{\textbf{Amortized time}} & \multicolumn{2}{c}{\textbf{Worst-case time}}\\
\cmidrule(lr){3-4}\cmidrule(l){5-6}
 & & \LowerBoundLabel & \UpperBoundLabel & \LowerBoundLabel & \UpperBoundLabel\\
\midrule
\multirow{2}{*}{Incremental} & List
 & $\sqrt{n/S}$\introtabref{\ref{thm:lb-amortized}}
 & $n/S$\introtabref{\ref{thm:smallspace}}
 & $n^2/S^2$\introtabref{\ref{thm:lb-list-wc}}
 & $n^2/S^2$\introtabref{\ref{thm:metropolis}}\\
 & Matrix
 & $n\sqrt{n/S}$\introtabref{\ref{thm:lb-amortized}}
 & $n^2/S$\introtabref{\ref{thm:smallspace}}
 & $n^2/S$\introtabref{\ref{thm:lb-matrix-wc}}
 & $n^2/S^{\,\dagger}$\introtabref{\ref{thm:largespace}}\\
\midrule
\multirow{2}{*}{Fully dynamic} & List
 & $n/S$\introtabref{\ref{cor:fd-lower}} & --- & --- & ---\\
 & Matrix
 & $n^2/S$\introtabref{\ref{cor:fd-lower}} & --- & --- & ---\\
\bottomrule
\end{tabular}
\end{table}

\paragraph{Scope}
To our knowledge, time--workspace tradeoffs for exact undirected
fixed-pair connectivity have not previously been studied in the dynamic
oracle setting defined above, in which the algorithm can probe the current
graph and all of its writable memory, persistent and temporary, is charged
to a single bound. Our claim of novelty is specific to this problem: we do
not claim novelty for the general read-only-input workspace paradigm,
which has substantial precedent. Appendix~\ref{sec:related} discusses
related work in detail, including static time--space tradeoffs,
deterministic small-space algorithms, compact connectivity indexes, lower
bounds for restricted machines, dynamic connectivity and its lower bounds,
streaming and sketching, sensitivity oracles, directed and quantum
connectivity, and models with read-only inputs and auxiliary memory.

\subsection{Technical Overview}\label{sec:overview}

We proceed to give some intuitive (but rather informal) description of a selection of our results.

\subsubsection*{Upper Bounds}

For the presentation of the upper bounds, we fix integers $n\ge2$ and a parameter $1\le S\le n$, two vertices $s,t\in[n]$, and an incremental update sequence $(e_i=u_iv_i)_{i=1}^M$ to an initially empty graph $G = ([n], E)$, which is fixed in advance and thus independent of the algorithm's randomness.
We denote by $G_k$ the graph after the first $k$ insertions, and update $i$ is processed with oracle access to $G_{i-1}$; after it, we need to report whether $s$ and $t$ are connected in $G_i$.

\paragraph{Overview of \cref{thm:largespace}.}
Without space constraints, a union-find structure on all vertices solves the problem: an insertion $\{u,v\}$ merges the classes of $u$ and $v$ in $\widetilde O(1)$ time, without querying the graph, and $s$ and $t$ are connected if and only if they lie in the same class.
With less space, a natural idea is to keep the same information only for a set $B$ of \emph{landmarks}, consisting of $s$, $t$, and $S$ uniformly random vertices, hoping that every vertex is close to $B$.
A union-find structure on $B$ then records which landmarks are connected.
At an insertion $\{u,v\}$, we look for a landmark near each endpoint by a breadth-first search (BFS) in the graph before the insertion, and merge the classes of the two landmarks found; if a BFS exhausts its (connected) component without finding one, nothing needs to be merged.

The hope is justified: $B$ intersects every fixed set of $r\approx(n/S)\log n$ vertices  with high probability, and, thus, $B$ intersects the first $r$ vertices that a BFS discovers, i.e., the $r$ vertices closest to its start, as they are determined by the graph alone.
Hence, a BFS discovers at most $r=\widetilde O(n/S)$ vertices, and, as it stores them, the space is $\widetilde O(S+n/S)$.
The time of a BFS is at most $r$ times the cost of scanning a discovered vertex, i.e., examining its neighbors.
In the matrix model, scanning a vertex means reading its row of $n$ entries, so an insertion takes $\widetilde O(rn)=\widetilde O(n^2/S)$ time.
In the list model, scanning a vertex examines at most $r+1$ of its neighbors, as each of them is discovered afterwards or ends the BFS, so an insertion takes $\widetilde O(r^2)=\widetilde O(n^2/S^2)$ time.

For the early insertions, we can do better in the list model, where the time of a BFS is also at most the total degree of the vertices it scans.
To keep this small, we also \emph{sample} the $i$-th insertion, i.e., add both of its endpoints to $B$, with probability $S/i$.
This adds only $\sum_iS/i=\widetilde O(S)$ vertices to $B$, but, as each earlier insertion was sampled with probability at least $S/i$, a BFS reaches $B$ before it has scanned vertices of total volume more than about $(i/S)\log n$: one of their edges was sampled with high probability.
Hence, the $i$-th insertion takes only $\widetilde O(1+i/S)$ time.
As there can be up to $\binom n2$ insertions, however, this beats $\widetilde O(n^2/S^2)$ only for roughly the first $n^2/S$ of them, so the worst case does not improve.
Sampling edges also makes $B$ dense among the at most $2i$ non-isolated vertices of $G_{i-1}$, so a BFS needs to discover only $\widetilde O(\min\{n,i\}/S)$ vertices before it reaches $B$.
Hence, it uses space $\widetilde O(S+\min\{n,i\}/S)$, which is $\widetilde O(S)$ while $i\lesssim S^2$, and, in the matrix model, time $\widetilde O(n(1+\min\{n,i\}/S))$, which beats $\widetilde O(n^2/S)$ for the first $o(n)$ insertions.

\paragraph{Overview of \cref{thm:smallspace}.}

The space bound $\widetilde O(S+\min\{n,i\}/S)$ of \cref{thm:largespace} depends on $S$ in an unusual way: as $S+n/S\ge2\sqrt n$, it does not allow $S$ to be arbitrarily small once $i\ge n$, and, for $S<\sqrt n$, the algorithm uses space $\widetilde O(\min\{n,i\}/S)$ rather than $\widetilde O(S)$ as soon as $i\gtrsim S^2$.
So, we aim to reduce the space to $\widetilde O(S)$.
The issue is the BFS: it stores up to $r\coloneqq \widetilde O(n/S)$ vertices, while everything else takes $\widetilde O(S)$ words.

The natural low-space counterpoint to replace the BFS with is a random walk.
So we keep $B$ and the union-find structure on it in our memory, and, at each insertion, we look for a landmark in the component of each endpoint by simple random walks, each of which stores only its current vertex and a counter of the length of the walk.

We use two facts about a simple random walk from a vertex $x$ in a (connected) component $C$
(see also \cref{lem:random-walk-facts,fact:resistance}):
\begin{enumerate*}[label=(\roman*)]
    \item the walk visits $x$ about $d(x)R(x,A)$ times before hitting a
    set $A$, where $R(x,A)$ is the \emph{effective
    resistance} from $x$ to $A$ (see \cref{sec:prelims} for its definition, in this overview it suffices to know that $R(x,A)\le\operatorname{dist}(x,A)$);
    \label{tool:visits}
    \item each return to $x$ takes $\operatorname{vol}(C)/d(x)$ steps on
    average.
    \label{tool:return}
\end{enumerate*}

Throughout, $i$ is the index of the current insertion $uv\coloneqq u_iv_i$.
For simplicity, we first aim for amortized time $\widetilde O(n/S)$ per insertion, and refine this bound at the end. However, even this poses three challenges:
\begin{enumerate}[label={(\arabic*)}]
   \item 

A walk revisits vertices, so its length depends on the total degree, not the number of the vertices it explores. With the sampled edges, as in \cref{thm:largespace}, a walk started while processing insertion $i$ hits a landmark
    before exploring vertices of total volume $\widetilde O(i/S)$.
    By~\ref{tool:visits}, it visits each explored vertex $y$ only about
    $d(y)R(y,B)\le d(y)\cdot\widetilde O(n/S)$ times before hitting $B$, so
    it stops after
    $\widetilde O(n/S)\cdot\widetilde O(i/S)=\widetilde O(ni/S^2)$ steps.
    Unfortunately, this bound is too much for every insertion by a factor of $i/S$, but we can still afford it for some ``important'' insertions, as for instance for the sampled ones, as they occur at rate $\approx S/i$.

\item  A walk does not notice when it has exhausted its (connected) component.
    By~\ref{tool:visits}, we know when to give up: if the component of $x$ contains a
    landmark, one lies within distance $\widetilde O(n/S)$ of $x$, as in
    \cref{sec:ub-large}, so the walk hits $B$ after about
    $d(x)R(x,B)=\widetilde O(d(x)n/S)$ visits to $x$. Hence, once the walk
    has visited $x$ more often than this without hitting a landmark, we
    declare the component \emph{landmark-free}.
    Such a walk costs no more than one that finds a landmark: if the
    component $C$ of $x$ is landmark-free, each return to $x$ takes $\operatorname{vol}(C)/d(x)$
    steps on average by~\ref{tool:return}, so the walk stops after
    $\widetilde O(d(x)n/S)\cdot\operatorname{vol}(C)/d(x)
    =\widetilde O((n/S)\operatorname{vol}(C))$ steps. Moreover,
    $\operatorname{vol}(C)=\widetilde O(i/S)$, as otherwise one of its edges
    would likely have been sampled, so this is again
    $\widetilde O(ni/S^2)$. As before, this is too much for most
    insertions, which the third challenge now addresses.

    \item There are two ways in which we can make an insertion
    $\{u,v\}$ cheaper:

    \begin{itemize}
        \item If $u$ and $v$ already lie in a common component $C$, no
        landmark classes need to be merged, so a walk from $u$ may also stop
        when it hits $v$. This can make walks much cheaper, for two reasons:
        \begin{enumerate*}[label=(\alph*)]
            \item by~\ref{tool:visits}, a walk from $u$ visits $u$ only about
            $d(u)R(u,v)$ times before reaching $v$;
            \label{reason:a}
            \item using a clever cycle-removal argument, we can show that the resistances $R(u,v)$ of all insertions {between two already connected nodes} sum to
            $\widetilde O(n)$ (\cref{lem:ub-cumulative-resistance}).
            \label{reason:b}
        \end{enumerate*}

        By~\ref{reason:b}, we can afford walks of length
        $T\approx(i/S)R(u,v)$ for the $i$-th insertion, as these costs sum to 
        $\sum_{i\le k}(i/S)R(u_i,v_i)=\widetilde O(kn/S)$ steps over the
        first $k$ insertions. So suppose we let a walk run only as long as we can afford it, i.e., we cut it off after $T$ steps.\footnote{As we know neither $R(u,v)$ nor
        whether $u$ and $v$ are connected, we in fact try $T=1,2,4,\ldots$, until a walk reaches the other endpoint, certifies a landmark-free
        component, or we found a landmark. Moreover, for technical reasons, rather than cutting a
        walk off after exactly $T$ steps, we kill it before each step with
        probability $1/(T+1)$, so that it makes about $T$ steps on
        average.} If it is cut off, rather than stopping at $v$ or at a
        landmark, then, by~\ref{reason:a}, it visited $u$ at most about
        $d(u)R(u,v)\approx d(u)TS/i$ times. Hence, it returned to $u$ only
        every $T/(d(u)TS/i)=i/(Sd(u))$ steps or so, which, by~\ref{tool:return},
        is typical for a landmark-free region of volume about $i/S$. In this case, we \emph{promote} the
        insertion, which means that we make $u$ and $v$ landmarks, as if it were
        sampled.\footnote{In fact, to boost the success probability, we
        run $O(\log n)$ such walks rather than one, and promote only if all
        of them are cut off.}  Thus, the insertion costs $\widetilde O((i/S)R(u,v))$
        steps, plus $\widetilde O(ni/S^2)$ if it is promoted or sampled, as then the new landmarks $u$ and $v$ each look for a landmark in their component by a walk without cut-off, which stops after $\widetilde O(ni/S^2)$ steps by the first two challenges.

        We can show that also promotions cannot happen too often: intuitively, each one covers a
        landmark-free region of volume about $i/S$, while each insertion
        adds only $2$ to the total volume. As these regions are fuzzy,
        overlap, and change with later insertions, we make it more formal by
        defining the potential $\sum_y d(y)p(y)$, where $p(y)$ is the probability
        that a walk of about $T$ steps from $y$ misses all landmarks: a more involved analysis shows that each
        promotion lowers the potential by $\Omega(i/S)$, while each insertion raises it
        by at most $2$ (\cref{lem:ub-root-potential}). Hence, like
        samplings, promotions occur at rate about $S/i$.

        \item If $u$ and $v$ lie in different components $C_u$ and $C_v$,
        say with $|C_u|\le|C_v|$, we run the same cut-off walks with the
        same $T$ for both endpoints; we need not know which component is smaller or how large it is, as, for $T=1,2,4,\ldots$, we run the walks from $u$ and then those from $v$, and stop at the first $T$ at which a walk ends the insertion, so that, as $T$ doubles, this last $T$ dominates the cost. We charge the cost of the
        insertion, apart from a possible promotion (paid as in the first
        case), to the smaller component $C_u$ only, even though the walks
        from $v$ explore $C_v$. This works because the cost can be bounded
        in terms of $C_u$ alone:
        \begin{itemize}
            \item If $C_u$ is landmark-free, a walk from $u$ certifies this
            by the second challenge once $T$ reaches about
            $(n/S)\operatorname{vol}(C_u)$, which ends the insertion. The
            walks from $v$ are cut off after the same number $T$ of steps,
            so they cost no more, however large $C_v$ is. Hence, the
            insertion costs $\widetilde O((n/S)\operatorname{vol}(C_u))$
            steps, which we charge as $\widetilde O(d(z)n/S)$ to every vertex
            $z\in C_u$.
            \item If $C_u$ contains a landmark, every walk, from $u$ or
            $v$, stops within $\widetilde O(ni/S^2)$ steps by the first
            challenge, so the insertion costs $\widetilde O(ni/S^2)$ steps,
            which we charge to a landmark in $C_u$.
        \end{itemize}
        As $|C_u\cup C_v|\ge2|C_u|$, each vertex lies in a charged component
        at most $\log_2n$ times. As the degrees sum to $2k$ after $k$
        insertions and there are only $\widetilde O(S)$ landmarks, the first
        $k$ insertions are charged $\widetilde O(kn/S)$ and
        $\widetilde O(S)\cdot\widetilde O(nk/S^2)=\widetilde O(kn/S)$ in
        the two cases, respectively.
    \end{itemize}
\end{enumerate}

Finally, as $G_{i-1}$ has at most $2i$ non-isolated vertices and the sampled edges make $B$ dense among them, if the component of $x$ contains a landmark, one lies within distance $\widetilde O(\min\{n,i\}/S)$ of $x$. Hence, we may replace $n$ by $\min\{n,i\}$ in the visit threshold $\widetilde O(d(x)n/S)$ of the second challenge. Moreover, as $G_k$ has at most $2k$ non-isolated vertices, the resistances of the first $k$ insertions sum to only $\widetilde O(\min\{n,k\})$, so the first $k$ insertions take time $\widetilde O(k(1+\min\{n,k\}/S))$. {Altogether, the algorithm of \cref{thm:smallspace} uses $\widetilde O(S)$ words and amortized time $\widetilde O(1+\min\{n,k\}/S)$ per insertion in the list model, and a factor $n$ more in the matrix model, where each step of a walk takes $O(n)$ probes.}

Simulating each list query by a row scan of the adjacency matrix, with $O(n)$ queries and $O(1)$ words, gives the matrix-model bound of \cref{thm:smallspace}.

\subsubsection*{Lower Bounds}

\paragraph{Amortized lower bounds (\cref{cor:fd-lower,thm:lb-amortized})} Here we focus on the fully dynamic amortized lower bound. Although our incremental amortized lower bound uses some different techniques, the high-level strategy of its proof is similar to our fully dynamic lower bound. To show this lower bound, we introduce a hard instance that uses $\Theta(n)$ insertions to build a base graph and then create $\Theta(n)$ \emph{queries}, each of which is one insertion and one deletion such that it is hard for an algorithm with these space and probe restrictions to answer the queries. Our base graph consists of two vertex-disjoint paths such that one of them contains $s$ and neither of them contains $t$. This way if we insert an edge between $t$ and a vertex on one of the two paths, $s$ and $t$ are connected if and only if that vertex is on the same path as $s$. Thus, we aim to make it hard for the algorithm to find out which path an arbitrary vertex is on. We introduce the construction in two steps:

As a first attempt, let the base graph be the following. We define $\lfloor (n-2)/2\rfloor$ fixed slots such that the $i$-th slot is the two fixed vertices $(r^0_i,r^1_i)$. Now for every $i\in[\lfloor (n-2)/2\rfloor-1]$, we take an independent fair bit $z_i$ and if $z_i=0$, we add the edges $\{r_i^0,r_{i+1}^0\}$ and $\{r_i^1,r_{i+1}^1\}$, and otherwise $\{r_i^0,r_{i+1}^1\}$ and $\{r_i^1,r_{i+1}^0\}$, i.e., we switch the tracks if $z_i=1$. We call these bits \emph{switch bits}. Also, the edge $\{s, r_1^0\}$ is in the graph.
In this graph if we insert the edge $\{r^0_i,t\}$, $s$ and $t$ will be connected if and only if $\bigoplus_{j\in[i-1]}z_j=0$. Thus, the problem is simply asking for the parity of prefix sums of $z=(z_1,\ldots,z_{\lfloor{(n-2)/2}\rfloor})$. Note that in this construction for each query we first insert $\{r^0_i,t\}$ for a random $i$ and then remove it after getting the answer of $st$-connectivity to undo the insertion and go back to the base graph. This construction was already used in the $\Omega(\log n/\log\log n)$ lower bound on the time per operation for dynamic connectivity in the cell probe model in 1998~\cite{HF98}, but on its own it is not hard enough: one can store the parity of the prefix of every $(n/B)$-th slot and then for each query $\{r^0_i,t\}$, find the switch bit $z_j$ for every $j$ between the nearest preceding slot with a recorded prefix parity and $i$. Since the vertices are known this can be done with $O(n/B)$ probes even in the adjacency matrix model (worst-case).

In our second attempt (see \cref{fig:construction}), we get around this obstruction by hiding where each switch happens.
We use $d=\lfloor(n-2)/8\rfloor=\Theta(n)$ slots, and we interrupt both tracks in the middle of every slot: slot $j$ now consists of the vertices $\{l^0_j,l^1_j,r^0_j,r^1_j\}$, with no edges between them, where intuitively, the tracks enter the slot at $l^0_j,l^1_j$ and leave it at $r^0_j,r^1_j$.
Consecutive slots are joined by the edges $\{r^0_j,l^0_{j+1}\}$ and $\{r^1_j,l^1_{j+1}\}$ for every $j<d$, and $s$ is adjacent only to $l^0_1$. 
The gap in each slot is bridged by an \emph{item}, a pair of
parallel edges: there are $d$ items, one per slot, where item $i$ consists of the edges $\{u^0_i,v^0_i\}$ and
$\{u^1_i,v^1_i\}$.
A uniformly random bijection $\pi$ decides which item fills which slot, and the switch bit decides whether the tracks pass through it straight or crossed: for every $i,j\in[d]$ with $\pi(i)=j$ and every $b\in\{0,1\}$, the edges $\{l^b_j,u^{b\oplus z_j}_i\}$ and $\{v^b_i,r^b_j\}$ are in the graph.
Now, if we insert the edge $\{v^0_i,t\}$, $s$ and $t$ are connected if and only if $\bigoplus_{j\in[\pi(i)]}z_j=0$. 

Since all vertices are known, the only probes with a nontrivial answer test whether a vertex of item $i$ is
adjacent to a vertex of slot $j$. Such a probe returns ``NO'' unless $\pi(i)=j$, in which case we
call it a \emph{hit}; a hit can reveal $z_j$. Our construction thus uses two ingredients, the permutation $\pi$ and the switch bits $z$, and both matter: Without the permutation, every switch bit can be read with one probe, as in the first attempt. One could also try to make the problem hard without switch bits, for example by hiding the order of the slots. Then, however, the answer is a function of the hidden order alone, and every ``NO'' answer reveals some information about this order, so an algorithm might deduce an answer without ever probing near the queried item, and it is unclear how to bound such deductions. The switch bits avoid this problem: a ``NO'' answer reveals nothing about $z$, as $z$ is independent of $\pi$, so the algorithm learns $z_j$ only by a hit at slot $j$; otherwise, $z_j$ is a fair coin.

We show that answering $\Theta(n)$ queries as explained above with $B$ bits of memory requires $\Omega(n^3/B)$
probes in total, which gives the amortized bound. To do so, we partition the queries into $\Theta(n/B)$ blocks
of $\Theta(B)$ queries each. We view the memory at the start of a block as a $B$-bit advice string, which may
depend arbitrarily on $(\pi,z)$, and show that, for a small enough constant $c>0$, an algorithm with such
advice answers all queries of a block correctly with fewer than $cn^2$ probes only with small probability.
Within a block, we even allow unlimited workspace, so we use the space restriction only for the information
that persists from one block to the next.

\emph{Why a block needs many hits.} Partition the slots into \emph{intervals} of length $\Theta(d/B)=\Theta(n/B)$, and
color them alternately red and blue. Since the queried items are random, with constant probability,
$\Omega(B)$ of the red intervals contain the slot of an item queried during the block. In each such red interval, pick one such slot. Since red and blue intervals alternate, if $j<j'$ are the
picked slots of two consecutive such red intervals, then the range $\{j+1,\dots,j'\}$ contains an entire blue
interval. Moreover, the answers to the queries for the items in slots $j$ and $j'$ are
$\bigoplus_{k\leq j}z_k$ and $\bigoplus_{k\leq j'}z_k$, so by XOR-ing them, a correct algorithm learns the
parity $\bigoplus_{k=j+1}^{j'}z_k$ of the switch bits in this range. Recall that the algorithm learns $z_k$
only by a hit at slot $k$. So if some slot of the range was not hit, its switch bit is a fair coin given
everything the algorithm has seen apart from its memory, and since the parity of the range is the XOR of its
switch bits, this parity is a fair coin as well. The $B$ bits of memory can store the parities of only $O(B)$ such
ranges, so, informally, the algorithm has to hit every slot in most of the $\Omega(B)$ ranges, that is, make
$\Omega(n)$ hits (since each range contains a blue interval of length $\Theta(n/B)$).

\emph{Checkpoints.} The main difficulty in turning this argument into a proof is that an adaptive algorithm decides which slots to hit, so we cannot fix in advance the slots it must hit. We therefore pick a uniformly random \emph{checkpoint} slot in each blue interval, which is
used only in the analysis and is hidden from the algorithm. Since the checkpoints are hidden, an algorithm with few hits hits only a small fraction of the checkpoints. However, intuitively, if the checkpoint of a range is not hit, then the parity of the range is a fair coin
given everything the algorithm has seen apart from its memory. So, informally, if most checkpoints are not
hit, a correct algorithm must guess the parities of most of the $\Omega(B)$ ranges correctly. These are more parities than the $B$ bits of memory can carry, and a second encoding argument shows that this ``correct guess'' happens
only with probability $2^{-\Omega(B)}$. 

\emph{Why hits are expensive.} Since $\pi$ is hidden, each probe is a hit only with probability about $1/d$,
so $L$ probes yield only about $L/d$ hits. Using an encoding argument, we show that even with $B$ bits of
advice, an algorithm with fewer than $cn^2$ probes makes $\Omega(n)$ hits only with probability
$2^{B-\Omega(n)}$. Hence, an algorithm that makes fewer than $cn^2$ probes during a block answers all of its queries correctly only with small probability.

\paragraph{The Worst-Case Lower Bound for List Model (\cref{thm:lb-list-wc})} We reduce from \textsc{Index}. Here, we describe the proof at a high level and defer the graph construction and the details to \cref{sec:wc-list-lb}. In the \textsc{Index} problem, Alice holds $x\in\{0,1\}^k$, Bob holds an index $i\in[k]$, and Bob must output $x_i$ after a single message from Alice. This message is known to require $\Omega(k)$ bits for Bob to succeed with probability at least $2/3$~\cite{kremerNiRo99}. We use $k=\Theta(B)$. In our reduction, Alice inserts the edges of a graph encoding $x$, and Bob inserts one edge such that $s$ and $t$ become connected if and only if $x_i=1$. Typically, in such reductions for streaming space lower bounds, Alice runs the algorithm on her insertions and sends its memory state, from which Bob continues the run, so the \textsc{Index} lower bound on the message length is a lower bound on the space. This standard approach fails here, as the algorithm's probes during Bob's insertion read the graph, which encodes $x$. So Alice, who does not know $i$, simulates all $k$ possible insertions of Bob and sends the state together with enough information to answer all of the probes of these simulations. Then, Bob continues from that state with the insertion of the edge corresponding to $i$ and uses Alice's message to answer the algorithm's probes. This run, which is identical to Alice's simulation for $i$, yields $x_i$ with probability at least $2/3$. Thus, the message accounts for both the space and the time of the algorithm, and it remains to show that it has only $O(B)$ bits, which is too few for \textsc{Index} if the constant in $k=\Theta(B)$ is large enough.

\emph{Why the message is short.} The state takes $B$ bits, so it remains to encode the probe answers compactly. For this, we hide the bit $x_j$ in the graph as well as possible. Bit $x_j$ is represented by a \emph{gadget} consisting of two cliques, on the vertices $A_{j,1},\dots,A_{j,a}$ and $B_{j,1},\dots,B_{j,a}$, where $a=\Theta(n/B)$. If $x_j=0$, the two cliques are disjoint. If $x_j=1$, we pick a pair $\{u,v\}\subseteq[a]$ uniformly at random among the $\Theta(n^2/B^2)$ pairs, and replace the edges $\{A_{j,u},A_{j,v}\}$ and $\{B_{j,u},B_{j,v}\}$ by the two \emph{cross edges} $\{A_{j,u},B_{j,v}\}$ and $\{A_{j,v},B_{j,u}\}$, which connect the two cliques (see~\cref{fig:wc-reduction}). Intuitively, the construction ensures that a probe reveals information about $x$ only if it finds a cross edge. Thus, besides the state, Alice only sends the indices $j$ of the gadgets in which some simulated probe found a cross edge. Informally, if every insertion takes at most $cn^2/B^2$ probes, the $k=\Theta(B)$ simulations make $O(n^2/B)$ probes, and since each cross edge is hidden among $\Theta(n^2/B^2)$ candidates, these find cross edges in only $O(B)$ gadgets with high probability, even given the $B$ bits of the state, so the message has only $O(B)$ bits.

\section{Preliminaries}
\label{sec:prelims}

\subsection*{Problem and Model Definitions.}

In the standard model of space-bounded computation, an algorithm runs on a word RAM whose input resides in a read-only memory and whose read--write working memory consists of $S$ words.
Its running time is the number of word operations, including reads of the input, and a time--space tradeoff asks how the running time must grow as $S$ decreases.

We study the dynamic counterpart of this model: the input changes through a sequence of updates, and the algorithm has to maintain the answer to a problem throughout these updates.
In our model the input remains read-only and outside of the working memory, which is the only state that the algorithm carries from one update to the next.
As the input changes over time, we require that any read access by the algorithm to the input must be performed by a query to an oracle that has access to the current input instance.

\begin{problembox}{\OracleModel}
    \textbf{Initialization.} The algorithm receives the space parameter $S$ together with some parameters of the problem instance $\mathcal I$, which are determined by the problem to be solved on $\mathcal I$ and include its size $n$.

    \textbf{Space.} The algorithm has a read--write working memory consisting of $S$ words of $\Theta(\log n)$ bits each.

    \textbf{Oracle.} The instance $\mathcal I$ is stored outside of the working memory, and the algorithm can inspect it only through \emph{queries} to a read-only oracle. The type of queries that the oracle supports is a parameter of the model.

    \textbf{Updates.} The updates to $\mathcal I$ arrive one at a time, each arrives after the algorithm has ended its processing of the prior update, and each update is communicated to the algorithm before it takes effect in the oracle. More specifically, while processing an update, the algorithm has oracle access to $\mathcal I$ as it was before the update, and it must report the answer of the problem for $\mathcal I$ as it is after the update. Once the algorithm has ended its processing of an update, the update takes effect in the oracle.
\end{problembox} 

We consider the following two models for the oracle:
\begin{itemize}
    \item \emph{Adjacency-list model} (or \emph{list model} for short). The neighbors of each vertex $v$ are listed in an arbitrary order as $w_1,\ldots,w_{d(v)}$. For $v\in[n]$, the query $\Degree(v)$ returns $d(v)$, and, for $j\in[d(v)]$, the query $\Neighbor(v,j)$ returns $w_j$. 
    \item \emph{Adjacency-matrix model} (or \emph{matrix model} for short). For $u,v\in[n]$, the query $\Adjacent(u,v)$ returns $1$ if $uv\in E$, and $0$ otherwise. Note that each query of the list model can be simulated by $n$ queries of the matrix model{, and that all incident edges of a vertex can be traversed using $n$ adjacency matrix queries.}
\end{itemize}

\paragraph{Complexity Measures.}
We measure the following three quantities of an algorithm:
\begin{itemize}
    \item \emph{Space}: the number $S$ of words of its working memory.
    \item \emph{Queries}: the number of queries that it makes to the oracle while processing an update.
    \item \emph{Time}: the number of word operations (including queries) that it performs while processing an update. A time bound $T$ can be \emph{worst-case} (every update costs at most $T$) or \emph{amortized} (for every $k\ge1$, the first $k$ updates cost at most $kT$ in total).
\end{itemize}

Note that, since (in our specific graph case) we always start with an empty graph, we do not allow initialization time as any such work could equally be done during the first update.
Moreover, a \emph{randomized} algorithm may additionally read independent, uniformly random words.
Next, we define the problem that we solve in this model, i.e., $s$-$t$ connectivity in simple undirected graphs.

\begin{problembox}{\STConn}
    \textbf{Initialization.} The instance $\mathcal I$ is a graph $G=([n],E)$, which is initially empty. Thus, at initialization, the algorithm is given the size $n\ge2$, two vertices $s,t\in[n]$, and $S$.

    \textbf{Updates.} Edge insertions and deletions in $G$. In the \emph{incremental} setting, all updates are insertions.

    \textbf{Output.} After every update, the algorithm outputs whether $s$ and $t$ are connected in the current graph.
\end{problembox}

\subsection*{Notation and Conventions.}

\paragraph{Graph Notation.}
Given a simple undirected graph $G=(V,E)$, we denote with $N_{G}(v)$ the neighborhood of $v$, and set $d_{G}(v) \coloneqq |N_G(v)|$. When $G$ is clear from the context, we may omit $G$ from the subscript.
For $U\subseteq V$, write $\operatorname{vol}(U)\coloneqq\sum_{v\in U}d(v)$.
For $U\subseteq V$ and $e\notin E$, let $G[U]$ be the subgraph of $G$ induced by $U$, and $G+e\coloneqq(V,E\cup\{e\})$; for $e\in E$, let $G-e\coloneqq(V,E\setminus\{e\})$.
For $x\in V$, let $C_G(x)\subseteq V$ be the vertex set of the connected component of $G$ containing $x$.
For $x\in V$ and $A\subseteq V$, let $\operatorname{dist}_G(x,A)$ be the length of a shortest path from $x$ to a vertex of $A$, {where we consider all edges of $G$ to have length 1} and $\operatorname{dist}_G(x,y)\coloneqq\operatorname{dist}_G(x,\{y\})$.

\paragraph{Random walks.}
Given a simple undirected graph  $G=(V,E)$, a \emph{simple random walk} $(X_j)_{j\ge0}$ moves to a
uniform neighbor at each step.
We write $\Pr_z$ and $\Ex_z$ for the probability and expectation given $X_0=z$.
By the strong Markov property, at any almost surely finite stopping time,
conditional on the current vertex, the future is a fresh walk independent
of the past.
For $A\subseteq V$ and a target node $x\in V$, define the following quantities for a random walk (which might start at any node in $V$): 
\[
    \tau(A)\coloneqq\inf\{j\ge0:X_j\in A\},
    \qquad \tau(x)\coloneqq\tau(\{x\}),
    \qquad \tau^+(x)\coloneqq\inf\{j\ge1:X_j=x\},
\]
where $\inf\emptyset=\infty$. 
Note that if the walk starts at $x$ itself, then $\tau(x) = 0$ and $\tau^+(x) > 0$, otherwise $\tau(x) = \tau^+(x)$.
For a walk started at $x\notin A$, let $V_x(A)\coloneqq\sum_{0\le j<\tau(A)}\mathbf{1}[X_j=x]$ count the visits to the starting node $x$ before hitting $A$, including the initial visit.

\paragraph{Effective resistance.}
Let $G = (V,E)$ be an undirected graph.
For $f:V\to\mathbb R$, we define the graph Laplacian and the energy by
\[
    (\mathcal L_G f)(z)\coloneqq\sum_{u\in N(z)}(f(z)-f(u)),
    \qquad
    \mathcal E_G(f)\coloneqq\sum_{\{u,v\}\in E}(f(u)-f(v))^2
    =\sum_{z\in V}f(z)(\mathcal L_G f)(z),
\]
where the last equality follows by summing over edges.
For a node $x \in V$ and a set $A\subseteq V\setminus\{x\}$ with $A\cap C_G(x)\neq\emptyset$, we also define
\[
    R_G(x,A)\coloneqq
    \Big(\textstyle \min_{\substack{f:V\to\mathbb R\\f(x)=1,\ f|_A=0}}
    \mathcal E_G(f)\Big)^{-1},
    \qquad R_G(x,y)\coloneqq R_G(x,\{y\}),
\]
where in the former definition $f:V\to\mathbb R$ ranges over all functions such that $f(x) = 1$ and $f(a) = 0$ for all $a \in A$ (the latter condition is expressed by the shorthand notation $f_{|A} = 0$).
For $x\in A$, we set $R_G(x,A)\coloneqq0$; in particular, $R_G(x,x)=0$.
Note that $R_G(x,y)=R_G(y,x)$.\footnote{To see this, observe that the map $f\mapsto 1-f$ is a bijection between the functions with $f(x)=1,\,f(y)=0$ and those with $f(y)=1,\,f(x)=0$, and it preserves $\mathcal E_G$ since every term $(f(u)-f(v))^2$ is unchanged, so the two minima coincide.}

\begin{fact}
\label{fact:resistance}
Let $G=(V,E)$ be a graph, $x\in V$, $A\subseteq V\setminus\{x\}$ with
$A\cap C_G(x)\ne\emptyset$, and $y\in C_G(x)\setminus\{x\}$.
\begin{enumerate}[label=(\alph*),ref=\alph*,nosep,topsep=3pt]
    \item \label{itm:resistance-rayleigh}
    If $H$ is a subgraph of $G$ with $V(H)=V$ and
    $A\cap C_H(x)\ne\emptyset$, then $R_G(x,A)\le R_H(x,A)$. In
    particular, $R_G(x,A)\le\operatorname{dist}_G(x,A)$.
    \item \label{itm:resistance-restrict}
    If $x\in U\subseteq V$ and no edge of $G$ has exactly one endpoint in
    $U$, then $R_G(x,y)=R_{G[U]}(x,y)$.
    \item \label{itm:resistance-bridge}
    If $e=ab\notin E$ with $b\notin C_G(a)$, then $R_{G+e}(x,y)=R_G(x,y)$.
\end{enumerate}
\end{fact}

\begin{proof}
Part~\eqref{itm:resistance-rayleigh} is Rayleigh's monotonicity law (see also \cite[Proposition~2.4]{lyonsperes2021}), but we prove it here for completeness.
As $E(H)\subseteq E$, every $f$ has
$\mathcal E_G(f)\ge\mathcal E_H(f)$, so the minimum defining $R_G(x,A)$
is at least that defining $R_H(x,A)$.
For the second claim of~\eqref{itm:resistance-rayleigh}, let $\ell\coloneqq\operatorname{dist}_G(x,A)$ and let $x=v_0,v_1,\dots,v_\ell=a$ be a shortest path from $x$ to $A$, so $a\in A$. Let $H$ be the subgraph of $G$ with $V(H)=V$ whose edges are the $\ell$ edges of this path, so $\mathcal E_G(f)\ge\mathcal E_H(f)=\sum_{i=1}^{\ell}(f(v_{i-1})-f(v_i))^2$ for every $f$. For $f$ with $f(x)=1$ and $f|_A=0$, the Cauchy--Schwarz inequality $(\sum_i a_ib_i)^2\le(\sum_i a_i^2)(\sum_i b_i^2)$ with $a_i\coloneqq f(v_{i-1})-f(v_i)$ and $b_i\coloneqq1$ for $i\in[\ell]$ gives
\[
    1=(f(x)-f(a))^2=\Big(\sum_{i=1}^{\ell}(f(v_{i-1})-f(v_i))\Big)^2\le\ell\sum_{i=1}^{\ell}(f(v_{i-1})-f(v_i))^2=\ell\,\mathcal E_H(f)\le\ell\,\mathcal E_G(f).
\]
Hence the minimum defining $R_G(x,A)$ is at least $1/\ell$, so $R_G(x,A)\le\ell$.

For \eqref{itm:resistance-restrict}, note that $y\in C_G(x)\subseteq U$.
For every $f$, $\mathcal E_G(f)\ge\mathcal E_{G[U]}(f|_U)$, with equality
if $f$ is constant outside $U$, as no edge of $G$ has exactly one
endpoint in $U$. So the minima defining $R_G(x,y)$ and $R_{G[U]}(x,y)$
coincide.

For \eqref{itm:resistance-bridge}, $R_{G+e}(x,y)\le R_G(x,y)$ by
\eqref{itm:resistance-rayleigh}. Conversely, as $C_G(a)\ne C_G(b)$ and
$y\in C_G(x)$, we may name the endpoints of $e$ so that
$x,y\notin C_G(b)$. For $f$ with $f(x)=1$ and $f(y)=0$, setting
$f\equiv f(a)$ on $C_G(b)$ keeps these constraints, does not increase
$\mathcal E_G(f)$, as no edge of $G$ has exactly one endpoint in
$C_G(b)$, and makes $f(a)=f(b)$, so
$\mathcal E_{G+e}(f)=\mathcal E_G(f)+(f(a)-f(b))^2=\mathcal E_G(f)$.
Hence $R_{G+e}(x,y)\ge R_G(x,y)$.
\end{proof}

We will use the following well-known properties of random walks and effective resistances.

\begin{lemma}[Random-walk properties, see {\cite[Section~2.2, equations~(2.4)--(2.5)]{lyonsperes2021}, \cite[Propositions~1.19 and~10.7]{levinperes2017}}]
\label{lem:random-walk-facts}
\label{lem:visits-before-hitting}
Let $C$ be a connected component of $G$ with $\nu=\operatorname{vol}(C)>0$,
let $x,y\in C$ be distinct, and let $A\subseteq V\setminus\{x\}$ be such that $A \cap C \neq \emptyset$.
Then
\[
    V_x(A)\sim\Geo\!\Big( \frac{1}{d(x)R_G(x,A)} \Big),
    \qquad \Ex\nolimits_x[\tau^+(x)]=\frac{\nu}{d(x)},
    \qquad \Ex\nolimits_y[\tau(x)]\le\nu R_G(y,x),
\]
where $\Geo(p)$ is the geometric distribution on $\{1,2,\ldots\}$ with
success probability $p$. In particular, we have $\Ex_x[V_x(A)]=d(x)R_G(x,A)$.
\end{lemma}

\paragraph{Metropolis walks.}
A \emph{Metropolis walk} $(Y_j)_{j\ge0}$ on a simple undirected graph $G=(V,E)$ moves from its current vertex $x$ by drawing a uniform neighbor $y$ of $x$ and moving to $y$ with probability $\min\{1,d(x)/d(y)\}$, staying at $x$ otherwise \cite{NOSY10,Kosowski13}.
Thus it moves from $x$ to each neighbor $y$ with probability $\min\{1/d(x),1/d(y)\}$, so the uniform distribution on each component is stationary.
We use $\Pr_z$, $\Ex_z$, $\tau(A)$, and the strong Markov property for Metropolis walks as well.
A Metropolis walk started in a connected component $C$ of $G$ stays in $C$ and is a Metropolis walk on $G[C]$, so the following bounds for connected graphs apply to each component.

\begin{lemma}[Metropolis-walk properties, see {\cite[Lemma~2]{Kosowski13}}]
\label{lem:metropolis-facts}
Let $C$ be a connected component of $G$ with maximum degree $\Delta$, and let $x\in C$. Then, for a Metropolis walk on $G$, every $A$ with $x\in A\subsetneq C$, and every $0<h<6|C|^2$,
\[
    \Ex\nolimits_x[\tau(C\setminus A)]\le(|A|+1)(6|A|+2\Delta),
    \qquad
    \Ex\nolimits_x\Big[\sum_{0\le j<h}\mathbf{1}[Y_j=x]\Big]\le5\sqrt h+2\Delta.
\]
In particular, $\Ex_x[\tau(y)]\le8|C|^2$ for every $y\in C\setminus\{x\}$, by the first bound with $A=C\setminus\{y\}$ and $\Delta\le|C|-1$.
\end{lemma}

\paragraph{Additional Data Structures.}

\begin{lemma}[Union--find]
\label{lem:union-find}
A disjoint-set union structure $D$ maintains a partition of an initially
empty set $B$ under the following operations:
\begin{itemize}[nosep,topsep=3pt]
    \item $\mathtt{MakeSet}(x)$, for $x\notin B$: insert $x$ as a singleton class.
    \item $\mathtt{Find}(x)$, for $x\in B$: return the representative of $x$'s class.
    \item $\mathtt{Union}(x,y)$, for $x,y\in B$: merge their classes, doing nothing
    if they already coincide.
\end{itemize}
Thus, $\mathtt{Find}(x)=\mathtt{Find}(y)$ if and only if $x,y$ belong to the
same class. For at most $k\ge1$ inserted elements, $D$ uses $O(k)$ words
and, given element handles, supports each operation in $O(\log(k+1))$
worst-case time and in $O(\alpha(k))$ amortized time, where $\alpha$ is the
inverse Ackermann function.
For vertex labels $B\subseteq[n]$, a balanced search tree maps labels to
handles in $O(\log(k+1))$ worst-case time per operation using $O(k)$ additional words.
\end{lemma}

\section{Fully Dynamic Amortized Lower Bound}
\label{sec:lower-bound}

In this section, we explain our lower bound on the amortized number of probes for the problem in the fully dynamic setting. We show a lower bound of $\Omega(n^2/B)$ amortized probes for the matrix model (which implies a lower bound of $\Omega(n/B)$ amortized probes in the list model). Formally, we show the following:

\fdlower*

We first give a high-level explanation of the proof. In \cref{sec:fd-construction}, we reduce the PPHP problem defined below to our connectivity problem, and in
   \cref{sec:fd-lb-proof}, we prove a lower bound for PPHP.

\begin{problembox}{Prefix Parity with Hidden Permutation (PPHP)}
\textbf{Instance.} A bijection $\pi\colon[d]\to[d]$ and a vector $z\in\{0,1\}^d$, both uniformly random and
independent. We call the elements of the domain of $\pi$ \emph{items} and the elements of its range
\emph{slots}: $\pi(i)=j$ means that item $i$ is placed in slot $j$, and $z_j$ is the \emph{switch bit} of
slot $j$. The algorithm knows neither $\pi$ nor $z$.

\textbf{Preprocessing.} The algorithm sees $X \coloneqq (\pi,z)$ once and stores a $B$-bit string $\calM(\pi,z)$.

\textbf{Queries.}
In the $i$-th query (called \emph{epoch}), where $i\in[N]$, the algorithm receives $I_i \sim\operatorname{Unif}([d])$ and must output
$y_{I_i}=\bigoplus_{k=1}^{\pi(I_i)} z_k$. While processing queries, we have $B$ bits of space and access to the following oracle: given $i,j\in[d]$, return $P_X(i,j)$, which is $z_j$ if $\pi(i)=j$ and $\bot$ otherwise.

A probe $P_X(i,j)$ with $\pi(i)=j$ is called a \emph{hit}. The cost of an algorithm is the number of oracle probes it makes. The algorithm succeeds if it answers all
$N$ queries correctly.
\end{problembox}
To reduce from the PPHP problem to $s$-$t$ connectivity we construct for every instance $(\pi,z)$ of PPHP a graph $\mathcal{G}_n(\pi,z)$ such that after inserting the edges of this graph, every query on the PPHP problem can be answered using $1$ insertion and $1$ deletion in the graph and querying then whether  the fixed vertices $s$ and $t$ are connected. Moreover, every adjacency-matrix probe in this graph can be answered with at most one oracle probe of PPHP.
Hence, an algorithm for $s$-$t$ connectivity yields an algorithm for PPHP with the same memory and at most
the same number of probes, so it suffices to prove the lower bound for PPHP.

To prove the lower bound for PPHP, we partition the queries into $\Theta(n/B)$ blocks of $\Theta(B)$ queries each. We model the memory available at the start of a block as a $B$-bit advice string, which may be an arbitrary function of the input, and show that, for a small enough constant $c_L>0$, the algorithm answers all queries of the block correctly with fewer than $c_Ld^2$ probes with probability at most $1/4$. This implies that producing the entire output sequence requires $\Omega(n^3/B)$ probes, that is, $\Omega(n^2/B)$ amortized probes per query. We use the space restriction only for the memory that persists between processing two consecutive blocks; while processing a block, the algorithm may use unlimited workspace, which only makes the lower bound stronger.
We show our lower bound for each block in two steps: first, in \cref{cor:cheapHighHit}, we show that it is ``unlikely'' to have a small number of probes, while having a lot of \emph{hit probes} (probes that do not result in a $\bot$ response). Then, in \cref{lem:correctLowHc}, we show that it is unlikely to answer queries correctly without having made a large number of hit probes.

\subsection{Reduction from PPHP} \label{sec:fd-construction}

We first describe the idea. Two vertex-disjoint paths, one of which contains $s$, run side by side through $d$
consecutive \emph{slots}. In slot $j$, they either continue on the same track or switch tracks, depending on
$z_j$, so the path of $s$ has switched tracks an even number of times up to slot $j$ if and only if
$\bigoplus_{k\in[j]}z_k=0$. However, the tracks are not connected inside the slots directly. Instead, each
slot has a gap in the middle, which is filled by one of $d$ \emph{items}, each consisting of two parallel
edges. The bijection $\pi$ decides which item fills which slot: item $i$ fills slot $\pi(i)$, and the tracks
pass through it straight if $z_{\pi(i)}=0$ and crossed if $z_{\pi(i)}=1$. Since the algorithm does not know
$\pi$, it must first find the item in slot $j$ before it can learn $z_j$.

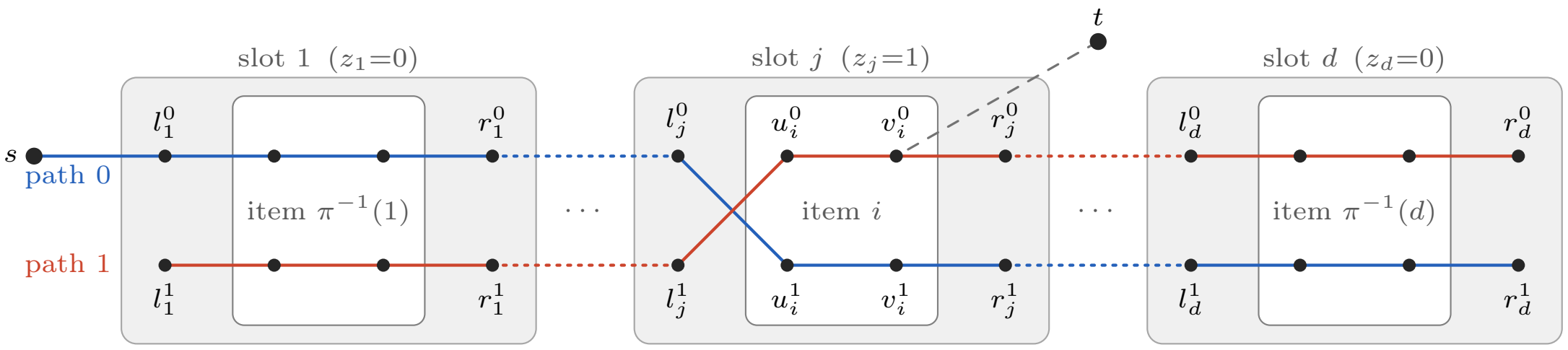
\begin{figure}[h]
    \centering
    \definecolor{arczero}{RGB}{36,96,186}
\definecolor{arcone}{RGB}{204,68,44}
\begin{tikzpicture}[x=1cm, y=1cm, line cap=round, line join=round,
  vtx/.style={circle, fill=black!85, inner sep=0pt, minimum size=3.4pt},
  big/.style={circle, fill=black!85, inner sep=0pt, minimum size=4.4pt},
  lbl/.style={font=\scriptsize, inner sep=1pt},
  up/.style={lbl, above=3.5pt},
  dn/.style={lbl, below=3.5pt},
  tag/.style={font=\scriptsize, text=black!65, inner sep=1pt},
  arc/.style={line width=0.8pt},
  hid/.style={line width=0.8pt, dotted},
  qry/.style={line width=0.6pt, dashed, black!55},
  slot/.style={rounded corners=4pt, fill=black!6, draw=black!35, line width=0.4pt},
  item/.style={rounded corners=3pt, fill=white, draw=black!50, line width=0.4pt},
]
\def\h{0.5}
\foreach \X/\J/\Z/\I in {1.2/1/0/{\pi^{-1}(1)}, 5.9/j/1/i, 10.6/d/0/{\pi^{-1}(d)}} {
  \draw[slot] (\X-0.4,-1.22) rectangle (\X+3.4,1.22);
  \draw[item] (\X+0.62,-1.05) rectangle (\X+2.38,1.05);
  \node[tag, above] at (\X+1.5,1.22) {slot $\J$ \,($z_{\J}{=}\Z$)};
  \node[tag] at (\X+1.5,0) {item $\I$};
}
\coordinate (s) at (0,\h);
\coordinate (t) at (9.75,1.55);
\foreach \X/\k in {1.2/1, 5.9/2, 10.6/3} {
  \foreach \b/\y in {0/\h, 1/-\h} {
    \coordinate (l\b-\k) at (\X,\y);
    \coordinate (u\b-\k) at (\X+1,\y);
    \coordinate (v\b-\k) at (\X+2,\y);
    \coordinate (r\b-\k) at (\X+3,\y);
  }
}
\draw[arc, arczero] (s) -- (l0-1) -- (u0-1) -- (v0-1) -- (r0-1);
\draw[arc, arcone]  (l1-1) -- (u1-1) -- (v1-1) -- (r1-1);
\draw[arc, arczero] (l0-2) -- (u1-2) -- (v1-2) -- (r1-2);
\draw[arc, arcone]  (l1-2) -- (u0-2) -- (v0-2) -- (r0-2);
\draw[arc, arczero] (l1-3) -- (u1-3) -- (v1-3) -- (r1-3);
\draw[arc, arcone]  (l0-3) -- (u0-3) -- (v0-3) -- (r0-3);
\draw[qry] (t) -- (v0-2);
\draw[hid, arczero] (r0-1) -- (l0-2);
\draw[hid, arcone]  (r1-1) -- (l1-2);
\draw[hid, arczero] (r1-2) -- (l1-3);
\draw[hid, arcone]  (r0-2) -- (l0-3);
\node[tag] at (5.05,0) {$\cdots$};
\node[tag] at (9.75,0) {$\cdots$};
\node[big] at (s) {};
\node[big] at (t) {};
\foreach \k in {1,2,3} \foreach \b in {0,1} \foreach \p in {l,u,v,r} \node[vtx] at (\p\b-\k) {};
\node[lbl, left=3pt]  at (s) {$s$};
\node[lbl, above=3pt] at (t) {$t$};
\node[tag, text=arczero, below left, yshift=-1pt] at (0.75,\h) {path 0};
\node[tag, text=arcone, left] at (0.75,-\h) {path 1};
\foreach \k/\J in {1/1, 2/j, 3/d} {
  \node[up] at (l0-\k) {$l^0_{\J}$};
  \node[dn] at (l1-\k) {$l^1_{\J}$};
  \node[up] at (r0-\k) {$r^0_{\J}$};
  \node[dn] at (r1-\k) {$r^1_{\J}$};
}
\node[up] at (u0-2) {$u^0_i$};
\node[dn] at (u1-2) {$u^1_i$};
\node[up] at (v0-2) {$v^0_i$};
\node[dn] at (v1-2) {$v^1_i$};
\end{tikzpicture}
    \caption{The graph $\mathcal{G}_n(\pi,z)$ without the isolated vertices $D$. The slots form a chain with two parallel tracks; slot $j$ holds item $i$ with $\pi(i)=j$, and the tracks pass through it straight if $z_j=0$ and crossed if $z_j=1$. The edges form path~0 (blue), which starts at $s$, and path~1 (red). The vertex $t$ is isolated; in every epoch $k$ with $I_k=i$, the dashed edge $\{t,v^0_i\}$ is inserted, so $s$ and $t$ are connected if and only if $v^0_i$ is on path~0.}
    \label{fig:construction}
\end{figure}

Formally, let $d=\lfloor\frac{n-2}{8}\rfloor$. For a bijection $\pi\colon[d]\to[d]$ and $z\in\{0,1\}^d$, the
graph $\mathcal{G}_n(\pi,z)=(V,E)$ is defined as follows.
\begin{description}
  \item[Vertices.] $V=\{s,t\}\cup\{l^0_j,l^1_j,r^0_j,r^1_j\mid j\in[d]\}\cup\{u^0_i,u^1_i,v^0_i,v^1_i\mid i\in[d]\}\cup D$,
    where $D$ is a set of isolated vertices such that $|V|=n$. The \emph{$j$-th slot} consists of the vertices
    $l^0_j,l^1_j,r^0_j,r^1_j$, and the \emph{$i$-th item} consists of the vertices $u^0_i,u^1_i,v^0_i,v^1_i$.
  \item[Fixed edges.] The edge $\{s,l^0_1\}$; the edges $\{r^b_j,l^b_{j+1}\}$ for $j\in[d-1]$ and $b\in\{0,1\}$,
    which join consecutive slots; and the edges $\{u^b_i,v^b_i\}$ for $i\in[d]$ and $b\in\{0,1\}$.
  \item[Variable edges.] For every item $i$, with $j=\pi(i)$, and every $b\in\{0,1\}$, the edges
    $\{l^b_j,u^{b\oplus z_j}_i\}$ and $\{v^b_i,r^b_j\}$. We say that item $i$ is \emph{placed} in slot $j$;
    the tracks pass through it straight if $z_j=0$ and crossed if $z_j=1$.
\end{description}
All vertices and all fixed edges are known to the algorithm; only $(\pi,z)$ and the edges that depend on them are hidden. Every vertex has degree at most $2$. Apart from the isolated vertices $t$ and $D$, the edges form two
   vertex-disjoint paths: \emph{path~0}, which contains $s$, and \emph{path~1}, which contains $l^1_1$ (see
   \Cref{fig:construction}).

We use the terms \emph{slot} and \emph{item} in both settings, with matching meanings. In PPHP, slots and items are elements of $[d]$: the items form the domain of $\pi$, the slots form its range, and item $i$ is placed in slot $\pi(i)$. In the graph $\mathcal{G}_n(\pi,z)$, slot $j$ and item $i$ are sets of four vertices each, defined above, and item $i$ is attached to slot $\pi(i)$.

We may assume that $B\geq 10$. Indeed, an algorithm that uses $B<10$ bits of memory also uses at most $10$
bits, so the lower bound for $10$ bits applies to it, and $n^2/10$ is within a constant factor of $n^2/B$.

\subsubsection{Input Stream}
The input stream is chosen as follows: first we choose permutation $\pi$ and binary vector $z$ uniformly at random and independently of each other. In the first $8d-1$ updates, edges of $\mathcal{G}_n(\pi,z)$ are inserted. Uniformly at random and independently sample $I_1,\ldots,I_N\sim\operatorname{Unif}([d])$ and let $\calI=(I_1,\ldots,I_N)$.  
Then, for the next $N$ epochs do the following: 
\begin{enumerate*}[label=(\arabic*), itemjoin={{; }}, itemjoin*={{; and }}]
\item insert edge $e_i=\{t, v_{I_i}^0\}$ 
\item query connectivity of $s$ and $t$ 
\item delete $e_i$. 
\end{enumerate*}
\subsubsection{Equivalence to PPHP}\label{sec:reduction}
In the described construction, note that the query asked at epoch $i$ is equivalent to asking which path $v_{I_i}^0$ is on, since $s$ and $t$ are connected if and only if $s$ and $v_{I_i}^0$ are on the same path, i.e., if and only if $v_{I_i}^0$ is in path 0. Equivalently, the query is asking $y_{I_i}=\bigoplus_{k\in[\pi(I_i)]}z_k$ since \cref{clm:path} shows that $y_{I_i}$ is $1$ if and only if $s$ and $t$ are not connected in the current graph. 
\begin{claim}
    \label{clm:path}
    For every item $i$, the vertex $v^0_i$ lies on path~$0$ if and only if $y_i=\bigoplus_{k\in[\pi(i)]}z_k=0$.
\end{claim}
\begin{proof}
    Let $p_j=\bigoplus_{k\in[j]}z_k$, with $p_0=0$. We show by induction over $j$ that for $b\in\{0,1\}$, the vertex $l^b_j$ lies on path~$b\oplus p_{j-1}$ and the vertex $r^b_j$ lies on path~$b\oplus p_j$. This shows the claim since $v^0_i$ is on the same path as $r^0_{\pi(i)}$, which is path~$p_{\pi(i)}=y_i$.

    For $j=1$, the claim for $l^0_1$ and $l^1_1$ holds since they are on paths~$0$ and~$1$, respectively, by the definition of the paths, and $p_0=0$. Also, if $z_1=0$, then $r^0_1$ and $r^1_1$ are on the same path as $l^0_1$ and $l^1_1$, respectively, and otherwise it is the opposite, which is exactly the claim since
    $p_1=z_1$.

    Now, let $j>1$ and let $i$ be the item placed in slot $j$. For any $b\in\{0,1\}$, $l^b_j$ is adjacent to $r^b_{j-1}$, so it is on the same path, which is path~$b\oplus p_{j-1}$ by the induction hypothesis. Also, $r^b_j$ is on the same path as $l^{b\oplus z_j}_j$, as they are connected through $u^b_i$ and $v^b_i$. By the previous sentence, this is path~$b\oplus z_j\oplus p_{j-1}=b\oplus p_j$, so the claim holds for $r^b_j$
    as well.
\end{proof}

Moreover, since all vertices and fixed edges are known, every probe has a trivial answer except for the six
pairs $\{l^b_j,u^{b'}_i\}$ and $\{v^b_i,r^b_j\}$ of an item $i$ and a slot $j$, for $b,b'\in\{0,1\}$. The pair
$\{l^b_j,u^{b'}_i\}$ is an edge if and only if $\pi(i)=j$ and $z_j=b\oplus b'$, and the pair $\{v^b_i,r^b_j\}$ is
an edge if and only if $\pi(i)=j$. Hence, each such probe can be answered by the single oracle probe
$P_X(i,j)$.

Thus, it suffices for us to show a lower bound for the PPHP problem.

\subsection{Proof of the Lower Bound}
\label{sec:fd-lb-proof}
\subsubsection{A Few Definitions and Tools for Analysis} \label{sec:fewDefs}

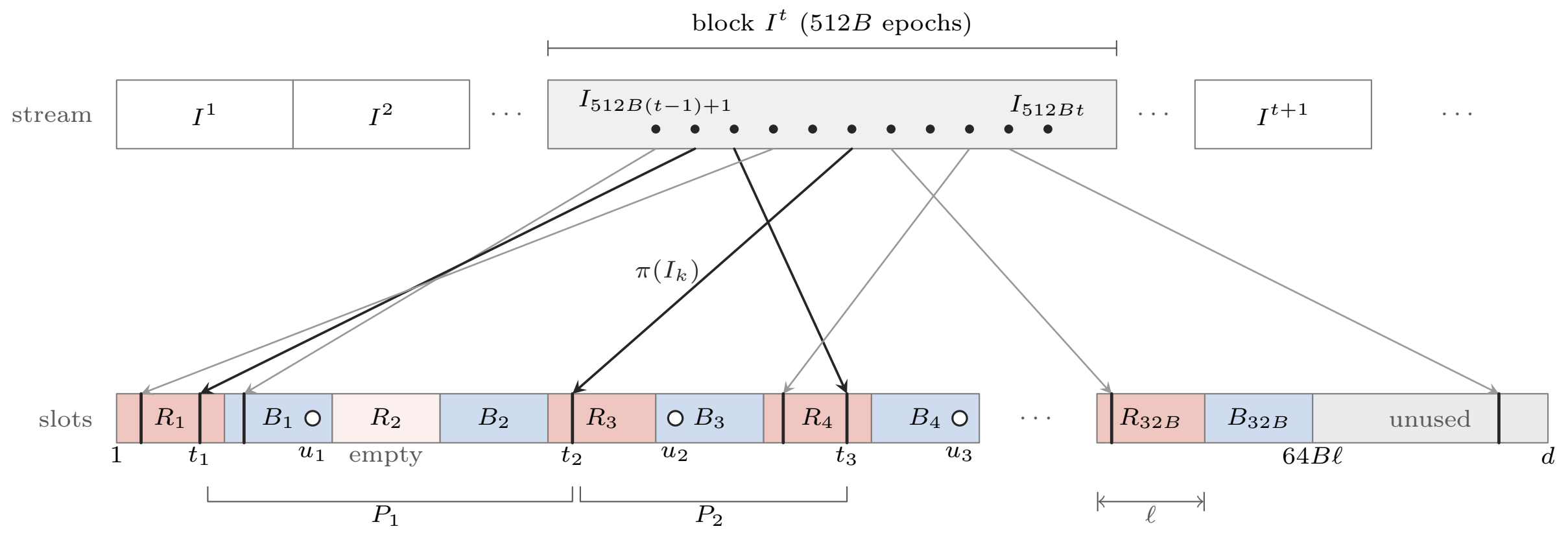
\begin{figure}[h]
    \centering
    \definecolor{arczero}{RGB}{36,96,186}
\definecolor{arcone}{RGB}{204,68,44}
\begin{tikzpicture}[x=1cm, y=1cm, line cap=round, line join=round,
  lbl/.style={font=\scriptsize, inner sep=1pt},
  tag/.style={font=\scriptsize, text=black!65, inner sep=1pt},
  box/.style={draw=black!50, line width=0.4pt},
  rint/.style={box, fill=arcone!30},
  rempty/.style={box, fill=arcone!8},
  bint/.style={box, fill=arczero!22},
  rest/.style={box, fill=black!8},
  blk/.style={box, fill=white},
  cur/.style={box, fill=black!6},
  map/.style={->, >=stealth, line width=0.5pt, black!40},
  first/.style={->, >=stealth, line width=0.7pt, black!85},
  hit/.style={line width=0.9pt, black!85},
  cp/.style={circle, draw=black!85, fill=white, line width=0.6pt, inner sep=0pt, minimum size=4.2pt},
  dim/.style={black!65, line width=0.4pt},
]
\def\w{1.1}
\foreach \x/\sty/\lab in {0/rint/{R_1}, 1/bint/{B_1}, 2/rempty/{R_2}, 3/bint/{B_2},
                          4/rint/{R_3}, 5/bint/{B_3}, 6/rint/{R_4}, 7/bint/{B_4}} {
  \draw[\sty] (\x*\w,0) rectangle (\x*\w+\w,0.5);
  \node[lbl] at (\x*\w+0.5*\w,0.25) {$\lab$};
}
\node[tag] at (9.4,0.25) {$\cdots$};
\draw[rint] (10.0,0) rectangle (11.1,0.5); \node[lbl] at (10.55,0.25) {$R_{32B}$};
\draw[bint] (11.1,0) rectangle (12.2,0.5); \node[lbl] at (11.65,0.25) {$B_{32B}$};
\draw[rest] (12.2,0) rectangle (14.6,0.5); \node[tag] at (13.4,0.25) {unused};
\node[tag, anchor=east] at (-0.2,0.25) {slots};
\node[lbl, below] at (0,0) {$1$};
\node[lbl, below] at (12.2,0) {$64B\ell$};
\node[lbl, below] at (14.6,0) {$d$};
\draw[dim, |<->|] (10.0,-0.6) -- (11.1,-0.6) node[midway, below, lbl] {$\ell$};
\draw[blk] (0,3.0) rectangle (1.8,3.7);   \node[lbl] at (0.9,3.35) {$I^1$};
\draw[blk] (1.8,3.0) rectangle (3.6,3.7); \node[lbl] at (2.7,3.35) {$I^2$};
\node[tag] at (4.0,3.35) {$\cdots$};
\draw[cur] (4.4,3.0) rectangle (10.2,3.7);
\node[tag] at (10.6,3.35) {$\cdots$};
\draw[blk] (11.0,3.0) rectangle (12.8,3.7); \node[lbl] at (11.9,3.35) {$I^{t+1}$};
\node[tag] at (13.7,3.35) {$\cdots$};
\node[tag, anchor=east] at (-0.2,3.35) {stream};
\foreach \x in {5.5,5.9,...,9.5} \fill[black!85] (\x,3.2) circle (1.3pt);
\node[lbl, above] at (5.5,3.3) {$I_{512B(t-1)+1}$};
\node[lbl, above] at (9.5,3.3) {$I_{512Bt}$};
\draw[dim] (4.4,3.95) -- (4.4,4.1) (4.4,4.025) -- (10.2,4.025) (10.2,3.95) -- (10.2,4.1);
\node[lbl, above] at (7.3,4.1) {block $I^t$ ($512B$ epochs)};
\draw[map]   (5.5,3.0) -- (1.3,0.5);
\draw[first] (5.9,3.0) -- (0.85,0.5);
\draw[first] (6.3,3.0) -- (7.45,0.5);
\draw[map]   (6.7,3.0) -- (0.25,0.5);
\draw[first] (7.5,3.0) -- (4.65,0.5) node[midway, left=2pt, lbl] {$\pi(I_k)$};
\draw[map]   (7.9,3.0) -- (10.15,0.5);
\draw[map]   (8.7,3.0) -- (6.8,0.5);
\draw[map]   (9.1,3.0) -- (14.1,0.5);
\foreach \x in {0.25,0.85,1.3,4.65,6.8,7.45,10.15,14.1} \draw[hit] (\x,0) -- (\x,0.5);
\node[lbl, below] at (0.85,0) {$t_1$};
\node[lbl, below] at (4.65,0) {$t_2$};
\node[lbl, below] at (7.45,0) {$t_3$};
\node[tag, below] at (2.75,0) {empty};
\foreach \x/\k in {2.0/1, 5.7/2, 8.6/3} {
  \node[cp] at (\x,0.25) {};
  \node[lbl, below] at (\x,0) {$u_\k$};
}
\foreach \a/\b/\k in {0.85/4.65/1, 4.65/7.45/2} {
  \draw[dim] (\a+0.08,-0.45) -- (\a+0.08,-0.6) -- (\b,-0.6) -- (\b,-0.45);
  \node[lbl, below] at ({(\a+\b)/2},-0.6) {$P_\k$};
}
\end{tikzpicture}
    \caption{The definitions of \Cref{sec:fewDefs}. Bottom: the first $64B\ell$ slots are split into red and blue intervals of length $\ell$; the remaining slots are unused. Top: the input stream is split into blocks of $512B$ epochs. Each epoch $I_k$ of the block $I^t$ is mapped to the slot $\pi(I_k)$. A red interval is occupied if some epoch of $I^t$ is mapped into it and empty otherwise. Dark arrows mark the earliest epoch mapped into each occupied red interval; their slots are $t_1<t_2<\cdots$, and $P_k=\{t_k+1,\ldots,t_{k+1}\}$ contains the blue interval following $t_k$. The checkpoint $u_k$ is a uniformly random position of that blue interval.}
    \label{fig:blocks}
\end{figure}

Assume $B\leq d/128$ is an integer and let $\ell:=\lfloor\frac{d}{64B}\rfloor\geq \frac{d}{128B}$. 
Now, partition the input stream into blocks of $512B$ epochs each (i.e., $I^t=I_{512B(t-1)+1},\ldots,I_{512Bt}$) and consider only the $t$-th block $I^t$, for an arbitrary $t$ (see \Cref{fig:blocks}).
At a high level, we will show the final result through showing that any algorithm that answers all the epochs in a block successfully after receiving the $B$-bit advice $\calM(\pi,z)$, needs at least $\Omega(d^2)$ probes for each block (in \cref{lem:blockFinal}), and then use that to show that the algorithm uses $\Omega(d^3/B)$ probes for $\Theta(d/B)$ blocks (of total length $\Theta(d)$ epochs).

Partition the slots $[d]$ into \emph{intervals} of $\ell$ consecutive slots. Color the first $64B \ge d/(2\ell)$ intervals alternately  \emph{red} and \emph{blue}, starting with red, resulting in intervals $R_1, B_1,R_2,B_2,\ldots,R_{32B},B_{32B}$, each of length $\ell$. Now each epoch  $I_k$ is mapped to the \emph{slot} $\pi(I_k)$.

We call a red interval $R_a$ \emph{occupied} if for some $I_k\in I^t$, $\pi(I_k)\in R_a$, and \emph{empty} otherwise. A fixed red interval is empty with probability $(1-\ell/d)^{512B}\leq e^{-512B\ell/d}\leq e^{-4}$. Thus the expected number of empty red intervals is at most $32e^{-4}B$. 
Let $\calE$ be the event that at least $16B$ red intervals (i.e., at least half of them) are occupied. By Markov's inequality, 
$\Pr_{I^t}[\bar\calE\mid\pi,z]=\Pr_{I^t}[\text{more than $16B$ red intervals are empty}\mid\pi,z]
\leq\frac{32e^{-4}B}{16B}=2e^{-4}\leq\frac{1}{16}$.

On $\calE$, take the first $8B+1$ occupied red intervals, take the earliest epoch whose queried item lies in each of them, and let the slots of these items be $t_1<t_2<\ldots<t_{8B+1}$. For every $k\in[8B]$, let $P_k=\{t_k+1,\ldots,t_{k+1}\}$. Note that if the algorithm is successful, by xoring every consecutive pair of answers for these epochs, we can calculate $Y_k=\bigoplus_{j\in P_k}z_j$ for every $k\in[8B]$, and also note that from construction, if $t_k\in R_q$, then $B_q\subseteq P_k$.

For each $k\in[8B]$, sample $\phi_k\sim\operatorname{Unif}([\ell])$ independently and uniformly at random and let $\Phi$ be the vector $(\phi_1,\ldots,\phi_{8B})$. Note that $\Phi$ is only for analysis and is not seen by the algorithm. If $t_k\in R_q$, let $u_k$ be the $\phi_k$-th slot of $B_q$. Call these points \emph{checkpoints}. If $H$ distinct slots are hit during the probes of this block and $H_c$ of them are checkpoints, then $E_\Phi[H_c]\leq H/\ell$. Call an execution \emph{low-hit} if $H\leq \frac{B\ell}{32}$, and \emph{high-hit} otherwise. For every low-hit execution, by Markov's inequality, $\Pr_\Phi[H_c\leq\frac{B}{2}]\geq\frac{15}{16}$.

\subsubsection{Deterministic Algorithms}
In this section, we prove the lower bound for any deterministic algorithm.

\begin{lemma} \label{lem:cheapHighHit}
    Let $\calA$ be an algorithm that receives any $B$-bit string $\calM(\pi,z)$, and does at most $L$ probes using the oracle $P_X(\cdot,\cdot)$ where $X=(\pi,z)$, then
    $$\Pr_{X}[H\geq h]\leq 2^B\frac{{\binom{L}{h}}(d-h)!}{d!}$$
    where $H$ is the number of distinct hits.
\end{lemma}
\begin{proof}
    We encode every $X=(\pi,z)$ such that $H\geq h$ using the following:
    \begin{enumerate}
        \item The advice string $\calM(\pi,z)$ \label{enum:CHH_Mpz}
        \item Position of the first $h$ distinct hits (among the $L$ probes), and their corresponding switch bits \label{enum:CHH_hits}
        \item The bijection and switch bits corresponding to all of the other $d-h$ slots and items. \label{enum:CHH_bijection}
    \end{enumerate}
    From the record mentioned above one can reconstruct $X$ by running the algorithm: First the algorithm is given $\calM(\pi,z)$. Then, for every probe, if it is among the $h$ chosen probes (the first $h$ distinct hit probes), its switch bit is given to the algorithm from the record. 
    We stop the simulation after the last of the $h$ chosen probes. Up to that point, every other probe is a hit if and only if it repeats the item and slot pair of an earlier chosen probe, since otherwise it would be one of the first $h$ distinct hits. In that case, we give the algorithm the recorded switch bit, and otherwise, we give it $\bot$. At the end, the chosen probes reveal the $h$ hit items, their slots, and their switch bits, and \cref{enum:CHH_bijection}, read as a bijection between the remaining $d-h$ items and the remaining $d-h$ slots (each in increasing order) together with their switch bits, gives the rest of $X$.
    Now, we count the number of encodings: for \cref{enum:CHH_Mpz}, it is at most $2^B$. For \cref{enum:CHH_hits}, since there are $L\choose h$ different ways to choose the positions and $2^h$ different binary vectors for switch bits, it is ${\binom{L}{h}}2^h$. For \cref{enum:CHH_bijection}, it is $(d-h)!2^{d-h}$. So in total the count is at most $2^B\cdot{L\choose h}2^h\cdot (d-h)!2^{d-h}=2^B\cdot{L\choose h}\cdot (d-h)!2^d$. Since the total count of pairs $(\pi,z)$ is $d!2^d$, the probability of $H\geq h$ is at most $\frac{2^B\cdot{L\choose h}\cdot (d-h)!2^d}{d!2^d}= 2^B\frac{{L\choose h}(d-h)!}{d!}$.
\end{proof}

\begin{corollary}\label{cor:cheapHighHit}
    There exist constants $c_L,\eta>0$ such that for every $10\leq B\leq\eta d$,
    $$\Pr_{\pi,z}[L<c_Ld^2\;\wedge\;\text{execution is high-hit}]<\frac{1}{16}.$$
\end{corollary}
\begin{proof}
    Truncate the algorithm after $c_Ld^2$ probes and apply \cref{lem:cheapHighHit} with $L=c_Ld^2$ and $h=\lfloor\frac{B\ell}{32}\rfloor+1$. Note that from the definition we have $d/4096\leq h\leq d/2$. Thus, we have:
    \begin{align*}
        \Pr[L<c_Ld^2\;\wedge\;\text{execution is high-hit}]\leq&2^B\frac{{L\choose h}(d-h)!}{d!}\\
        \leq&2^B\frac{(eL/h)^h}{(d-h)^h}\\
        \leq&2^B\big(\frac{eL}{h(d-h)}\big)^h\\
        <&2^B\big(\frac{ec_Ld^2}{d/4096\cdot d/2}\big)^{d/4096}
    \end{align*}
    where the second inequality holds since ${L\choose h}\leq(eL/h)^h$ and $d(d-1)\ldots (d-h+1)\geq (d-h)^h$. Thus, by choosing small enough $c_L$, we have:
    \begin{align*}
        \Pr[L<c_Ld^2\;\wedge\;\text{execution is high-hit}]< 2^{B-d/4096}&
    \end{align*}
    So by taking $\eta$ small enough, since $B\leq\eta d$, for large enough $d$, the probability above will be smaller than $1/16$.
\end{proof}
\begin{lemma}\label{lem:correctLowHc}
    For every fixed block $I^t$, $\Pr_{\pi,z,\Phi}[\calE\wedge \text{($Y$-correct)}\wedge H_c\leq\frac{B}{2}]\leq2^{-30}$, where $Y$-correct is the event that $Y_k$ is calculated correctly from the output (for all $k\in[8B]$).
\end{lemma}
\begin{proof}
    We prove this lemma with an encoding argument as well. We encode every tuple $(\pi,z,\Phi)$ for which $\calE\wedge \text{($Y$-correct)}\wedge H_c\leq\frac{B}{2}$ holds. We record the following:
    \begin{enumerate}
        \item The advice string $\calM(\pi,z)$ \label{enum:CLHC_mpz}
        \item The bijection $\pi$ \label{enum:CLHC_bijection}
        \item All of the checkpoint indices $\Phi$ \label{enum:CLHC_checkpoint}
        \item Switch bits for every position other than checkpoints. \label{enum:CLHC_otherSwitch}
        \item Positions (among the $8B$ checkpoints) and switch bits of all hit checkpoints.\label{enum:CLHC_hitSwitch}
    \end{enumerate}
    Using this encoding, one can determine $(\pi,z,\Phi)$ by running the algorithm as a blackbox. $\pi$ and $\Phi$ are directly in the records and for $z$ we do the following: Run the algorithm with $\calM(\pi,z)$ given to it according to the record. Then for every probe, we can determine if it is a hit from $\pi$. Then, if it is a hit, its corresponding switch bit is available in the record, and otherwise, $\bot$ is returned. Finally, for every unhit checkpoint $u_k$ (whose switch bits are the only ones not in the record), $z_{u_k}$ is the only unknown bit in the xor used to calculate $Y_k$, we xor $Y_k$ (from the output of the algorithm) and the other switch bits in the interval $P_k$ to determine $z_{u_k}$.

    Now, we count the number of such encodings. We count for each one of the items above and multiply the counts. For \cref{enum:CLHC_mpz,enum:CLHC_bijection,enum:CLHC_checkpoint,enum:CLHC_otherSwitch} the counts are $2^B$, $d!$, $\ell^{8B}$, and $2^{d-8B}$, respectively. Now, we upper bound the count for \cref{enum:CLHC_hitSwitch}. Since for choosing the position of hit checkpoints (among the checkpoints) and choosing their switch bits if we have $j$ hit checkpoints, there are ${8B\choose j}2^j$ ways, we need to bound $\sum^{\lfloor B/2\rfloor}_{j=0}{8B\choose j}2^j$. These summands are increasing in $j$. So the count is at most $(\lfloor{B/2}\rfloor+1){8B\choose \lfloor{B/2}\rfloor}2^{\lfloor{B/2}\rfloor}$. Since ${a\choose b}\leq (ea/b)^b$, we have:
    \begin{align*}
        (\lfloor{B/2}\rfloor+1){8B\choose \lfloor{B/2}\rfloor}2^{\lfloor{B/2}\rfloor}\leq& (\lfloor{B/2}\rfloor+1)(\frac{e\cdot8B}{B/2})^{B/2}2^{B/2}\\
        =&(\lfloor{B/2}\rfloor+1)(32e)^{B/2}\\
        \leq& 2^{B/2}\cdot2^{7B/2}=2^{4B}
    \end{align*}
    Thus, there are at most $2^B\cdot d!\cdot\ell^{8B}\cdot2^{d-8B}\cdot2^{4B}=d!\cdot\ell^{8B}\cdot2^{d-3B}$. Since the total number of tuples $(\pi,z,\Phi)$ is $d!2^d\ell^{8B}$, we have:
    $$\textstyle\Pr_{\pi,z,\Phi}[\calE\wedge \text{($Y$-correct)}\wedge H_c\leq\frac{B}{2}]\leq \frac{d!\ell^{8B}2^{d-3B}}{d!2^d\ell^{8B}}=2^{-3B}\leq 2^{-30}.$$
\end{proof}
\begin{lemma} \label{lem:blockFinal}
    For each block, $\Pr_{\pi,z,I^t}[\text{correct}\wedge L<c_Ld^2]\leq\frac14$ where $c_L$ is according to \cref{cor:cheapHighHit}. 
\end{lemma}
\begin{proof}
    In this lemma, we upper bound the probability of correctness of the algorithm while making at most $c_Ld^2$ probes by a set of probabilities bounded previously in this section.
    \begin{align*}
        \Pr[\text{correct}\wedge L<c_Ld^2]\leq&\Pr[\text{correct}\wedge L<c_Ld^2\wedge\calE]+\Pr[\text{correct}\wedge L<c_Ld^2\wedge\neg\calE]\\
        \leq&\Pr[\text{correct}\wedge L<c_Ld^2\wedge\calE]+\Pr[\neg\calE]\\
        \leq&\Pr[\text{correct}\wedge L<c_Ld^2\wedge\calE]+1/16
    \end{align*}
    where the last inequality is because of the bound on $\Pr[\neg\calE]$ shown in \cref{sec:fewDefs}. Next, we have:
    \begin{align*}
        \Pr[\text{correct}\wedge L<c_Ld^2\wedge\calE]\leq&\Pr[\text{correct}\wedge L<c_Ld^2\wedge\calE\wedge H>\frac{B\ell}{32}]+\\&\Pr[\text{correct}\wedge L<c_Ld^2\wedge\calE\wedge H\leq\frac{B\ell}{32}]\\
        \leq& 1/16+ \Pr[\text{correct}\wedge L<c_Ld^2\wedge\calE\wedge H\leq\frac{B\ell}{32}]
    \end{align*}
    where the second inequality comes from \cref{cor:cheapHighHit}. Now, we bound the second summand:
    \begin{align*}
        \MoveEqLeft \Pr[\text{correct}\wedge L<c_Ld^2\wedge\calE\wedge H\leq\frac{B\ell}{32}] \\
        &\leq\Pr[\text{correct}\wedge\calE\wedge H\leq\frac{B\ell}{32}] \\
        &\leq \Pr[\text{correct}\wedge\calE\wedge H\leq\frac{B\ell}{32}\wedge H_c\leq B/2]+
        \Pr[\text{correct}\wedge\calE\wedge H\leq\frac{B\ell}{32}\wedge H_c> B/2]\\
        &\leq \Pr[\text{correct}\wedge\calE\wedge H_c\leq B/2]+
        \Pr[\calE\wedge H\leq\frac{B\ell}{32}\wedge H_c> B/2]\\
        &\leq \Pr[\text{$Y$-correct}\wedge\calE\wedge H_c\leq B/2]+
        \Pr[H_c> B/2\mid\calE\wedge H\leq\frac{B\ell}{32}]
        \leq 2^{-30}+1/16
    \end{align*}
    
    where the last inequality comes from \cref{lem:correctLowHc} and arguments stated in the end of \cref{sec:fewDefs}. Thus, we have:
    $$\Pr[\text{correct}\wedge L<c_Ld^2]\leq1/16+1/16+2^{-30}+1/16\leq1/4$$
\end{proof}
\begin{corollary}[Deterministic version] \label{cor:deterministic}
    There exists constant $c_B>0$ such that 
    $\Pr_{\pi,z,\calI}[\text{correct}_\text{total}\wedge L_\text{total}\leq c_Bc_L\frac{d^3}{B}]\leq\frac13$, where $c_L$ is a constant chosen according to \cref{cor:cheapHighHit}, $\text{correct}_\text{total}$ is the event of output of the algorithm being correct for all of the blocks, and $L_\text{total}$ is the total number of probes done by the algorithm (on all of the blocks altogether).
\end{corollary}
\begin{proof}
    Assume there are $4c_Bd/B$ blocks in the input stream (for some constant $c_B>0$). Also, let $E_i$ be the event of $(\text{correct}\wedge L< c_Ld^2)$ holding for the $i$-th block. 
    Fix the epochs of the first $i-1$ blocks. Then the state of the $B$ bits of memory at the beginning of block $i$ is a function $\mathcal{M}(\pi, z)$ of $(\pi, z)$ alone, and the epochs of block $i$ are independent of $(\pi, z)$ and of the earlier epochs. So \cref{lem:blockFinal} gives $\mathbb{P}[E_i \mid \text{epochs of the first } i-1 \text{ blocks}] \le 1/4$, and averaging over these epochs gives $\mathbb{P}[E_i] \le 1/4$ for every $i \in [4c_B d/B]$.
    Thus, we have that the expected number of blocks $i$ that $E_i$ occurs is at most $c_Bd/B$. Thus, by Markov's inequality:
    $$\Pr[\text{number of blocks $i$ that $E_i$ occurs}\geq 3c_Bd/B]\leq\frac13.$$
    Therefore, for all of the blocks we have:
    $$\Pr[\text{correct}_\text{total}\wedge L_\text{total}\leq c_Bc_L\frac{d^3}{B}]\leq\Pr[\text{number of blocks $i$ that $E_i$ occurs}\geq 3c_Bd/B]\leq\frac13.$$
    Note that the first inequality above holds since if for less than $3c_Bd/B$ blocks the number of probes is less than $c_Ld^2$, then for more than $c_Bd/B$ blocks at least $c_Ld^2$ probes are done and hence, $L_\text{total}$ is greater than $c_Bc_L\frac{d^3}{B}$.
\end{proof}
\subsubsection{Randomized Algorithms}
\begin{theorem}\label{thm:lb-pphp}
    For every randomized algorithm $\calA$, there exists an input such that 
    $$\Pr_{\calA}[\text{correct}_\text{total}\wedge L_\text{total}\leq c_Bc_L\frac{d^3}{B}]\leq\frac13,$$
    where $c_B$ and $c_L$ are constants chosen according to \cref{cor:deterministic} and \cref{cor:cheapHighHit}, respectively.
\end{theorem}
\begin{proof}
     Let $\text{Acc}_{\calR,\text{inp}}$ be a random variable that is $1$ if the algorithm $\calA$ given the random tape $\calR$ and input $\text{inp}$ (the bijection $\pi$, the switch bits $z$, and the epochs $\calI$) correctly answers all of the epochs and does at most $c_Bc_Ld^3/B$ probes, and $0$ otherwise. We have:
    \begin{align*}
        \mathbb{E}_\text{inp}[\Pr_{\calR}[\text{Acc}_{\calR,\text{inp}}=1]]&=\mathbb{E}_\text{inp}[\mathbb{E}_{\calR}[\text{Acc}_{\calR,\text{inp}}]]\\
        &=\mathbb{E}_{\text{inp},\calR}[\text{Acc}_{\calR,\text{inp}}]\\
        &=\mathbb{E}_\calR[\mathbb{E}_{\text{inp}}[\text{Acc}_{\calR,\text{inp}}]]\\
        &=\mathbb{E}_\calR[\Pr_{\text{inp}}[\text{Acc}_{\calR,\text{inp}}=1]]\leq1/3
    \end{align*}
    where the probability over $\text{inp}$ is taken over the set of inputs described in \cref{sec:fd-construction} with input streams consisting of $4c_Bd/B$ blocks, and the last inequality comes from \cref{cor:deterministic} since $\calA$ given the fixed random tape $\calR$ is a deterministic algorithm. Thus, there exists an input $\text{inp}$ such that $\Pr_{\calR}[\text{Acc}_{\calR,\text{inp}}=1]\leq1/3$. 
\end{proof}

From the reduction stated in \cref{sec:reduction}, \cref{thm:lb-pphp} implies the following.
Indeed, the reduction turns an update sequence $\sigma$ of length $|\sigma|=\Theta(d)=\Theta(n)$ into an input of \cref{thm:lb-pphp} such that each adjacency probe of $\calA$ is answered by at most one probe $P_X(i,j)$, and $\calA$ is correct at all time steps only if $\text{correct}_\text{total}$ holds; we choose $c$ so small that $BT\le cn^2$ implies $4|\sigma|T\le c_Bc_Ld^3/B$.
If $\calA$ makes at most $|\sigma|T$ probes on every execution, then $L_\text{total}\le c_Bc_Ld^3/B$ always, and if it does so only in expectation, then $\Pr_\calA[L_\text{total}>c_Bc_Ld^3/B]\le1/4$ by Markov's inequality; in both cases, $\calA$ is correct at all time steps with probability at most $1/3+1/4<2/3$.

\fdlower*

 Note that the lower bound of the adjacency-list model holds since every probe of the adjacency-list model can be simulated by $n$ probes of the adjacency matrix model.

\section{Worst-Case Incremental Lower Bound for Matrix Model}
In this section, we use a roll-back argument on the lower bound in \cref{cor:fd-lower} to show an $\widetilde{\Omega}(n^2/B)$ lower bound on the worst-case number of probes for algorithms that solve our problem in the incremental setting (for the adjacency matrix model).
\matrixwclb*
\begin{proof}
    Let $T$ be the worst-case number of probes per insertion of the algorithm. First, we amplify the success probability of the algorithm to $1-n^{-3}$. To do so, run $O(\log{n})$ independent copies of the algorithm on the same update sequence and take their majority answer. This amplified algorithm uses space $B'=O(B\log{n})$ and $T'=O(T\log(n))$ probes per insertion.

    Now, to solve an instance of the PPHP problem using an incremental algorithm, we do the following: first, just like the fully dynamic case, insert the edges of $\mathcal{G}_n(\pi,z)$. Now, store the state of the algorithm. Then, for the $i$-th query $I_i$, insert the edge $e_i=\{t, v_{I_i}^0\}$, record the answer of the algorithm to the connectivity problem and answer that query of PPHP accordingly, then roll the state of the algorithm back to its state after insertion of $\mathcal{G}_n(\pi,z)$ (instead of deleting $e_i$) and then do the same for the next query and so on. This way, the $i$-th execution will have the edge set of $\mathcal{G}_n(\pi,z)$ plus the edge $e_i$, which is the same as the graph that the fully dynamic algorithm used to answer the $i$-th query.

    Since there are $\Theta(n)$ queries and the algorithm has failure probability $n^{-3}$, a union bound over the correctness probabilities of all of these executions of that algorithm shows that the failure probability is at most $O(n^{-2})<1/3$.

    This simulation uses at most $T'$ probes per query and since saving the state doubles the space uses at most $2B'=O(B\log{n})$ bits of space. Thus, applying \cref{thm:lb-pphp} shows the desired lower bound. 
\end{proof}

\section{Worst-Case Incremental Lower Bound for List Model} \label{sec:wc-list-lb}
In this section, we give our worst-case lower bound for \STConn in the adjacency-list model for incremental input.
\listwclb*

\paragraph{The \idx Communication Problem}
We show the lower bound using a reduction from the \idx problem defined below.
\begin{definition}[Public Coin Index]
    Two parties, named Alice and Bob, both have access to a common random string. Alice receives a $k$-bit string $x \in \{0, 1\}^k$, and Bob receives a single index $i \in [k]$. Alice and Bob engage in a single round of communication, in which Alice is allowed to send an arbitrary message of $m$ bits to Bob. (Bob is not allowed to send any message to Alice.) We say that Alice and Bob succeed if, after the communication round, Bob outputs $x_i$, i.e., the $i$th-indexed bit of $x$. 
    \label{defn:index}
\end{definition}

We use the following classic result implied by \cite[Theorem $3.7$]{kremerNiRo99} (see also Section $3$ of \cite{jayramKuSi08} for a concise, direct proof).

\begin{theorem}[\idx Communication Complexity Lower Bound]
    To succeed with probability at least $2/3$ for \idx, Alice must send more than $k/64$ bits in her message to Bob.
    \label{thm:index_lb}
\end{theorem}

\paragraph{Proof Overview of Theorem \ref{thm:lb-list-wc}} The proof of Theorem \ref{thm:lb-list-wc} utilizes the public-coin \idx problem \cite{kremerNiRo99}. Alice receives an input $x \in \{0, 1\}^k$  (we will choose $k = cB$, for a constant $c$ of our choice) and is tasked with sending a single message to Bob. Bob, on the other hand, receives an index $i \in [k]$, and is tasked with outputting $x_i$, the $i$th index in Alice's input. Theorem \ref{thm:index_lb} tells us, that even with shared randomness, Alice must send $\Omega(k)$ bits to achieve $2/3$ success probability. 

Let $\mathcal{A}$ denote an arbitrary incremental connectivity algorithm for the list model. In subsection \ref{sec:wc_lb_con}, we present a construction of an incrementally-constructed family of graphs. In particular, we construct a random graph $G$ that satisfies the following property: $s$ and $t$ are initially not connected. Furthermore, for every index $i$, there exists an edge $e_i$, such that after the edge $e_i$ is inserted $s$ and $t$ will be connected if and only if $x_i = 1$. This motivates the following reduction from $\idx$ to incremental connectivity. If Bob can figure out how to simulate $\mathcal{A}$ internally, %
upon the given edge insertion, then he can determine the value $x_i$. Simulating $\mathcal{A}$ is not a priori possible for Bob, since he does not possess the entire graph $G$ (which is a function of Alice's input $x$), an adjacency-list oracle for $G$, or the starting state of algorithm $\mathcal{A}$ before edge insertion $e_i$. The necessary information (stated earlier) to simulate $\mathcal{A}$ will therefore have to be included (implicitly) in Alice's message to Bob.

To complete the proof, we utilize that Theorem \ref{thm:index_lb} gives us a lower bound on Alice's message. This ultimately gives us a lower bound on the worst-case update time of $\mathcal{A}$. Any algorithm $\mathcal{A}$ with worst-case update time at most $\binom{a}{2}/c^2=\Theta(n^2/B^2)$ (where $a$ is defined in \secref{wc_lb_con}) would allow Alice to ensure (with probability at least $2/3$) that she can send a short message (less than $k/64$ bits) to Bob allowing him to correctly solve $\idx$. This would contradict Theorem \ref{thm:index_lb}.

In \secref{wc_lb_con}, we present the graph construction. In \secref{wc_lb_bob}, we discuss Bob's simulation of $\mathcal{A}$. In particular, what information does Bob need from Alice to complete the simulation? Naively, he would require both the starting state of $\mathcal{A}$, which consists of $B$ bits, and the answers to any potential adjacency-list query $\mathcal{A}$ might make. In \secref{wc_lb_alice}, we prove that Alice can send this information compactly and complete the proof. 

\subsection{The Graph Construction}
\label{sec:wc_lb_con}

    Given a binary string $x \in \{0, 1\}^k$, we construct the following graph $G$ with $n$ vertices. The graph consists of a vertex $s$, a vertex $t$, $k$ distinct cliques $A_j$, and $k$ distinct cliques $B_j$ (for $j\in[k]$). The cliques each contain $a$ vertices, $A_{j, 1}, A_{j, 2}, \ldots, A_{j, a}$ and $B_{j, 1}, B_{j, 2}, \ldots, B_{j, a}$ respectively, where $a = \lfloor (n-2)/(2k)\rfloor$. We assume that $B\leq (n-2)/(6c)$, for the constant $c$ fixed in \secref{wc_lb_bob}, so that $a\geq 3$; this holds for large $n$ if the constant $\eta$ in \cref{thm:lb-list-wc} satisfies $\eta<1/(6c)$. In addition to the edges internal to the cliques, there is an edge between $t$ and $B_{j, 1}$ for every $j \in [k]$. 

\begin{figure}[t]
    \centering
    \newif\ifshowdeleted
\showdeletedtrue

\definecolor{swapblue}{RGB}{24,91,150}

\begin{tikzpicture}[
    vertex/.style={circle,fill=black,inner sep=1.65pt},
    kept/.style={draw=black!60,line width=0.55pt},
    added/.style={draw=swapblue,line width=1.15pt},
    removed/.style={draw=black!45,densely dashed,line width=0.65pt},
    every node/.style={font=\small},
    line cap=round
]
    \begin{scope}[xshift=-2.5cm]
        \foreach \i/\angle in {2/36,5/-36,3/-108,4/180,1/108}
            \coordinate (u\i) at (\angle:1.4cm);
    \end{scope}
    \begin{scope}[xshift=2.5cm,xscale=-1]
        \foreach \i/\angle in {2/36,5/-36,3/-108,4/180,1/108}
            \coordinate (v\i) at (\angle:1.4cm);
    \end{scope}

    \foreach \side in {u,v}
        \foreach \i/\j in {1/2,1/3,1/4,1/5,2/3,2/4,3/4,3/5,4/5}
            \draw[kept] (\side\i) -- (\side\j);

    \ifshowdeleted
        \draw[removed] (u2) -- (u5);
        \draw[removed] (v2) -- (v5);
    \fi

    \draw[added] (u2) -- (v5);
    \draw[added,preaction={draw=white,line width=3.2pt}]
        (u5) -- (v2);

    \foreach \side in {u,v}
        \foreach \i in {1,...,5}
            \node[vertex] at (\side\i) {};

    \node[above=5pt] at (u2) {$u_2$};
    \node[below=5pt] at (u5) {$u_5$};
    \node[above=5pt] at (v2) {$v_2$};
    \node[below=5pt] at (v5) {$v_5$};

    \node at (-2.5,-1.8) {$K_5-u_2u_5$};
    \node at ( 2.5,-1.8) {$K_5-v_2v_5$};

    \ifshowdeleted
        \draw[removed] (-1.65,-2.4) -- (-1.05,-2.4);
        \node[anchor=west] at (-0.95,-2.4) {removed};
        \draw[added] (0.65,-2.4) -- (1.25,-2.4);
        \node[anchor=west] at (1.35,-2.4) {added};
    \fi
\end{tikzpicture}
    \caption{A swap operation on a gadget consisting of two $K_5$ cliques.}
    \label{fig:wc-gadget}
\end{figure}
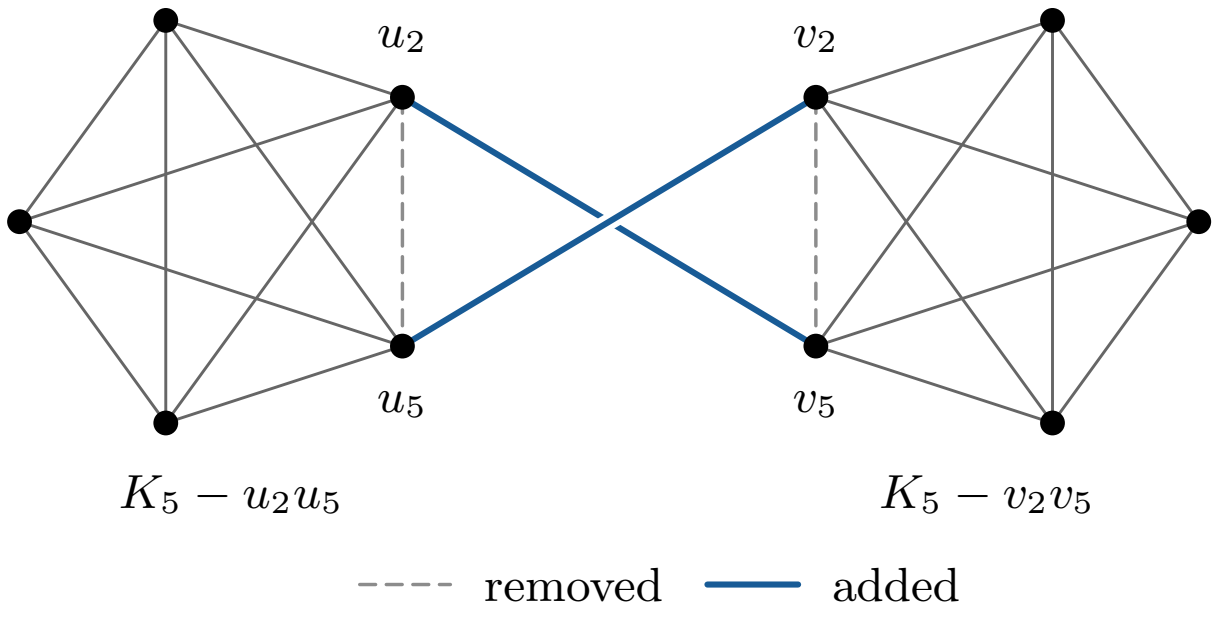
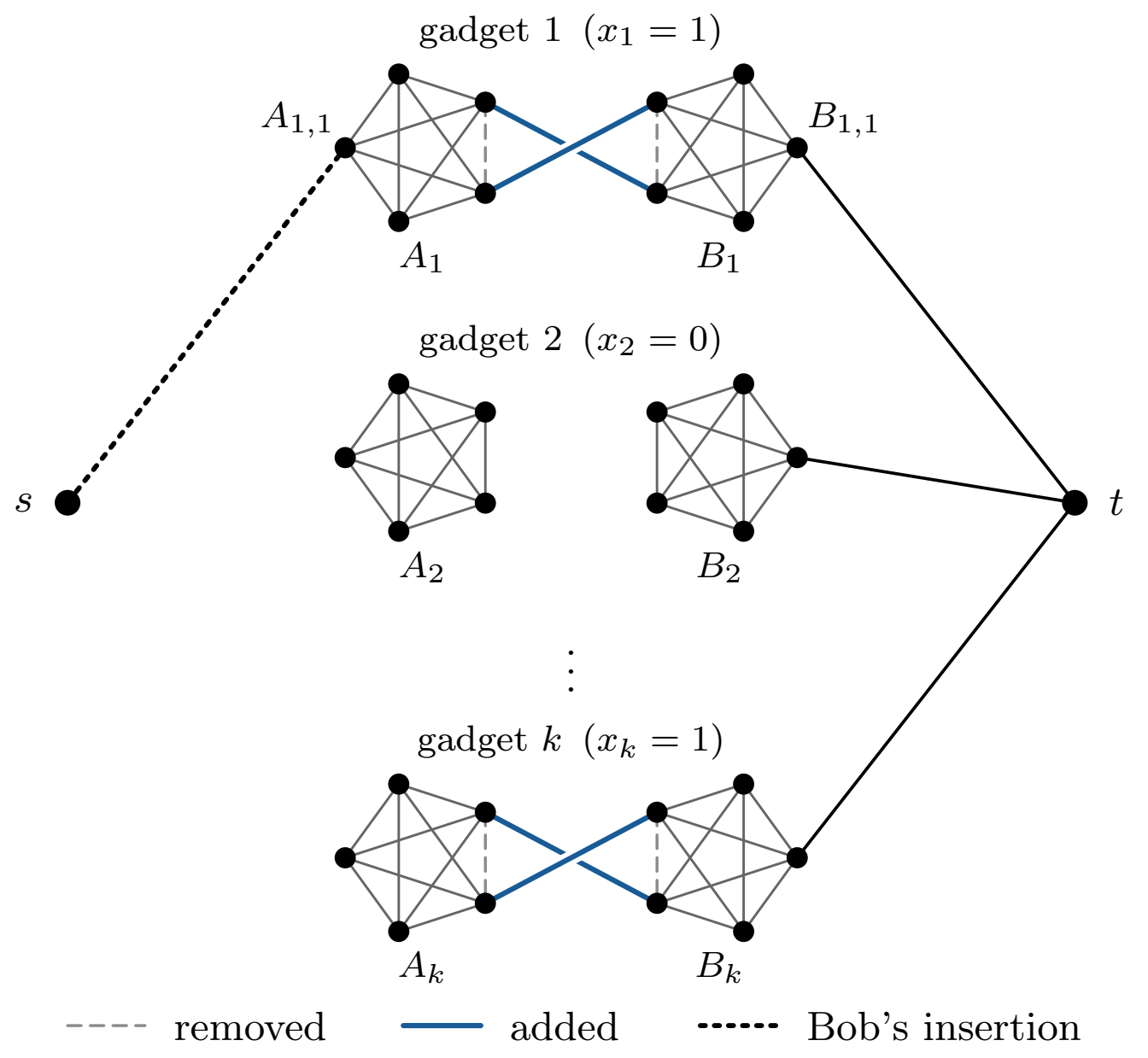
\begin{figure}
    \centering
    \newif\ifshowdeleted
\showdeletedtrue

\definecolor{swapblue}{RGB}{24,91,150}

\begin{tikzpicture}[
    vertex/.style={circle,fill=black,inner sep=1.65pt},
    kept/.style={draw=black!60,line width=0.55pt},
    added/.style={draw=swapblue,line width=1.15pt},
    removed/.style={draw=black!45,densely dashed,line width=0.65pt},
    attach/.style={draw=black,line width=0.7pt},
    bob/.style={draw=black,dotted,line width=1.1pt},
    every node/.style={font=\small},
    line cap=round
]
    \foreach \g/\y/\x/\lab in {g1/0/1/1, g2/-2.4/0/2, gk/-5.5/1/k} {
        \begin{scope}[shift={(-1.15,\y)}]
            \foreach \i/\angle in {2/36,5/-36,3/-108,1/180,4/108}
                \coordinate (\g A\i) at (\angle:0.6cm);
        \end{scope}
        \begin{scope}[shift={(1.15,\y)},xscale=-1]
            \foreach \i/\angle in {2/36,5/-36,3/-108,1/180,4/108}
                \coordinate (\g B\i) at (\angle:0.6cm);
        \end{scope}
        \foreach \side in {A,B}
            \foreach \i/\j in {1/2,1/3,1/4,1/5,2/3,2/4,3/4,3/5,4/5}
                \draw[kept] (\g\side\i) -- (\g\side\j);
        \ifnum\x=0
            \draw[kept] (\g A2) -- (\g A5);
            \draw[kept] (\g B2) -- (\g B5);
        \else
            \ifshowdeleted
                \draw[removed] (\g A2) -- (\g A5);
                \draw[removed] (\g B2) -- (\g B5);
            \fi
            \draw[added] (\g A2) -- (\g B5);
            \draw[added,preaction={draw=white,line width=3.2pt}] (\g A5) -- (\g B2);
        \fi
        \node[font=\footnotesize] at (0,\y+0.9) {gadget $\lab$ \,($x_{\lab}=\x$)};
        \node[font=\footnotesize] at (-1.15,\y-0.85) {$A_{\lab}$};
        \node[font=\footnotesize] at ( 1.15,\y-0.85) {$B_{\lab}$};
    }
    \node at (0,-3.95) {$\vdots$};

    \coordinate (s) at (-3.9,-2.75);
    \coordinate (t) at ( 3.9,-2.75);
    \foreach \g in {g1,g2,gk}
        \draw[attach] (t) -- (\g B1);
    \draw[bob] (s) -- (g1A1);

    \foreach \g in {g1,g2,gk}
        \foreach \side in {A,B}
            \foreach \i in {1,...,5}
                \node[vertex] at (\g\side\i) {};
    \node[vertex,inner sep=2pt] at (s) {};
    \node[vertex,inner sep=2pt] at (t) {};
    \node[left=4pt] at (s) {$s$};
    \node[right=4pt] at (t) {$t$};
    \node[font=\footnotesize,anchor=south east,inner sep=2pt] at (g1A1) {$A_{1,1}$};
    \node[font=\footnotesize,anchor=south west,inner sep=2pt] at (g1B1) {$B_{1,1}$};

    \ifshowdeleted
        \draw[removed] (-3.9,-6.8) -- (-3.3,-6.8);
        \node[anchor=west] at (-3.2,-6.8) {removed};
    \fi
    \draw[added] (-1.3,-6.8) -- (-0.7,-6.8);
    \node[anchor=west] at (-0.6,-6.8) {added};
    \draw[bob] (1.0,-6.8) -- (1.6,-6.8);
    \node[anchor=west] at (1.7,-6.8) {Bob's insertion};
\end{tikzpicture}
    \caption{The graph $G$ with Bob's insertion. Each gadget $j$ consists of two cliques $A_j$ and $B_j$, and
$t$ is adjacent to $B_{j,1}$ for every $j$. If $x_j=1$, the two edges of the hidden pair are removed (dashed)
and replaced by the two crossing edges (blue), which connect $A_j$ and $B_j$. For illustration, the hidden
pair is the same in every gadget. After Bob's insertion $\{s,A_{i,1}\}$ (dotted, here $i=1$), $s$ and $t$ are
connected if and only if $x_i=1$.}
    \label{fig:wc-reduction}
\end{figure}

    We then perform the following \defn{swap} operation on the graph. For every pair of cliques $A_j, B_j$, for $j \in [k]$, we independently pick a uniformly random edge $e_{A_j} = \{A_{j, u}, A_{j, v}\}$ from the clique $A_j$, and we denote the corresponding edge in $B_j$ by  $e_{B_j} = \{B_{j,u}, B_{j,v}\}$. 
    If $x_j = 1$, we remove (swap out) the edges $e_{A_j}$ and $e_{B_j}$ and add the edges $\{A_{j, u}, B_{j, v}\}$ and $\{A_{j, v}, B_{j, u}\}$. Otherwise, if $x_j = 0$, we make no change. After performing this swap, we refer to $(A_j, B_j)$ as the $j$th \defn{gadget} in $G$. Notably, $A_j$ and $B_j$ are connected if and only if $x_j = 1$.

    The resulting graph $G$ consists of an isolated vertex $s$, $k$ disjoint gadgets $(A_j, B_j)$, and a vertex $t$ that is connected to each of the $B_j$.\footnote{This results in a graph with $\Theta(n)$ vertices, and exactly $2ak + 2 \leq n$ vertices. We can add additional isolated vertices to achieve exactly $n$ vertices. Without loss of generality, these vertices need not be considered.}

    $G$ serves as our starting graph. %
    We observe that if edge $\{s, A_{i, 1}\}$ is inserted in the graph $G$, then $s$ and $t$ are connected if and only if $x_i = 1$. This fact will play a role in allowing Bob to solve $\idx$ with a single $s$-$t$ incremental connectivity update in \secref{wc_lb_bob}.

    \begin{lemma}
        Let $G$ be a graph constructed as described above, given an arbitrary binary string $x \in \{0, 1\}^k$. Consider the graph $G'$ resulting from adding the edge $\{s, A_{i, 1}\}$. Vertices $s$ and $t$ are connected in $G'$ if and only if $x_i = 1$.
        \label{lem:s_t_equivalence}
    \end{lemma}

    \begin{proof}
        If $x_i = 0$, then $s$ is connected to the vertices of $A_i$, but $A_i \cup s$ constitutes a (maximal) connected component. Thus $s$ and $t$ are not connected. On the other hand, if $x_i = 1$, then $s$ is connected to the clique $A_i$, which stays connected after the swap since $a\geq 3$, and which is connected to $B_i$ through the $(A_i, B_i)$ gadget. $B_i$ is connected to $t$ by the edge $\{t, B_{i, 1}\}$. Thus $s$ and $t$ are connected.
    \end{proof}

    \paragraph{Adjacency list ordering of $G$} In our bounded-space dynamic model, the order of adjacency lists is arbitrary (i.e., adversarial). We define the adjacency lists to be sorted lexicographically by reversed vertex label. (E.g., if a vertex is connected to vertices $A_{i, 1}$, $B_{i, 2}, A_{i, 3}$, they would be sorted in that order by their least significant character $1 < 2 < 3$.) The vertices $s$ and $t$ are placed at the end of every list containing them. This has the effect that swapped edges are swapped in-place in their respective adjacency lists. 

    Alice generates the graph $G$ locally upon receiving input $x$. She uses the shared random string to choose the swaps; thus Bob also knows which edges were chosen by Alice (but not whether the edges for a gadget $i$ were actually swapped, i.e., whether $x_i = 1$).

    \paragraph{Splitting the shared randomness} Alice and Bob have a common random string, which they will use for two different tasks in our reduction. The first use case, described above, is to choose the swap edges in each gadget. The second task for which shared randomness is required, described below, is the simulation of the randomized algorithm $\mathcal{A}$. We ensure these sources of randomness are separate and independent. This can be implemented, e.g., by using the even channel for the former and the odd channel for the latter. This has the effect of blinding the algorithm $\mathcal{A}$ to the random choices used in the construction of $G$; even though Bob possesses this information, he does not share it with $\mathcal{A}$, and instead gives $\mathcal{A}$ a separate source of common random bits.

\subsection{Bob's Simulation}
\label{sec:wc_lb_bob}

Bob, upon receiving input $i$, will internally simulate the incremental connectivity algorithm $\mathcal{A}$ during edge insertion $\{s, A_{i, 1}\}$. $\mathcal{A}$ satisfies worst-case space bound $B = k/c$, where we fix the constant $c=2048$. Suppose, for the sake of contradiction, that further $\mathcal{A}$ has worst-case update time $T \leq \binom{a}{2}/c^2 = \Theta(n^2/B^2)$. To solve $s$--$t$ connectivity after the insertion, it is sufficient for Bob to receive the state of $\mathcal{A}$ before the insertion, which we will call the \emph{advice string}, as well as answers to the set of probes that $\mathcal{A}$ performs during the insertion. (We will see that Alice will not explicitly send this set in her message, but actually a compacted message that is sufficient for Bob to simulate the adjacency-list probes made by $\mathcal{A}$.)
Then he can simply locally simulate $\mathcal{A}$ for the insertion. He starts from the received internal state, and runs $\mathcal{A}$, while answering any adjacency-list oracle queries made by $\mathcal{A}$ using the message from Alice.

In order to facilitate Bob's simulation, Alice will locally construct the state of $\mathcal{A}$, by starting with the empty graph and incrementally inserting every edge in $G$. She includes this full initial algorithm state, which consists of at most $B$ bits, in her message to Bob. This part of the message allows Bob to initialize $\mathcal{A}$'s state on graph $G$.

Bob still lacks access to the oracle, since during his simulation of $\mathcal{A}$, it will attempt to make adjacency-list probes to the graph $G$, which Bob does not have access to. Alice can herself simulate $\mathcal{A}$ on each of the $k$ possible edge insertions corresponding to Bob's possible index. Indeed, she can even simulate $\mathcal{A}$ with the same random string as Bob, since they share a common random string. The total number of adjacency-list probes for any update is at most $T$, so this sums to $Tk$ probes. The naive approach of including all $Tk$ possible probes in her message, however, would make Alice's message too long.

\paragraph{Not every probe is special} We observe that the outcomes of adjacency-list probes made by $\mathcal{A}$ to $G$ are highly predictable by Bob, even though he doesn't possess the underlying graph. For example, Bob can exactly simulate any degree query, since he knows that $s$ has degree $0$ in $G$, $t$ has degree $k$, and most vertices in cliques have degree $a-1$ (except those connected to either $t$, which Bob knows have degree $a$). Thus, he can perfectly simulate degree queries even without access to an adjacency-list oracle. Bob can also perfectly simulate $j$th neighbor queries of a vertex $v$, conditioned on the $j$th neighbor not being an edge that was swapped in a gadget. This is because Bob mostly knows the lexicographical layout of the adjacency lists. His only missing information is the identities of the swapped edges, which modify the adjacency lists in place. We call an adjacency-list neighbor probe that finds a swapped edge (i.e., an edge across two cliques in a gadget) a \defn{successful} probe.

As a consequence of splitting the shared randomness, the algorithm $\mathcal{A}$ is not aware of the random bits used to choose the swap edges, even though Bob himself is aware through his shared randomness with Alice. Thus, Bob's simulation of $\mathcal{A}$ cannot make many successful probes, even though Bob himself could.

\begin{lemma}[Successful Probe Tail Bound]
    For any fixed advice string $z \in \{0, 1\}^B$, the total number of distinct gadgets in which Bob makes successful probes (across all $k$ possible inputs), when simulating $\mathcal{A}$ with initial state $z$, is more than $2B$ with probability at most $2^{-2B}$.
    \label{lem:wc_lb_unsuccessful_bob}
\end{lemma}

\begin{proof}
We prove that, for any arbitrary advice string, and any input index $i \in [k]$, Bob's simulation of $\mathcal{A}$ is unlikely to make many distinct successful probes. This bounds the total number of successful adjacency-list probes that Bob might have to answer across all of his potential inputs $i$.
Summing over all $k$ possible input indices, Bob performs at most $Q = kT$ total adjacency-list probes. Consider an arbitrary fixed advice string $z$ of $B$ bits. For any fixed advice string $z$, the choices of edge swaps remain uniform and independent, as does the shared random string that Alice/Bob use to generate $G$ (since, for example, we are not conditioning on $z$ being the actual advice string that is the state of $\mathcal{A}$). The $Q$ probes result in at most $2Q/ \binom{a}{2}$ gadgets experiencing at least $0.5 \binom{a}{2}$ probes. We denote Bob's first $0.5 \binom{a}{2}$ probes to a gadget (across the $k$ simulations) as \defn{light probes}. The total number of gadgets receiving successful probes is at most $2Q/\binom{a}{2}$ plus the number of distinct successful light probes.

Bob's light probes, conditioned on being made to a gadget in which his simulation of $\mathcal{A}$ has not yet made a successful probe, each succeed with probability at most $2/(0.5\binom{a}{2}) = 4/\binom{a}{2}$. The location of a swapped edge in such a gadget, in which no swapped edge has yet been found, remains uniform among the remaining edges even after conditioning on previous probes. Thus, the number of successful light probes is stochastically dominated by $X$, the sum of $Q$ i.i.d.\ indicator variables, each $1$ with probability $4/\binom{a}{2}$. The expected number of light probes finding selected edges is thus at most $\E[X] = 4Q/\binom{a}{2}$. 
    
Applying a Chernoff bound, we find that $\Pr[X \geq B] \leq 4^{-B} \leq 2^{-2B}$ (as $c=2048$). Thus, the number of such successful light probes exceeds $B$ with probability at most $2^{-2B}$. Therefore the total number of distinct gadgets receiving successful probes exceeds $2Q/\binom{a}{2} + B$ with probability at most $2^{-2B}$. We finish the proof by observing $2Q/\binom{a}{2} + B \leq 2k/c = 2B$ (recall $T \leq \binom{a}{2}/c^2$).
\end{proof}

Union bounding over all $2^B$ possible advice strings yields a tail bound probability of $2^{-B}$. In particular, this implies that starting from the actual initial state of $\mathcal{A}$ for $G$, with probability at least $1-2^{-B}$, Bob successfully probes (across all $k$ possible inputs) in at most $2B$ distinct gadgets.

\subsection{The Communication Protocol}
\label{sec:wc_lb_alice}
    Alice receives an input $x \in \{0, 1\}^k$, and generates a graph $G$ following the construction of \secref{wc_lb_con} using shared randomness. She then runs $\mathcal{A}$, inserting edges one at a time, starting from an empty graph, until she has the state of $\mathcal{A}$ for graph $G$. She includes this state in her message, at the cost of $B$ bits.
    In addition to the initial state of $\mathcal{A}$, Alice would also like to send Bob the set of his potential probes that are successful from the $kT$ probes he may make. Sending even this set of successful probes explicitly, however, (e.g., using $\Omega(\log n)$ bits for each of the at most $2B$ gadgets containing successful probes) is too expensive. We now see how Alice can represent this compactly in her message to Bob.%
    
    \paragraph{Compacting the successful probes} 
    It is sufficient for Alice to send Bob the indices of the at most $2B$ gadgets for which he makes successful probes. Since Alice and Bob used shared randomness to generate $G$, Bob already knows which edges within a gadget were potentially swapped, so the indices of the gadgets in which $\mathcal{A}$ makes successful probes are sufficient information for Bob to fully simulate an adjacency-list oracle as he runs $\mathcal{A}$: he answers a probe to a potentially swapped edge of gadget j as swapped if and only if j is among the indices sent by Alice. This is correct since, by induction over the probes, Bob's run coincides with Alice's simulation for $i$, so every probe of his run that finds a swapped edge was a successful probe in Alice's simulation, and hence its gadget is among the indices sent by Alice.
    Thus, Alice can compress her successful probe message by encoding it as a choice of subset of $[k]$ of cardinality at most $2B = 2k/c$. 
    Since $2B\leq k$, there are at most $\sum_{h=0}^{2B}\binom{k}{h}\leq (ek/(2B))^{2B}=(ec/2)^{2B}$ such subsets. So, together with one bit indicating whether she aborts and the $B$ bits of the state, Alice's message has at most $B+2+2B\log_2(ec/2)$ bits. For $c=2048$, this is at most $B+2+23B<32B=k/64$ for every $B\geq 1$.
    If the tail bound offered by Lemma \ref{lem:wc_lb_unsuccessful_bob} fails, then Alice simply aborts; this only reduces the success probability to $3/4 - 2^{-B} > 2/3$ (for $B\geq 4$).
    
    Thus, for $c=2048$, we have a communication protocol between Alice and Bob in violation of Theorem \ref{thm:index_lb}. This contradicts our assumption that $T \leq \binom{a}{2}/c^2  = \Theta(n^2/B^2)$, completing the proof of Theorem \ref{thm:lb-list-wc}.
\section{Incremental Amortized Lower Bound}
In this section, we show our $\Omega(n^{1.5}/\sqrt{B})$ lower bound on the amortized number of probes made by an algorithm with $B$ bits of storage and oracle access to the adjacency matrix of the current graph in the incremental setting. This lower bound implies a lower bound of $\Omega(\sqrt{n/B})$ in the adjacency list model. Formally, we show the following:
\begin{restatable}{theorem}{ioamortizedlb}
    \label{thm:io-amortized-lb}
    There are absolute constants $c, c_0>0$ such that for large enough $n$ and $B\leq c_0n$, for every $B$-bit adjacency matrix algorithm with total probe budget $q_{\text{total}}<cn^{5/2}/\sqrt{B}$ there is a fixed sequence of $m=\Theta(n)$ insertions on which the algorithm answers all queries correctly with probability strictly less than $2/3$. Since the budget is an upper bound on every execution on every input of length $m$, success probability at least $2/3$ requires amortized $\Omega(n^{3/2}/\sqrt{B})$ probes. 

    Since an adjacency list probe can be simulated with $O(n)$ probes in the adjacency matrix model, an amortized lower bound of $\Omega(\sqrt{n/B})$ probes holds for the adjacency list model.
\end{restatable}

We first explain the idea of the construction. As in \cref{sec:lower-bound}, the graph consists of two
vertex-disjoint paths, one containing $s$ and the other containing $t$, which run side by side and switch
tracks according to hidden random bits. Without deletions, however, we cannot repeatedly connect $t$ to a
vertex and remove the edge again. Instead, every round asks a new question by inserting two edges between the vertices of a fresh pair of positions. Depending on these edges, they either connect each path to itself, or connect
the two paths, which connects $s$ and $t$. To answer, the algorithm must therefore know how the two endpoints
of the edges lie relative to each other on the paths.

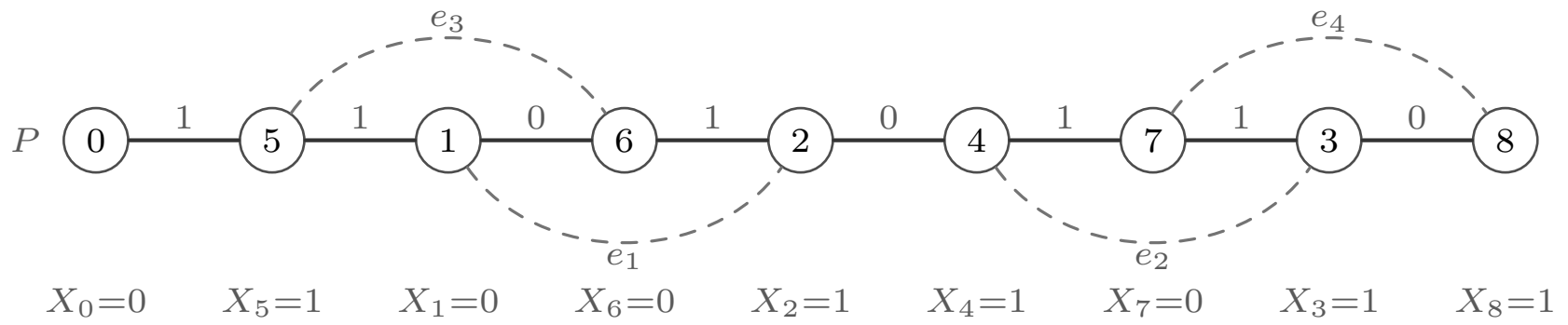
\begin{figure}[h]
    \centering
    \begin{tikzpicture}[x=1cm, y=1cm, line cap=round, line join=round,
  nd/.style={circle, draw=black!70, fill=white, line width=0.5pt, inner sep=0pt, minimum size=13pt, font=\scriptsize},
  tag/.style={font=\scriptsize, text=black!65, inner sep=1pt},
  pth/.style={line width=0.8pt, black!80},
  mat/.style={line width=0.6pt, dashed, black!55},
]
\def\s{1.25}
\foreach \v [count=\k from 0] in {0,5,1,6,2,4,7,3,8} \node[nd] (p\v) at (\k*\s,0) {$\v$};
\foreach \a/\b/\p in {0/5/1, 5/1/1, 1/6/0, 6/2/1, 2/4/0, 4/7/1, 7/3/1, 3/8/0}
  \draw[pth] (p\a) -- node[tag, above=1pt] {$\p$} (p\b);
\draw[mat] (p5) to[bend left=55]  node[tag, above] {$e_3$} (p6);
\draw[mat] (p7) to[bend left=55]  node[tag, above] {$e_4$} (p8);
\draw[mat] (p1) to[bend right=55] node[tag, below] {$e_1$} (p2);
\draw[mat] (p4) to[bend right=55] node[tag, below] {$e_2$} (p3);
\foreach \v/\x [count=\k from 0] in {0/0,5/1,1/0,6/0,2/1,4/1,7/0,3/1,8/1}
  \node[tag] at (\k*\s,-1.15) {$X_{\v}{=}\x$};
\node[tag, left] at (-0.35,0) {$P$};
\end{tikzpicture}
    \caption{The graph on $V_N$ (with $N=9$ and more matching edges than in the construction). The
hidden path $P$ is drawn in its own order, and each of its edges $\{u,v\}$ is labeled with $X_u\oplus X_v$.
The dashed edges form the public matching $M$; in round $j$, a gadget on $e_j=\{u_j,v_j\}$ is inserted. The
parity $\beta_j=X_{u_j}\oplus X_{v_j}$ is the XOR of the labels along $P$ between $u_j$ and $v_j$; for
example, $\beta_4=X_7\oplus X_8=0\oplus1=1$.}
    \label{fig:io-vn}
\end{figure}

Concretely, we take a uniformly random Hamiltonian path $P$ on a set $V_N$ of $N$ positions, starting at
position~$0$, and a uniformly random bit $X_i$ for every position $i$, with $X_0=0$ (\cref{fig:io-vn}). Every
position $i$ is represented by two vertices $i^0$ and $i^1$. Along every edge $\{u,v\}$ of $P$, the two tracks
pass straight if $X_u\oplus X_v=0$ and cross otherwise, so that $i^b$ lies on the path of $s$ if and only if
$b=X_i$ (\cref{fig:io-v}). In round $j$, we insert the two edges $\{u_j^0,v_j^b\}$ and $\{u_j^1,v_j^{1\oplus b}\}$ for a bit $b\in\{0,1\}$, where $\{u_j,v_j\}$ are
the two positions of the $j$-th edge of a fixed public matching $M$. Whether it connects the two paths depends on whether or not $b=X_{u_j}\oplus X_{v_j}$. Unlike in \cref{sec:lower-bound}, there are no slots in a fixed order:
the order of the positions along the paths is itself hidden. Informally, to learn $X_{u_j}\oplus X_{v_j}$, the algorithm
has to find a chain of edges of $P$ and earlier matching edges between $u_j$ and $v_j$, and, as $P$ is random,
each probe finds an edge of $P$ only with probability $O(1/N)$.

In \cref{sec:io_construction}, we define the construction formally. In \cref{sec:io_block}, as in the fully
dynamic case, we bound the number of probes in a block of $\Theta(B)$ rounds, given a $B$-bit advice string.
In \cref{sec:io_final}, we combine the blocks to prove \cref{thm:io-amortized-lb}.

\subsection{Construction}
\label{sec:io_construction}
Assume $n$ is large enough, and let $N=\lfloor\frac{n}2\rfloor$ and $V_N=\{0,\ldots, N-1\}$. Let $M$ be a matching on $V_N$ such that $M=\{e_j=\{u_j=2j-1, v_j=2j\}|j\in[R]\}$ where $R=\lfloor\frac{N-1}8\rfloor$.
$P=(0,\pi(1),\ldots,\pi(N-1))$ is a Hamiltonian path of $V_N$ chosen uniformly at random among all Hamiltonian paths such that the first position is $0$ and avoids the edges of $M$, i.e., $E(P)\cap M=\emptyset$.

\begin{figure}[h]
    \centering
    \definecolor{arczero}{RGB}{36,96,186}
\definecolor{arcone}{RGB}{204,68,44}
\definecolor{gadzero}{RGB}{34,139,34}
\begin{tikzpicture}[x=1cm, y=1cm, line cap=round, line join=round,
  vtx/.style={circle, fill=black!85, inner sep=0pt, minimum size=3.4pt},
  big/.style={circle, fill=black!85, inner sep=0pt, minimum size=4.4pt},
  lbl/.style={font=\scriptsize, inner sep=1pt},
  up/.style={lbl, above=3.5pt},
  dn/.style={lbl, below=3.5pt},
  tag/.style={font=\scriptsize, text=black!65, inner sep=1pt},
  arc/.style={line width=0.8pt},
  qry/.style={line width=0.6pt, dashed, black!55},
]
\def\h{0.55}
\def\s{1.25}
\foreach \v [count=\k from 0] in {0,5,1,6,2,4,7,3,8} {
  \coordinate (a\v) at (\k*\s,\h);
  \coordinate (b\v) at (\k*\s,-\h);
}
\draw[arc, arczero] (a0) -- (b5) -- (a1) -- (a6) -- (b2) -- (b4) -- (a7) -- (b3) -- (b8);
\draw[arc, arcone]  (b0) -- (a5) -- (b1) -- (b6) -- (a2) -- (a4) -- (b7) -- (a3) -- (a8);
\draw[qry] (a1) .. controls +(0.5,1.2) and +(-0.5,1.2) .. node[tag, above] {round $1$ of $I_1$} (a2);
\draw[qry] (b1) .. controls +(0.5,-1.2) and +(-0.5,-1.2) .. (b2);
\draw[qry, gadzero] (a1) -- (b2);
\draw[qry, gadzero] (b1) -- (a2);
\node[tag, text=gadzero, fill=white, align=center, inner sep=0.5pt, font=\tiny] at (3.55,0) {round $1$\\of $I_0$};
\draw[qry] (a7) .. controls +(0.5,1.2) and +(-0.5,1.2) .. node[tag, above] {round $4$ of $I_4$} (a8);
\draw[qry] (b7) .. controls +(0.5,-1.2) and +(-0.5,-1.2) .. (b8);
\foreach \v in {5,1,6,2,4,7,3,8} { \node[vtx] at (a\v) {}; \node[vtx] at (b\v) {}; }
\node[big] at (a0) {}; \node[big] at (b0) {};
\node[lbl, left=3pt] at (a0) {$s=0^0$};
\node[lbl, left=3pt] at (b0) {$t=0^1$};
\foreach \v in {5,6,4,3} { \node[up] at (a\v) {$\v^0$}; \node[dn] at (b\v) {$\v^1$}; }
\node[lbl, above left=2pt and 0pt]  at (a1) {$1^0$};
\node[lbl, below left=2pt and 0pt]  at (b1) {$1^1$};
\node[lbl, above right=2pt and 0pt] at (a2) {$2^0$};
\node[lbl, below right=2pt and 0pt] at (b2) {$2^1$};
\node[lbl, above left=2pt and 0pt]  at (a7) {$7^0$};
\node[lbl, below left=2pt and 0pt]  at (b7) {$7^1$};
\node[lbl, above right=2pt and 0pt] at (a8) {$8^0$};
\node[lbl, below right=2pt and 0pt] at (b8) {$8^1$};
\end{tikzpicture}
    \caption{The graph on $V$ for the same $P$ and $X$ as in \cref{fig:io-vn}. Position $v$ is represented by
$v^0$ (top) and $v^1$ (bottom). Along every edge $\{u,v\}$ of $P$, the two tracks pass straight if
$X_u\oplus X_v=0$ and cross otherwise, so $v^b$ lies on the path of $s$ (blue) if and only if $b=X_v$. The
dashed edges show gadgets inserted in different input sequences. In round~$1$ of $I_0$, the gadget
$g(1,2,1)$ (green) has the correct parity $\beta_1=X_1\oplus X_2=1$, so each of its edges joins two vertices
on the same path. In round~$1$ of $I_1$ and round~$4$ of $I_4$, the gadgets $g(1,2,0)$ and $g(7,8,0)$ (gray)
have the wrong parity, since $\beta_1=\beta_4=1$, so each of them connects the two paths, and $s$ and $t$
become connected.}
    \label{fig:io-v}
\end{figure}
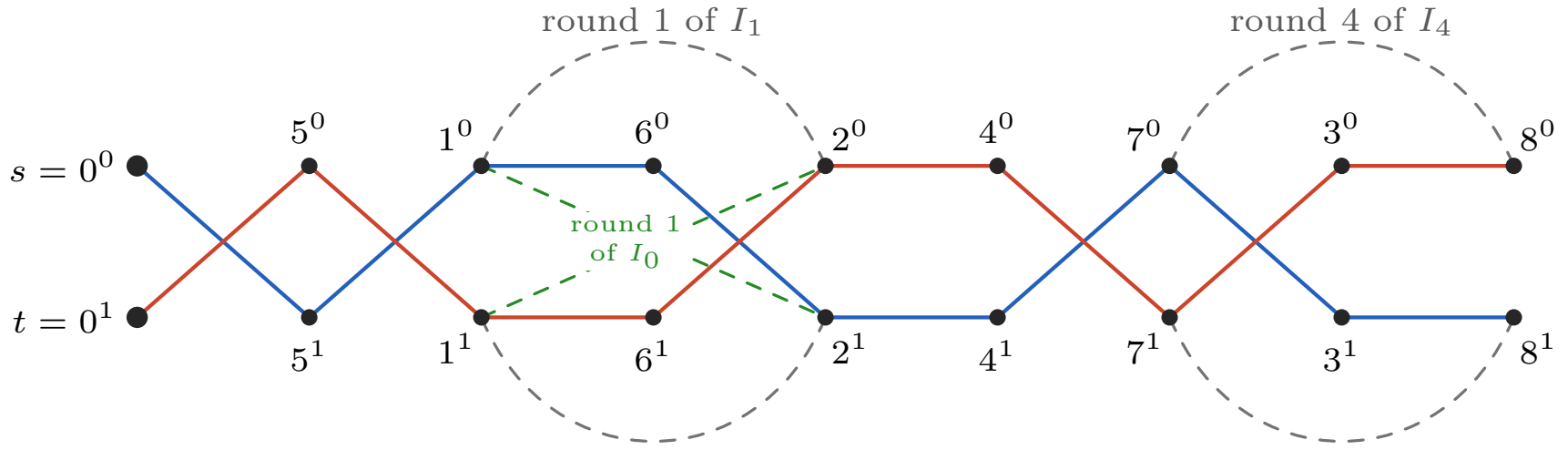

Now, let the vertex set of the graph be $V=\{i^b|i\in V_N,\,b\in\{0,1\}\}$ such that $s=0^0$ and $t=0^1$. If $n$ is odd, the remaining vertex is isolated and will receive no insertions throughout the stream. For any pair $u,v\in V_N$ and any bit $b$ define $g(u,v,b)=\{\{u^0,v^b\},\{u^1,v^{1\oplus b}\}\}$ (as shown in \cref{fig:io-v}). Now, we introduce the input sequences corresponding to the path $P$ as stated above and the binary vector $X$ such that $X_0=0$ and for any other $i\in V_N$, $X_i$ is chosen uniformly at random and independently. Let $I_0$ be as follows: in the first $2(N-1)$ insertions, insert $g(u,v,X_u\oplus X_v)$ for any two consecutive vertices $u$ and $v$ of $P$. Then, for $R$ \emph{rounds}, in round $j\in[R]$, insert the edges of $g(u_j,v_j,\beta_j)$ where, as above, $u_j=2j-1$ and $v_j=2j$, and $\beta_j:=X_{u_j}\oplus X_{v_j}$. In the input sequence $I_j$ (for any $j\in[R]$), we have insertions identical to $I_0$ with the one exception that in round $j$, instead of $g(u_j,v_j,\beta_j)$, we insert the edges of $g(u_j,v_j,1\oplus \beta_j)$. Note that edges inserted in the rounds $j\in[R]$ correspond to the matching $M$ on $V_N$. See \cref{fig:io-v} for an example. From the construction, we have the following:
\begin{observation}
    Every insertion sequence $I_j$ ($0\leq j\leq R$) contains $2(N-1)+2R=\Theta(n)$ insertions. 
\end{observation}

\begin{claim}
    In the sequence $I_0$, $s$ and $t$ are not connected throughout the whole sequence. In $I_j$ for any $j>0$ however, $s$ and $t$ are not connected until round $j$, but they become connected in round $j$ (and stay connected thereafter).
\end{claim}
\begin{proof}
    After the first $2(N-1)$ insertions, the graph consists of two vertex-disjoint paths; let path~$0$ be the
    one containing $s=0^0$ and path~$1$ the one containing $t=0^1$. From the construction, it can be shown by
    induction on $i$ that for $w=\pi(i)$, if $X_w=0$, then $w^0$ is on path~$0$ and $w^1$ is on path~$1$, and
    if $X_w=1$, then $w^0$ is on path~$1$ and $w^1$ is on path~$0$. Thus, for every $w\in V_N$ and
    $b\in\{0,1\}$, $w^b$ is on path~$b\oplus X_w$. Thus, the insertion of $g(u_j,v_j,\beta_j)$ only inserts
    edges whose endpoints are on the same path, and after the insertion of $g(u_j,v_j,1\oplus\beta_j)$, the
    two paths, and hence $s$ and $t$, are in the same connected component.
\end{proof}

Define the following oracle for every $\{u,v\}\in\binom{V_N}{2}$:
\[
\mathcal{O}_{P,X}(u,v) =
\begin{cases}
X_u \oplus X_v, & \text{if } \{u,v\}\in E(P),\\
\bot,          & \text{otherwise}.
\end{cases}
\]
If the algorithm is in the $j$-th round, and has free access to $(\beta_i)_{i<j}$ and the matching $M$ (which only makes the lower bound stronger), for any probe to the pair $\{u^b,v^{b'}\}$ the algorithm either already knows the answer to the probe (from freely available information), or it can find the answer using $\mathcal{O}_{P,X}(u,v)$, under these assumptions, it is enough to show the bound for algorithms that (instead of adjacency matrix) have access to this oracle.

\subsection{Block Lower Bound}
\label{sec:io_block}
\subsubsection{High-level Idea}
In this subsection we give an intuition of the proof. As in the fully dynamic case, we split the rounds into blocks of $k=\Theta(B)$
rounds and view the memory at the start of a block as $B$ bits of advice. Call round $j$ \emph{determined}
if $u_j$ and $v_j$ are connected by edges of $P$ revealed so far and earlier matching edges, since then the
parities of these edges determine $\beta_j$. If round $j$ is not determined, then $\beta_j$, and hence the
correct answer, is a fair coin even given everything the algorithm has seen. Informally, the $B$ bits of
advice can account for only $O(B)$ correct answers, so a correct algorithm needs $\Omega(k)$ determined rounds
in a block.

Every determined round closes a cycle in the graph formed by $M$ and the revealed edges of $P$, so the number
of determined rounds is at most the cycle rank of this graph. Two facts keep this cycle rank small. First,
since $P$ is random, each probe reveals a new edge of $P$ only with probability $O(1/N)$
(\cref{lem:edge_prob}), so few probes reveal few edges (\cref{lem:unlikely_halt}). Second, a newly revealed
edge closes a cycle only if both of its endpoints already lie in the part of the graph touched by earlier
revealed edges and their matching partners. With $r$ revealed edges, this part has only $O(r)$ vertices, so
informally, a new edge closes a cycle with probability $O(r/N)$, and $r$ revealed edges close only
$O(r^2/N)$ cycles in total (\cref{lem:rank_bound}). Hence, informally, $\Omega(k)$ determined rounds require
$r=\Omega(\sqrt{Nk})$ revealed edges, and thus $\Omega(N\sqrt{Nk})$ probes per block of $k=\Theta(B)$ rounds,
which is amortized $\Omega(N^{3/2}/\sqrt{B})$ probes per round. 

Two technical points remain. First, in $I_0$, the correct answer is always ``not connected'', so an algorithm that always answers ``not
connected'' is correct on $I_0$. What is hard is to tell $I_0$ apart from $I_j$, that is, to tell whether the
gadget inserted in round $j$ has parity $\beta_j$, which keeps $s$ and $t$ disconnected, or $1\oplus\beta_j$,
which connects them. Therefore, for every round $j$, we analyze an experiment that runs a copy of the
algorithm on a gadget of uniformly random parity $\alpha_j$ and checks whether its answer is correct. Second, to
simplify the analysis, we give the algorithm some information for free, which only makes the lower bound
stronger: the matching $M$, the parities $\beta_i$ of all earlier rounds, the last vertex of $P$, and, for
every vertex that has been probed more than $\tau=\Theta(N)$ times in the current block, its two neighbors in
$P$ together with the parities of these edges. Fixing the last vertex of $P$ makes $P$ uniform among
Hamiltonian paths with fixed endpoints, as \cref{lem:edge_prob} requires. Revealing the neighbors of
frequently probed vertices prevents any vertex from collecting many $\bot$ probe answers, which would make the
distribution of its neighbors far from uniform.

\subsubsection{Formal Proof}
Define a block to be $k=512(B+2)$ consecutive rounds $J=\{\ell+1,\ldots,\ell+k\}$. Let $Z$ be the $B$-bit arbitrary advice that the algorithm receives at the beginning of the block (internal state of the algorithm). $Z$ can arbitrarily depend on $(P,X)$. The algorithm may use unlimited working space throughout the block and we only use the $B$-bit limit for the memory handed to the algorithm at the beginning of a block.

In the rest of this section, we assume the following information is publicly available during an input sequence, so the algorithm can use them for free. 
\begin{itemize}
    \item The matching $M$.
    \item The parity $\beta_i$ for all of the previously answered rounds, i.e., $(\beta_i)_{i<j}$.
    \item The last vertex of the path $P$, i.e., $\pi(N-1)$.
    \item For every vertex that has been probed in the oracle during this block for more than $\tau=\lfloor N/1000\rfloor$ times, its incident edges in $P$ and their parities. The edges are revealed sequentially in uniformly random order.
\end{itemize}

When entering a block, the values $\pi(N-1)$ and $(\beta_i)_{i\leq\ell}$ are fixed and all of the probabilities and expectations are conditioned on these values.

\paragraph{Experiments} For an algorithm $\calA$ define the experiments $\calE_j$ for $j\in[R]$ such that in experiment $\calE_j$, we take fair bit $\alpha_j$ (independent of everything else) and give the algorithm the following input sequence: for all the rounds except round $j$ the input is identical to $I_0$, and in round $j$ we insert $g(u_j,v_j,\alpha_j)$. This way, if $\alpha_j=\beta_j$, we are running the algorithm on $I_0$ and otherwise on $I_j$. 
We run these experiments using the same execution of the algorithm on $I_0$: for every $j$, before giving the algorithm the actual insertions corresponding to $\beta_j$, we take a copy of the current state of the algorithm and insert the two edges corresponding to the $j$-th round of the input sequence corresponding to $\alpha_j$. Then, we record the output of the algorithm in this experiment (let it be $A_j$ which is $1$ if the algorithm answers ``connected’' and $0$ otherwise), discard this copy of the algorithm, and continue with the input sequence $I_0$. Note that although these experiments are not independent (since they use the same execution on $I_0$), our arguments will still hold because we only need the linearity of expectations on the correctness indicators $\mathbf{1}\{A_j=\beta_j\oplus \alpha_j\}$ (which is equal to $1$ if and only if the algorithm succeeds in the $j$-th round of experiment $\calE_j$).

Let the set $U$ contain all of the distinct edges of $P$ that are revealed, either by oracle probes, or from the last item above. For the sake of analysis, assume that the oracle stops answering the queries when $|U|=r$ (where $1\leq r=\Theta(\sqrt{Nk})\leq N/4$ will be fixed later in the proof), even if it is in the middle of revealing a neighborhood (call this event \texttt{HALT}). If \texttt{HALT} occurs, for the analysis, set the rest of the answers of the algorithm to be all $0$. We call these answers \emph{truncated answers}, denoted by $\widetilde{A}_j$ for $j\in J$. Note that if \texttt{HALT} does not occur, $A_j=\widetilde{A}_j$. 

For each block $J$, we analyze the execution on $I_0$ together with the copies used to obtain $A_j$ for $j\in J$. Every oracle probe made by either the execution on $I_0$ or one of these copies counts toward the block's total probe count. The counters used for the threshold $\tau$, the set $U$, and the transcript used in the analysis accumulate information from all these probes; discarding a copy does not erase this information from the analysis. These counters and $U$ start fresh at the beginning of each block. The only initial advice is
the $B$-bit state of the execution on $I_0$ at the block's start.

In the following lemma, for a random Hamiltonian path containing a set $H$ of limited cardinality and avoiding a set $F$ of limited maximum degree, we bound the probability of any single edge being in the path. This lemma will be used later in the proof.
\begin{lemma}
\label{lem:edge_prob}
For $N\geq 32$, let $H,F\subseteq\binom{V_N}{2}$ and $P$ be a uniformly random Hamiltonian path on vertices of $V_N$ starting from $0$ and finishing at $\pi(N-1)$ satisfying $H\subset E(P)$ and $E(P)\cap F=\emptyset$. If $|H|\leq N/4$ and $\Delta(F)\leq N/100$, then for every $e\in\binom{V_N}{2}\setminus H$, we have $\Pr[e\in E(P)]\leq8/N.$ 
\end{lemma}
\begin{proof}
    If $e\in F$, then $\Pr[e\in E(P)]=0$. Otherwise, fix a path $P$ that satisfies the constraints and contains $e$, and let $e=\{u,v\}$ such that $u$ comes before $v$ in $P$. Let $\{x,y\}\in E(P)$ be such that $x$ comes before $y$ in $P$, and $\{x,y\}\cap \{u,v\}=\emptyset$. Now, replace the edges $\{u,v\}$ and $\{x,y\}$ by $\{x,u\}$ and $\{y,v\}$ to get to $P'$, which is also a Hamiltonian path (see \cref{fig:edge_prob}). Note that $P'$ has the same endpoints as $P$, contains $H$ since $\{x,y\}\notin H$, and avoids $F$ since $\{x,u\},\{y,v\}\notin F$.
    \begin{figure}[htbp]
    \centering
    \includegraphics[width=0.5\linewidth]{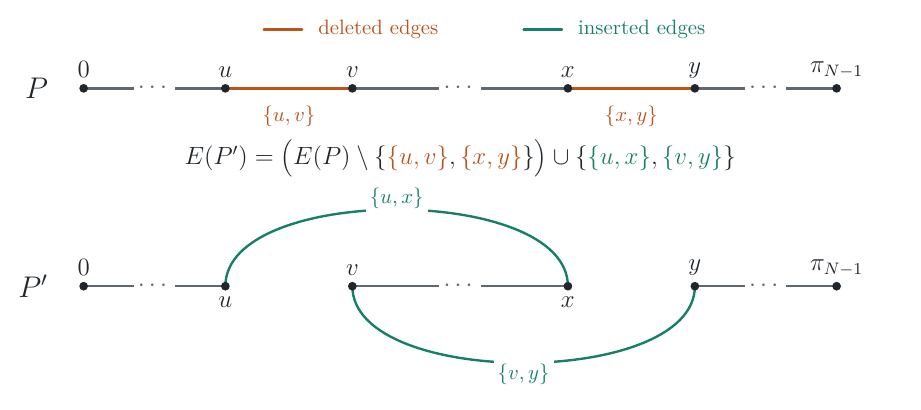}
    \caption{Construction of $P'$ by replacing $\{u,v\}$ and
    $\{x,y\}$ with $\{u,x\}$ and $\{v,y\}$.}
    \label{fig:edge_prob}
\end{figure}
    Now, we count how many options we have for the choice of $\{x,y\}$ (and hence $P'$ which does not include $\{u,v\}$). There are $N-1$ edges in $P$ in total, at most $3$ of which are incident to $v$ or $u$. At most $|H|$ of the remaining options are in $H$ and at most $2N/100$ of them are edges incident to $u$ or $v$ in $F$. Thus, the total number of options is at least: 
    $$(N-1) - 3 - |H| - 2N/100\geq N/2.$$
    Now, for the paths $P'$ that are made like this, each one of them has at most $4$ pre-images since $x$ is one of the two neighbors of $u$ in $P'$, and $y$ is a neighbor of $v$ in $P'$. Therefore, we have:
    $$\Pr[e\in E(P)]= \frac{\#\{P:e\in E(P)\}}{(\#\{P:e\in E(P)\})+(\#\{P:e\notin E(P)\})}\leq\frac{\#\{P:e\in E(P)\}}{\#\{P:e\notin E(P)\}}\leq \frac{4}{N/2}=\frac8N.$$
\end{proof}
In the next lemma, we show that if the number of probes made during this block is small enough, then we can bound the probability of \texttt{HALT}.
\begin{lemma}
    \label{lem:unlikely_halt}
    Let $q$ be an upper bound on the combined number of oracle probes made during block $J$, including those made by the experiment copies. If $q\leq \frac{r\tau}{8}$, then:
    $$\Pr[\texttt{HALT}]\leq2^B(\frac{16eq}{Nr})^{\lceil r/2\rceil}.$$
\end{lemma}
\begin{proof}
    Fix a candidate initial state $\zeta\in\{0,1\}^B$, and let $\Pr_\zeta$ denote probability
when $(P,X)$ is sampled from its original distribution
under the fixed public information, and the algorithm
is initialized with $\zeta$ and stopped after $q$ probes.

    Before $|U|=r$ (occurrence of \texttt{HALT}), apply \cref{lem:edge_prob} with $H=U$, and let $F$ be the set $M$ and all pairs that oracle answers $\bot$. For any vertex whose neighborhood has been fully revealed (by passing the $\tau$ threshold), remove its incident edges from $F$ because its incident edges in $H$ make the edges in $F$ redundant. Now, every vertex has degree at most $\tau+2\leq N/100$ in $F$. Since $\Delta(F)\leq N/100$ and $|H|\leq r\leq N/4$, \cref{lem:edge_prob} can be applied (note that $P$ is uniform subject to the revealed edge constraints).

    Note that since $q\leq r\tau/8$ probes are made, at most $r/4$ nodes have reached the $\tau$ threshold (two nodes per probe), and thus, the neighborhood of these vertices contribute at most $r/2$ edges to $U$ (at most two edges for each node). Thus, for $|U|$ to reach $r$, there must be at least $\lceil r/2\rceil$ distinct edges that were discovered by the non-$\bot$ probes to the oracle. Now, since the probability of each probe being successful (meaning it does not get a $\bot$ for a previously unrevealed edge) is at most $8/N$ (from \cref{lem:edge_prob}), we have:
    $$\Pr_\zeta[|U|=r]\leq\binom{q}{\lceil r/2\rceil}(\frac{8}{N})^{\lceil r/2\rceil}\leq(\frac{16eq}{Nr})^{\lceil r/2\rceil}.$$
    Where the probability above is for a fixed $\zeta$. Thus, by a union bound over all choices of $\zeta$, the lemma will be concluded.
\end{proof}
In the analysis, we use the definition of \emph{cycle rank} of a graph, defined as $\operatorname{rank}(G):=|E(G)|-|V(G)|+\operatorname{cc}(G)$, where $\operatorname{cc}(G)$ represents the number of connected components of $G$. The following lemma bounds the probability of cycle rank of $(V_N, M\cup U)$ being high.
\begin{lemma}
    \label{lem:rank_bound}
    Let $U$ be the final revealed set of the block $J$. For every integer $h\geq1$,
    $$\Pr[\operatorname{rank}(V_N, M\cup U)\geq 2h]\leq 2^{B+1}(\frac{64er^2}{Nh})^h.$$
\end{lemma}
\begin{proof}
    Take $\zeta$ and $\Pr_\zeta$ similar to the proof of \cref{lem:unlikely_halt}. Also, note that $P$ is uniform subject to the revealed edge constraints ($U$ and the probes that resulted in $\bot$), thus, again, if we remove the edges of $F$ for which an endpoint has reached the $\tau$ threshold and its neighborhood is fully revealed, we can apply \cref{lem:edge_prob}. Throughout the proof, $U$ stands for the edges revealed so far.

    Since $\operatorname{rank}(V_N, M)=0$, we can start by this graph and add the revealed edges one by one and count the edges that add $1$ to the cycle rank. There are two types of revealed edges: the ones that are revealed because one of their endpoints reached the $\tau$ threshold (\emph{threshold type}), and those that were probed directly (\emph{probe type}). If $\operatorname{rank}(V_N,M\cup U)\geq 2h$, one of these two types has contributed at least $h$ to the rank, thus it is enough to bound the probability of each type contributing $h$ and add the bounds (by union bound).

    First, we bound the threshold type. Every edge adds one to the rank only if its endpoints are in the same component of $M\cup U$. On the other hand, since $M$ is a matching and $|U|\leq r$, every component of $M\cup U$ has at most $2r+2$ vertices. Thus, for every such vertex $v$ whose neighborhood is revealed, there are at most $2r+1$ vertices $u\in V_N\setminus\{v\}$ for which $\{u,v\}$ adds one to the cycle rank. Thus, by union bound, the probability of each such edge adding to the rank is at most $\frac{8}{N}\cdot(2r+1)\leq\frac{32r}{N}$. There are at most $r$ edges in this type, so the probability of them contributing at least $h$ to the rank is at most:
    $$\binom{r}{h}(\frac{32r}{N})^h\leq (\frac{32er^2}{Nh})^h.$$

    Next, we bound the probe type. Let $\calM(U)$ be the set of endpoints of $U$ and their partners in $M$ (if they exist). Since $|V(U)|\leq 2r$, $|\calM(U)|\leq 4r$. Insertion of an edge from $E(P)\setminus U$ increases the cycle rank only if at the time of its insertion both of its endpoints are in $\calM(U)$ because each component outside of $\calM(U)$ is either an isolated vertex or an edge of $M$. Consider only the distinct probes for the previously unrevealed edges. Since $\calM(U)$ grows with $U$ and its final size is at most $4r$, there are at most $\binom{4r}{2}\leq8r^2$ vertex pairs whose probe could increase the cycle rank, and each one of them is an edge in $E(P)$ with probability at most $8/N$ (conditional on the transcript so far). Hence, the probability of edges of this type contributing at least $h$ is at most:
    $$\binom{8r^2}{h}(\frac8N)^h\leq (\frac{64er^2}{Nh})^h.$$
    Thus, by union bound we have:
    $$\Pr_\zeta[\operatorname{rank}(V_N, M\cup U)\geq 2h]\leq 2(\frac{64er^2}{Nh})^h$$
    and by a union bound over different options of $\zeta$, we have:
    $$\Pr[\operatorname{rank}(V_N, M\cup U)\geq 2h]\leq 2^{B+1}(\frac{64er^2}{Nh})^h.$$
\end{proof}
For any $j\in J$, let $D_j$ be $1$ if $u_j$ and $v_j$ are connected in $(V_N, U_j\cup \{e_i:1\leq i<j\})$ and $0$ otherwise, where $U_j$ is the set $U$ immediately before $A_j$ is recorded. Also, let $D=\sum_{j\in J}D_j$. In the lemma below, we bound $D$ by the cycle rank of $(V_N, M\cup U)$.
\begin{lemma} \label{lem:D_bound}
    $D\leq\operatorname{rank}(V_N, M\cup U)$.
\end{lemma}
\begin{proof}
    We start by the graph $G_\ell=(V_N,U\cup\{e_i|i\leq\ell\})$ and insert the edges $e_{\ell+1},\ldots,e_{\ell+k}$ one by one and name these graphs $G_{\ell+1},\ldots,G_{\ell+k}$. For every edge $e_j$ that is counted in $D$, its two endpoints are in the same connected component of $G_{j-1}$, so their insertion adds one to the cycle rank. Note that the edge set of $G_j$ for any $j\in J$ is a subset of $M\cup U$, thus monotonicity of cycle rank under insertions gives the correctness of the lemma.
\end{proof}

\begin{lemma}
    \label{lem:C_bound_small_D}
    Let $\widetilde C_j$ be $1$ if the recorded truncated connectivity answer $\widetilde{A}_j$ is correct and $0$ otherwise, and $\widetilde C=\sum_{j\in J}\widetilde C_j$. Then, for every $0\leq d\leq k$ and $\lambda\geq 0$:
    $$\Pr[\widetilde C\geq k/2 + d/2 + \lambda\,\wedge\,d\geq D]\leq 2^Be^{-2\lambda^2/k}.$$
\end{lemma}
\begin{proof}
    If $d$ is non-integer, replacing it by $\lfloor d\rfloor$ enlarges the event since $D$ is an integer. Also, the bound is trivial for $d=k$ since $\widetilde C\leq k$. For the rest of this proof, we assume that $d$ is an integer and $d<k$.
    
    Take $\zeta$ and $\Pr_\zeta$ similar to the proof of \cref{lem:unlikely_halt,lem:rank_bound}.

    For each $j\in J$, let $T_j$ be the transcript of the algorithm after outputting the answer $\widetilde A_j$ and before revelation of $\beta_j$. Conditioned on $T_j$, $X$ is uniform subject to $X_0=0$ and the parity equations of the edges of the graph $(V_N, U_j\cup \{e_i:i<j\})$. If $D_j=0$, then $u_j$ and $v_j$ belong to different components and at least one of them is in a component that does not include $0$ (w.l.o.g let it be $u_j$). Changing the parity of all of the vertices in the component of $u_j$ preserves all of the known parity equations and changes $\beta_j$, so conditioned on $T_j$, $\beta_j$ and therefore $\widetilde C_j$ are uniform on $\{0,1\}$.

    Now, take the correctness indicators ($\widetilde C_j$) for the first $k-d$ rounds with $D_j=0$ (in the event that fewer such rounds occur, pad the list with independent fair bits). Since they are all independent fair bits conditioned on the transcript ($T_j$) before revelation of their correct parity ($\beta_j$), by successive conditioning, together with the independent padding bits, the completed list consists of $k-d$ independent fair bits.
    Hence, if $C'$ is the sum of these indicators, then $C'\sim\operatorname{Bin}(k-d,1/2)$. Moreover, there are at most $d$ other rounds; thus, $\widetilde C\leq C'+d$. Therefore, we have:
    $$
        \Pr_\zeta[\widetilde C\geq k/2 + d/2 + \lambda\,\wedge\,d\geq D]\leq\Pr_\zeta[C'\geq \frac{k-d}{2}+\lambda]
        \leq e^{-2\lambda^2/(k-d)}\leq e^{-2\lambda^2/k}
    $$
    where the second inequality is from Hoeffding's inequality: for a binomial random variable $Y\sim\operatorname{Bin}(K,1/2)$, we have $\Pr[Y\geq K/2 + \gamma]\leq e^{-2\gamma^2/K}$ for any $\gamma\geq0$. The lemma is proven by the statement above and a union bound over the $2^B$ choices for the internal state.
\end{proof}

Now, we are ready to show our desired lower bound for each block. First, we bound the probability of the total number of probes in the experiments and the number of correct recorded answers both being large.
\begin{corollary}
    \label{cor:initial_block_cor}
    Let $L=2^{16}$, $q=\lfloor\frac{N\sqrt{Nk}}{L^2}\rfloor$, and assume that $k\leq N/L^2$. If $T$ is the total number of oracle probes in the block and $C$ is the number of correct recorded answers $A_j$ in this block, then: $$\Pr[C\geq5k/8\,\wedge\,T\leq q]\leq\frac{1}{48}.$$
\end{corollary}
\begin{proof}
    For analysis, truncate before probe $q+1$, and follow the same completion procedure as explained for \texttt{HALT} in the case of truncation before probe $q+1$. Set $r=\lceil{\sqrt{Nk}/L}\rceil$ and $h=k/16$. With these assignments, we  have $q\leq Nr/L$ and $q\leq \frac{r\tau}{8}$, so \cref{lem:unlikely_halt} gives:
    $$\Pr[\texttt{HALT}]\leq 2^{B-r}.$$
    Also, $r^2\leq2Nk/L^2$, so \cref{lem:rank_bound} gives:
    $$\Pr[\operatorname{rank}(V_N, M\cup U)\geq k/8]\leq2^{B+1}(2048e/L^2)^{k/16}\leq2^{B+1-k}.$$
    Applying \cref{lem:C_bound_small_D} with $d=k/8$ and $\lambda=k/16$ implies that $\Pr[\widetilde C\geq 5k/8\,\wedge\,d\geq D]\leq2^Be^{-k/128}$. 
    Now consider the case that $\neg\texttt{HALT}$ and $\operatorname{rank}(V_N, M\cup U)< k/8$ and $q\geq T$. Note that in the case of $\neg\texttt{HALT}$ and $T\leq q$, $C=\widetilde C$.
    Thus, since by \cref{lem:D_bound} we have $D<k/8$, a union bound gives:
    $$\Pr[C\geq5k/8\,\wedge\,T\leq q]\leq2^{B-r}+2^{B+1-k}+2^Be^{-k/128}\leq2^{B+2-k/128}=2^{-3B-6}<\frac{1}{48}.$$
\end{proof}

Now, we bound the expected value of number of probes made only on $I_0$:

\begin{corollary}
    \label{cor:io_block}
    Use the parameters and assumptions of \cref{cor:initial_block_cor}. For a block $J$, let $Q_J$ be the number of adjacency matrix probes made by the algorithm on block $J$ of input $I_0$. Then,
    $$\frac{\mathbb{E}[C]}{k}\leq \frac{31}{48}+\frac{4\mathbb{E}[Q_J]}{q}.$$
    Expectations are conditioned on the public information at the block’s start and include all input and
    algorithmic randomness, as well as the randomness used by the experiments.
\end{corollary}
\begin{proof}
    For this proof, consider the following auxiliary computation of the experiments of block $J$: Each copied experiment is allowed at most $b=\lfloor q/(2k)\rfloor$ matrix probes across its two insertions. If it requested probe $b+1$, take its output to be ``connected'' and discard the copy, otherwise it runs normally. Note that this can only reduce the correctness when $\alpha_j=\beta_j$. Let $\widehat{C}$ be the number of correct answers recorded in these auxiliary experiments.

    Let $Q_j$ be the number of matrix probes made by $I_0$ on round $j$, so $Q_J=\sum_{j\in J}Q_j$. Conditioned on $\alpha_j=\beta_j$, the copy runs exactly as the original execution on $I_0$ until the cutoff, so the probability of the cutoff equals $\Pr[Q_j>b]$. Thus, $\Pr[\alpha_j=\beta_j\,\wedge\,\text{the copy is cut off}]=\frac12\Pr[Q_j>b]$. Therefore, we have:
    $$\mathbb{E}[C]\leq\mathbb{E}[\widehat C]+\frac12\sum_{j\in J}\Pr[Q_j>b]\leq\mathbb{E}[\widehat C]+\frac{\mathbb{E}[Q_J]}{2b},$$
     where the second inequality comes from Markov's inequality.   

    Each matrix probe requires at most one oracle probe, so if $\widehat{T}$ is the combined oracle probe count for this auxiliary computation (including the copies), we have $\widehat T\leq Q_J+kb\leq Q_J+q/2$. Therefore, by Markov's inequality we have:
    $$\Pr[\widehat T> q]\leq \frac{2\mathbb{E}[Q_J]}{q}.$$

    By applying \cref{cor:initial_block_cor} to this auxiliary computation, we have:
    \begin{align*}
        \mathbb{E}[\widehat C]\leq& \frac{5k}{8}+k(\Pr[\widehat{C}\geq 5k/8\,\wedge\,\widehat{T}\leq q]+\Pr[\widehat{T}>q])\\
        \leq&\frac{31k}{48}+\frac{2k\mathbb{E}[Q_J]}{q}.
    \end{align*}
    Since $b\geq q/(4k)$, combining the statement above and $\mathbb{E}[C]\leq\mathbb{E}[\widehat C]+\frac{\mathbb{E}[Q_J]}{2b}$ proves the claim.
\end{proof}

\subsection{Proof of \cref{thm:io-amortized-lb}}
\label{sec:io_final}
\ioamortizedlb*
\begin{proof}
    Use $k$ and $q$ from \cref{cor:initial_block_cor}. Choose $c_0$ small enough that $k\leq N/L^2$ and $k\leq R/2$, and for $W=\lfloor R/k\rfloor$, partition the first $Wk$ rounds into $W$ blocks $J_1,\dots,J_W$.

    Let $C^t$ be the original count of correct answers in block $J_t$. Apply \cref{cor:io_block} to each block, average over their initial public information, and sum. By definition $\sum Q_{J_t}\leq q_{\text{total}}$. So,
    $$\frac{1}{Wk}\sum_{t=1}^W\mathbb{E}[C^t]\leq \frac{31}{48}+\frac{4q_{\text{total}}}{Wq}.$$
    Since $W=\Theta(n/k)$ and $k=\Theta(B)$, we have $Wq=\Theta(n^{2.5}/\sqrt{B})$. So if $c$ is small enough that $q_{\text{total}}<Wq/192$, the right-hand side of the inequality above becomes strictly less than $2/3$.

    By linearity of expectation, this is the average success probability of the original experiments. Each experiment chooses $I_0$ or $I_j$ with equal probability (conditioned on $(P,X)$). Therefore, for some fixed $(P,X)$, round $j$, and choice of input, the success probability is less than $2/3$. The probability that answers of all rounds are correct is no greater, so the proof is completed. 
\end{proof}
\section{Upper Bounds for Incremental Algorithms}
\label{sec:upper-bound}

In this section, we prove \cref{thm:upper}.
Throughout, we fix integers $n\ge2$ and $1\le S\le n$, two vertices $s,t\in[n]$, and an incremental update sequence $(e_i=u_iv_i)_{i=1}^M$ to an initially empty graph $G = ([n], E)$, which is fixed in advance and thus independent of the algorithm's randomness.
As in \cref{sec:prelims}, $G_k$ is the graph after the first $k$ insertions, and update $i$ is processed with oracle access to $G_{i-1}$; after it, we report whether $s$ and $t$ are connected in $G_i$.
Let $C_k(v)$ be the vertex set of the component of $v$ in $G_k$.

\paragraph{Output-adaptive adversaries.}
The guarantees of
\cref{thm:upper}
also hold against \emph{output-adaptive adversaries}, whose updates
and adjacency-list orderings may depend on previous connectivity
answers but on no other information about the algorithm's execution.
Indeed, connectivity answers are uniquely determined by the graph,
and any adversary gets from the connectivity answers no information about the internal state of the algorithm.
Thus, the reduction discussed by Aamand et al.\
\cite[p.~6:2, footnote~2]{AKLPRT23} applies directly to our formulation, which means that we may assume that the update sequence and
adjacency-list orderings have been fixed in advance, i.e., that the adversary is oblivious.

\subsection{Proof of \texorpdfstring{\cref{thm:largespace}}{Theorem \getrefnumber{thm:largespace}}}
\label{sec:ub-large}

We maintain a set $B \subseteq V$ of \emph{landmarks} and a disjoint-set union structure
$D$ on $B$ as in \cref{lem:union-find}. Initially, $B_0=\{s,t\}$ and
each landmark is in its own class. Write $B_k$ for the landmark set
after update $k$, so $B_{i-1}$ is the set before update $i$ begins.
The unsubscripted $B$ denotes the current set, which can grow during an
update, so $B_{i-1}\subseteq B\subseteq B_i$.
A (connected) component is \emph{landmark-free} if it contains no landmark in the set $B$.

\SetKwFunction{Uniform}{Uniform}
\SetKwFunction{Bernoulli}{Bernoulli}
\SetKwFunction{UBLocate}{Locate}
\SetKwFunction{UBAddRoot}{AddLandmark}
\SetKwFunction{UBRootEdge}{LandmarkEdge}
\SetKwFunction{UBUpdate}{Update}
\SetKwProg{UBFn}{Function}{:}{}

\begin{algorithm}[htb]
\caption{Helper functions}
\label{alg:ub-large-helpers}
\UBFn{\UBLocate{$x$}}{
    run a breadth-first search from $x$ in $G_{i-1}$, stopping as soon as it discovers a landmark\nllabel{ln:ub-large-locate-bfs}\;
    \lIf{it discovered a landmark $w$\nllabel{ln:ub-large-locate-landmark}}{\Return $w$}
    \Return $\mathsf{LandmarkFree}$\nllabel{ln:ub-large-locate-free}\;
}
\BlankLine
\UBFn{\UBAddRoot{$x$}}{
    \lIf{$x\in B$\nllabel{ln:ub-large-addroot-existing}}{\Return}
    $a\leftarrow$ \UBLocate{$x$}\nllabel{ln:ub-large-addroot-locate}\;
    $\mathtt{MakeSet}(x)$; $B\leftarrow B\cup\{x\}$\nllabel{ln:ub-large-addroot-insert}\;
    \lIf{$a\in[n]$\nllabel{ln:ub-large-addroot-union}}{$\mathtt{Union}(x,a)$}
}
\BlankLine
\UBFn{\UBRootEdge{$u,v$}}{
    \UBAddRoot{$u$}; \UBAddRoot{$v$}; $\mathtt{Union}(u,v)$\nllabel{ln:ub-large-rootedge}\;
}
\end{algorithm}

\begin{algorithm}[!htb]
\caption{\texttt{Update}$(i,u,v)$}
\label{alg:ub-large-update}
\lIf{$i\le S$\nllabel{ln:ub-large-update-initial}}{\UBAddRoot{\Uniform{$[n]$}}; \UBRootEdge{$u,v$}; \Return}
\lIf{\Bernoulli{$S/i$} $ = 1$\nllabel{ln:ub-large-update-sample}}{\UBRootEdge{$u,v$}; \Return}
$c_x\leftarrow$ \UBLocate{$x$} for $x\in\{u,v\}$\nllabel{ln:ub-large-update-locate}\;
\lIf{$c_u\ne\mathsf{LandmarkFree}$ and $c_v\ne\mathsf{LandmarkFree}$\nllabel{ln:ub-large-update-union}}{$\mathtt{Union}(c_u,c_v)$}
\end{algorithm}

On update $i$, we call \UBUpdate{$i,u_i,v_i$}, using the functions
of \cref{alg:ub-large-helpers,alg:ub-large-update}. The operations
$\mathtt{MakeSet}$, $\mathtt{Union}$, and $\mathtt{Find}$ act on $D$. More specifically, our functions do the following:
\begin{description}
    \item[\Bernoulli{$p$}:] draws a Bernoulli random variable with parameter $p$.
    \item[\Uniform{$U$}:] draws an element uniformly from the set $U$. For $U=N(z)$, it draws $j$ uniformly from $[\Degree(z)]$ and returns $\Neighbor(z,j)$.
    \item[\UBLocate{$x$}:] runs a breadth-first search from $x$ in
    $G_{i-1}$, keeping the set of discovered vertices in a balanced search
    tree. It enumerates $N(v)$ as
    $\Neighbor(v,1),\ldots,\Neighbor(v,\Degree(v))$ in the list model, and
    as the $w\in[n]$ with $\Adjacent(v,w)=1$ in the matrix model. The
    search stops as soon as it discovers a landmark (returning it) or has
    exhausted $C_{i-1}(x)$ (returning $\mathsf{LandmarkFree}$, which
    declares that $C_{i-1}(x)\cap B=\emptyset$).
    \item[\UBAddRoot{$x$}:] returns immediately if $x$ is already a landmark.
    Otherwise, searches for a landmark from $x$, adds $x$ to $B$, and joins it to the
    class of the landmark found, or creates a new class if the search returns
    $\mathsf{LandmarkFree}$.
    \item[\UBRootEdge{$u,v$}:] adds the endpoints \emph{one after the other} to $B$ and then joins their
    classes using the new edge. Note that the search from $v$ uses the old graph
    $G_{i-1}$ but the landmark set already includes $u$.
    Note that this order keeps the invariant that the classes of $D$ coincide with the components of $G_{i-1}$ on the current $B$, after every single addition: if $u$ and $v$ lie in a common component of $G_{i-1}$ without landmarks, then $u$ opens a new class for it, and the search from $v$ finds $u$ and joins $v$ to this class instead of opening a second one.
    
    \item[\UBUpdate{$i,u,v$}:] If $i\le S$, adds an independent uniform
    vertex as a landmark, calls \UBRootEdge{$u,v$}, and returns. If $i>S$, then
    with probability $S/i$ the update is \emph{sampled}: it calls
    \UBRootEdge{$u,v$} and returns.
    Otherwise, it searches for a landmark from each endpoint and joins the
    classes of the two landmarks found; if a search returns
    $\mathsf{LandmarkFree}$, no two landmark classes need to be merged.
    After \UBUpdate{$i,u,v$} returns, the algorithm reports whether
    $\mathtt{Find}(s)=\mathtt{Find}(t)$.
\end{description}

We now show that this algorithm satisfies \cref{thm:largespace}, which we restate for convenience.

\begin{theorem}[\cref{thm:largespace}]
Let $c>0$ be a constant, and let $n\ge2$ and $1\le S\le n$ be integers.
There is a randomized algorithm for \STConn in the incremental setting such that, for every update sequence, with probability at least $1-n^{-c}$, all of its answers are correct and, during its $i$-th update, it uses space $\widetilde O(S+\min\{n,i\}/S)$ and takes worst-case time $\widetilde O(n(1+\min\{n,i\}/S))$ in the matrix model and $\widetilde O(\min\{n^2/S^2,1+i/S\})$ in the list model.
\end{theorem}

\subsubsection*{Correctness Analysis}
\begin{proof}[Proof of \cref{thm:largespace} (correctness)]
We show by induction on $k$ that, after update $k$, the classes of $D$ coincide with the components of $G_k$ on $B_k$, i.e., $\mathtt{Find}(x)=\mathtt{Find}(y)$ if and only if $C_k(x)=C_k(y)$ for all $x,y\in B_k$.
As $s,t\in B_k$, this makes every answer correct.
For $k=0$, it holds as $G_0$ is empty and each initial landmark has its own class.
Now assume it for $k=i-1$.
During update $i$, searches run in $G_{i-1}$, so \UBLocate{$x$} returns a landmark of $C_{i-1}(x)$ if there is one, and $\mathsf{LandmarkFree}$ otherwise.
Hence \UBAddRoot{$x$} joins $x$ to the class of a landmark of its component, or opens a new class if there is none, which preserves the claim for $G_{i-1}$ and the current $B$.
Finally, the update calls $\mathtt{Union}(a,b)$ for landmarks $a\in C_{i-1}(u_i)$ and $b\in C_{i-1}(v_i)$, namely $u_i,v_i$ or $c_u,c_v$, which merges exactly the classes of the two components joined by $u_iv_i$, or it ends because $c_u$ or $c_v$ is $\mathsf{LandmarkFree}$, in which case $u_iv_i$ connects no two landmarks that were not connected before.
Hence the claim holds for $k=i$.
\end{proof}

\subsubsection*{Time and Space Analysis}
We bound only the time, as the number of queries never exceeds it.
Both this analysis and that of \cref{thm:metropolis} use the following fact about the landmarks.

\begin{fact}
\label{fact:ub-landmarks}
We have the two following:
\begin{enumerate}[label=(\alph*),ref=\alph*,nosep,topsep=3pt]
    \item \label[ubsamplingpart]{itm:ub-landmarks-size}
    For each constant $c>0$, with probability at least $1-n^{-c}$, we have $|B_k|=\widetilde O(S)$ for all $k$.
    \item \label[ubsamplingpart]{itm:ub-landmarks-hit}
    Let $i>S$, and let $Q$ be a set of vertices that depends only on the update sequence.
    Then $Q\cap B_{i-1}=\emptyset$ with probability at most $e^{-S|Q|/n-S\operatorname{vol}(Q)/(2i)}$, where $\operatorname{vol}$ is taken in $G_{i-1}$.
    In particular, if no vertex of $Q$ is isolated in $G_{i-1}$, this is at most $e^{-S|Q|/(2\min\{n,i\})}$.
\end{enumerate}
\end{fact}
\begin{proof}
As \UBAddRoot always adds its vertex, $B_k$ consists of $s$, $t$, the uniform landmarks drawn by updates $j\le\min\{k,S\}$, and the endpoints of the edges $e_j$, $j\le k$, with $j\le S$ or update $j$ sampled; in particular, $B_k$ does not depend on the searches or walks.
Hence $|B_k|\le2+3S+2X$ for the number $X$ of sampled updates, as defined in the update function.
For~\eqref{itm:ub-landmarks-size}, for $k>S$, $X$ is a sum of independent indicators with mean $\sum_{S<j\le M}S/j=O(S\log n)$, as $M<n^2$, so $X=O_c(S\log n)$ except with probability $n^{-c}$ by a Chernoff bound.
For~\eqref{itm:ub-landmarks-hit}, the $S$ uniform landmarks miss $Q$ with probability $(1-|Q|/n)^S\le e^{-S|Q|/n}$.
Moreover, at least $\operatorname{vol}(Q)/2$ edges $e_j$, $j<i$, have an endpoint in $Q$, and each of them, independently with probability $\min\{1,S/j\}\ge S/i$, has $j\le S$ or is sampled.
As $Q$ does not depend on these random choices, $\Pr[Q\cap B_{i-1}=\emptyset]\le e^{-S|Q|/n}(1-S/i)^{\operatorname{vol}(Q)/2}$, which gives the first bound; the second follows as $\operatorname{vol}(Q)\ge|Q|$ and $1/n+1/(2i)\ge1/(2\min\{n,i\})$.
\end{proof}

\begin{proof}[Proof of \cref{thm:largespace} (time and space)]
Fix a constant $\alpha$, sufficiently large in terms of $c$, put $q_i\coloneqq2\alpha i\log(2n)/S$, and, during update $i$, put $r\coloneqq\lceil\alpha\min\{n,i\}\log(2n)/S\rceil$.
For $i>S$ and $z\in[n]$, let the BFS from $z$ in $G_{i-1}$, run without landmarks, and suppose the BFS discovers the vertices of $C_{i-1}(z)$ in an order $v_1=z,v_2,\ldots$ that depends only on the update sequence, as the adjacency lists are fixed in advance; put $W_t\coloneqq\{v_1,\ldots,v_t\}$.
Let $\mathcal E$ be the event that $|B_k|=\widetilde O(S)$ for every $k$ and that, for all $i>S$ and $z\in[n]$, $B_{i-1} \cap W_r \neq \emptyset$ if $|C_{i-1}(z)|\ge r$, and $B_{i-1} \cap W_t \neq \emptyset$ if $t$ is minimal with $\operatorname{vol}(W_t)\ge q_i$ in $G_{i-1}$.
For fixed $(i,z)$, \cref{fact:ub-landmarks}\eqref{itm:ub-landmarks-hit} bounds the probability that $B_{i-1} \cap W_r = \emptyset$ or $B_{i-1} \cap W_t = \emptyset$ by $e^{-Sr/(2\min\{n,i\})}\le(2n)^{-\alpha/2}$, as no vertex of $W_r$ is isolated when $|C_{i-1}(z)|\ge r\ge\alpha\log(2n)\ge2$, or by $e^{-Sq_i/(2i)}\le(2n)^{-\alpha}$, respectively.
By \cref{fact:ub-landmarks}\eqref{itm:ub-landmarks-size} with exponent $c+1$ and a union bound over the fewer than $n^3$ pairs $(i,z)$, $\Pr[\mathcal E]\ge1-n^{-(c+1)}-2n^3(2n)^{-\alpha/2}\ge1-n^{-c}$.

\begin{subclaim}
\label{clm:ub-large-search}
On $\mathcal E$, every call \UBLocate{$x$} during an update $i$ discovers at most $r$ vertices and examines at most $\min\{r^2,q_i\}$ neighbors.
\end{subclaim}
\begin{subproof}
We may assume $x\notin B$, as otherwise \UBLocate{$x$}  discovers only $x$.
If $i\le S$, all earlier updates called \UBRootEdge, so $x$ is isolated in $G_{i-1}$, and \UBLocate{$x$}  discovers only $x$.
If $i>S$, then \UBLocate{$x$}  discovers the vertices of $C\coloneqq C_{i-1}(x)$ and scans them, i.e., examines their neighbors, in the order $v_1=x,v_2,\ldots$ until it stops, at the latest when it discovers a vertex of $B_{i-1}\subseteq B$.
If $|C|\ge r$, then $W_r$ contains such a vertex on $\mathcal E$; hence, in any case, we have that \UBLocate{$x$}  discovers at most $r$ vertices.
Thus each scanned vertex examines at most $r$ neighbors, as these are distinct discovered vertices, and \UBLocate{$x$} at most $r^2$.
If $t$ is minimal with $\operatorname{vol}(W_t)\ge q_i$, then $W_t$ contains such a vertex on $\mathcal E$, so \UBLocate{$x$} stops before it scans $v_t$, after examining at most $\operatorname{vol}(W_{t-1})<q_i$ neighbors; if there is no such $t$, it examines at most $\operatorname{vol}(C)<q_i$ neighbors.
\end{subproof}

\proofparagraph{Time.}
Assume $\mathcal E$.
An update calls \UBLocate{$\cdot$} at most three times and takes $\widetilde O(1)$ further time.
In the list model, a search takes $\widetilde O(1)$ time per examined neighbor, so $\widetilde O(1+\min\{r^2,q_i\})=\widetilde O(\min\{n^2/S^2,1+i/S\})$ time by \cref{clm:ub-large-search}.
In the matrix model, it reads one row of $n$ entries per discovered vertex, so it takes $\widetilde O(rn)=\widetilde O(n(1+\min\{n,i\}/S))$ time.

\proofparagraph{Space.}
On $\mathcal E$, the landmarks and $D$ occupy $\widetilde O(S)$ words, as $|B|\le|B_k|=\widetilde O(S)$ during update $k$.
Each execution of \UBLocate{$\cdot$} stores its at most $r=\widetilde O(1+\min\{n,i\}/S)$ discovered vertices and $O(1)$ further words.
\end{proof}

\subsection{Proof of \texorpdfstring{\cref{thm:metropolis}}{Theorem \getrefnumber{thm:metropolis}}}
\label{sec:ub-metro}

In this section, we work in the list model and prove \cref{thm:metropolis}.
We keep the notation and algorithm of \cref{sec:ub-large} and only replace its breadth-first searches by Metropolis walks (see \cref{sec:prelims}).

Choose a sufficiently large constant $\beta>0$, and, during update $i$, define $r\coloneqq\lceil\beta\min\{n,i\}\log(2n)/S\rceil$.

\begin{algorithm}[htb]
\caption{\texttt{Locate}$(x)$ with Metropolis walks}
\label{alg:ub-metro-locate}
\UBFn{\UBLocate{$x$}}{
    \lIf{$x\in B$\nllabel{ln:ub-metro-locate-landmark}}{\Return $x$}
    \lIf{$d(x)\ge r$\nllabel{ln:ub-metro-locate-heavy-x}}{\Return a vertex of $B$ among $\Neighbor(x,1),\ldots,\Neighbor(x,r)$, or $\mathsf{LandmarkFree}$ if there is none}
    \lIf{$d(x)=0$\nllabel{ln:ub-metro-locate-isolated}}{\Return $\mathsf{LandmarkFree}$}
    \For(\tcp*[f]{trial $j$}){$j=1,\ldots,\lceil\beta\log(2n)\rceil$\nllabel{ln:ub-metro-locate-trials}}{
        $z\leftarrow x$\nllabel{ln:ub-metro-locate-start}\;
        \For{$\ell=1,\ldots,32r^2$\nllabel{ln:ub-metro-locate-steps}}{
            $y\leftarrow$ \Uniform{$N(z)$}\nllabel{ln:ub-metro-locate-propose}\;
            \lIf{$y\in B$\nllabel{ln:ub-metro-locate-found}}{\Return $y$}
            \lIf{$d(y)\ge r$\nllabel{ln:ub-metro-locate-heavy-y}}{\Return a vertex of $B$ among $\Neighbor(y,1),\ldots,\Neighbor(y,r)$, or $\mathsf{LandmarkFree}$ if there is none}
            \lIf{\Bernoulli{$\min\{1,d(z)/d(y)\}$} $=1$\nllabel{ln:ub-metro-locate-accept}}{$z\leftarrow y$}
        }
    }
    \Return $\mathsf{LandmarkFree}$\nllabel{ln:ub-metro-locate-free}\;
}
\end{algorithm}

On update $i$, we call \UBUpdate{$i,u_i,v_i$} of \cref{alg:ub-large-update}, using the functions of \cref{alg:ub-large-helpers}, except that \UBLocate is replaced by \cref{alg:ub-metro-locate}, which does the following:
\begin{description}
    \item[\UBLocate{$x$}:] returns $x$ if it is a landmark, returns a landmark among $\Neighbor(x,1),\ldots,\Neighbor(x,r)$ if $d(x)\ge r$, and returns $\mathsf{LandmarkFree}$ if $x$ is isolated; if there is no landmark among these $r$ neighbors, here or in \cref{ln:ub-metro-locate-heavy-y}, it returns $\mathsf{LandmarkFree}$, which, w.h.p., never happens (\cref{lem:ub-metro-landmarkfree}). Otherwise, it runs $\lceil\beta\log(2n)\rceil$ \emph{trials}, one at a time. A trial runs a Metropolis walk from $x$ in $G_{i-1}$ for at most $32r^2$ steps, storing only its current vertex $z$: it draws a uniform neighbor $y$ of $z$, returns $y$ if it is a landmark, returns a landmark among $\Neighbor(y,1),\ldots,\Neighbor(y,r)$ if $d(y)\ge r$, and otherwise moves to $y$ with probability $\min\{1,d(z)/d(y)\}$. Hence the walk never moves to a landmark or to a vertex of degree at least $r$. If no trial returns, \UBLocate{$x$} returns $\mathsf{LandmarkFree}$, which declares that $C_{i-1}(x)\cap B=\emptyset$.
\end{description}

We now show that this algorithm satisfies \cref{thm:metropolis}, which we restate for convenience.

\begin{theorem}[\cref{thm:metropolis}]
Let $c>0$ be a constant, and let $n\ge2$ and $1\le S\le n$ be integers.
In the list model, there is a randomized algorithm with space $\widetilde O(S)$ for \STConn in the incremental setting such that, for every update sequence, with probability at least $1-n^{-c}$, all of its answers are correct and its $i$-th update takes worst-case time $\widetilde O(1+\min\{n,i\}^2/S^2)$.
\end{theorem}

\subsubsection*{Correctness Analysis}

\begin{lemma}
\label{lem:ub-metro-landmarkfree}
For every $c>0$, there is $K=K(c)$ such that, for $\beta\ge K$, with probability at least $1-n^{-c}$, every $\mathsf{LandmarkFree}$ declaration is correct throughout the processing of all updates, i.e., for all $i \in [1,{n \choose 2}]$  every call \UBLocate{$x$} during an update $i$ returns $\mathsf{LandmarkFree}$ only if $C_{i-1}(x)\cap B=\emptyset$.
\end{lemma}
\begin{proof}
For $i\le S$, all earlier updates called \UBRootEdge, so every vertex outside $B_{i-1}$ is isolated in $G_{i-1}$; hence \UBLocate{$x$} returns $x$ if $x\in B$, and $\mathsf{LandmarkFree}$ otherwise, which is correct.
Thus let $i>S$.
Call a vertex $v$ \emph{heavy} if $d_{G_{i-1}}(v)\ge r$, and \emph{light} otherwise, and let $H_{i-1}$ arise from $G_{i-1}$ by redirecting each edge from a light $z$ to a heavy $y$ to a new vertex, the \emph{leaf} of $z$ for $y$, and deleting the heavy vertices.
Thus light vertices keep their degrees, $H_{i-1}$ has maximum degree less than $r$, and $H_{i-1}$ depends only on the update sequence.

\begin{subclaim}
\label{clm:ub-metro-samples}
There is an event $\mathcal E$ of probability at least $1-n^{-(c+1)}$, depending only on $(B_{i-1})_{i>S}$, on which the following hold for all $i>S$ simultaneously:
\begin{enumerate}[label=(\alph*),ref=\alph*,nosep,topsep=3pt]
    \item \label[ubsamplingpart]{itm:ub-metro-scan}
    Every heavy vertex $y$ of $G_{i-1}$ has a vertex of $B_{i-1}$ among $\Neighbor(y,1),\ldots,\Neighbor(y,r)$.
    \item \label[ubsamplingpart]{itm:ub-metro-hit}
    For every light $x$ with $|C_{H_{i-1}}(x)|\ge r$, given $B_{i-1}$, a Metropolis walk on $H_{i-1}$ from $x$ has probability at most $1/4$ that its first $r^2$ positions contain at least $r/30$ distinct vertices but no leaf or vertex of $B_{i-1}$.
\end{enumerate}
\end{subclaim}
\begin{subproof}
For \eqref{itm:ub-metro-scan}, fix $(i,y)$: the first $r$ neighbors of $y$ form a set of $r$ non-isolated vertices that is fixed in advance, so \eqref{itm:ub-metro-scan} fails with probability at most $e^{-Sr/(2\min\{n,i\})}\le(2n)^{-\beta/2}$ by \cref{fact:ub-landmarks}\eqref{itm:ub-landmarks-hit}.
For \eqref{itm:ub-metro-hit}, fix $(i,x)$ and let $f$ be this probability, a function of $B_{i-1}$.
Fix an outcome of the first $r^2$ positions with no leaf and at least $r/30$ distinct vertices; these are non-isolated vertices of $G_{i-1}$, as $|C_{H_{i-1}}(x)|\ge r\ge\beta\log(2n)\ge2$, so $B_{i-1}$ misses them with probability at most $e^{-Sr/(60\min\{n,i\})}\le(2n)^{-\beta/60}$ by \cref{fact:ub-landmarks}\eqref{itm:ub-landmarks-hit}.
As $B_{i-1}$ does not depend on the walk (see the proof of \cref{fact:ub-landmarks}), averaging over it gives $\Ex[f]\le(2n)^{-\beta/60}$, so $\Pr[f>1/4]\le4(2n)^{-\beta/60}$ by Markov's inequality.
A union bound over the $O(n^3)$ pairs $(i,y)$ and $(i,x)$ gives the claim for sufficiently large $\beta$.
\end{subproof}
In particular, as $B_{i-1}\subseteq B$, on $\mathcal E$, \cref{ln:ub-metro-locate-heavy-x,ln:ub-metro-locate-heavy-y} of \cref{alg:ub-metro-locate} always find a vertex of $B$ during updates $i>S$ by \cref{clm:ub-metro-samples}\eqref{itm:ub-metro-scan}.

\begin{subclaim}
\label{clm:ub-metro-range}
Let $H$ be a graph of maximum degree at most $r$, and let $x$ be a vertex of $H$ with $|C_H(x)|\ge r$. Then, with probability at least $1/2$, the first $r^2$ positions of a Metropolis walk on $H$ started at $x$ visit at least $r/30$ distinct vertices.
\end{subclaim}
\begin{subproof}
Put $h\coloneqq r^2$, let $U\coloneqq\{Y_0,\ldots,Y_{h-1}\}$ be the set of visited vertices, and let $a_v$ count the visits to $v$ among $Y_0,\ldots,Y_{h-1}$. Then $\sum_{v\in U}a_v=h$ and
\[
    \sum_{v\in U}a_v^2=h+2\sum_{0\le j<h}\;\sum_{j<j'<h}\mathbf{1}[Y_{j'}=Y_j].
\]
Given $Y_j$, the inner sum counts, by the Markov property, the visits to the starting vertex of a Metropolis walk from $Y_j$ among its positions $1,\ldots,h-j-1$, so its expectation is at most $5\sqrt h+2r=7r$ by \cref{lem:metropolis-facts}, as $h<6|C_H(x)|^2$.
Hence $\Ex[\sum_{v\in U}a_v^2]\le15hr$, and $\sum_{v\in U}a_v^2\le30hr$ with probability at least $1/2$ by Markov's inequality.
Then the Cauchy--Schwarz inequality $h^2=(\sum_{v\in U}a_v)^2\le|U|\sum_{v\in U}a_v^2$ gives $|U|\ge h/(30r)=r/30$.
\end{subproof}

Recall that a \emph{trial} of \UBLocate{$x$} is one iteration of the loop in \cref{ln:ub-metro-locate-trials}, i.e., a Metropolis walk from $x$ of at most $32r^2$ steps that returns as soon as it draws a landmark or a heavy vertex.

\begin{subclaim}
\label{clm:ub-metro-trial}
On $\mathcal E$, consider a trial of a call \UBLocate{$x$} during an update $i>S$ with $C_{i-1}(x)\cap B\ne\emptyset$. Conditioning on all the choices of the algorithm before the trial, it returns a landmark with probability at least $1/4$.
\end{subclaim}
\begin{subproof}
As long as the trial runs, the current vertex $z$ is light, not a landmark, and not isolated.
Conditioning on all the choices of the algorithm before the trial, let $(Y_j)$ be a Metropolis walk on $H_{i-1}$ from $x$ that uses the trial's coins until the trial returns, and fresh coins afterwards.
Until the trial returns, $Y_j$ is its current vertex after $j$ steps.
Indeed, from a current vertex $z$, the trial draws a uniform $y\in N_{G_{i-1}}(z)$, and $(Y_j)$ draws the corresponding vertex of $N_{H_{i-1}}(z)$, which is $y$ if $y$ is light and the leaf of $z$ for $y$ if $y$ is heavy.
If $y$ is light, both move to $y$ with the same probability $\min\{1,d_{G_{i-1}}(z)/d_{G_{i-1}}(y)\}=\min\{1,d_{H_{i-1}}(z)/d_{H_{i-1}}(y)\}$, as light vertices have the same degree in $G_{i-1}$ and $H_{i-1}$; if $y$ is heavy, the trial returns, while $(Y_j)$ moves to the leaf, which has degree $1$.
On $\mathcal E$, \cref{ln:ub-metro-locate-heavy-y} always finds a vertex of $B$, so the trial fails to return a landmark only if it does not return, and then $Y_0,\ldots,Y_{32r^2}$, being its current vertices, contain no leaf and no landmark.

If $|C_{H_{i-1}}(x)|<r$, let $w$ be a leaf of $C_{H_{i-1}}(x)$ if there is one; otherwise {if there's no leaf}, no vertex of $C_{H_{i-1}}(x)$ has a heavy neighbor in $G_{i-1}$, so $C_{H_{i-1}}(x)=C_{i-1}(x)$, and we let $w$ be a landmark in $C_{i-1}(x)$
(which exists as $C_{i-1}(x)\cap B\ne\emptyset$). 
By \cref{lem:metropolis-facts}, $\Ex_x[\tau(w)]\le8|C_{H_{i-1}}(x)|^2<8r^2$, so the walk visits $w$ within $32r^2$ steps with probability at least $3/4$ by Markov's inequality.
If $|C_{H_{i-1}}(x)|\ge r$, then the first $r^2$ positions of the walk visit fewer than $r/30$ distinct vertices with probability at most $1/2$ by \cref{clm:ub-metro-range}, and they visit at least $r/30$ distinct vertices but no leaf or vertex of $B_{i-1}\subseteq B$ with probability at most $1/4$ by \cref{clm:ub-metro-samples}\eqref{itm:ub-metro-hit}.
In both cases, the trial fails with probability at most $3/4$.
\end{subproof}

Consider a call \UBLocate{$x$} during an update $i>S$ with $C_{i-1}(x)\cap B\ne\emptyset$.
On $\mathcal E$, it returns $\mathsf{LandmarkFree}$ only if all its trials fail, as $x$ is not isolated and \cref{ln:ub-metro-locate-heavy-x} always finds a vertex of $B$; by \cref{clm:ub-metro-trial}, this has probability at most $(3/4)^{\lceil\beta\log(2n)\rceil}\le(2n)^{-\beta/4}$.
As each update $i>S$ calls \UBLocate at most twice, a union bound over these at most $2M\le n^2$ calls and $\Pr[\mathcal E]\ge1-n^{-(c+1)}$ prove the lemma for sufficiently large $\beta$.
\end{proof}

\begin{proof}[Proof of \cref{thm:metropolis} (correctness)]
For $\beta$ large enough, by \cref{lem:ub-metro-landmarkfree} with exponent $c+1$, with probability at least $1-n^{-(c+1)}$, every call \UBLocate{$x$} returns a landmark of $C_{i-1}(x)$ if there is one, and $\mathsf{LandmarkFree}$ otherwise, as it reaches each landmark it returns along edges of $G_{i-1}$.
The proof of correctness of \cref{thm:largespace} uses only this property of \UBLocate and thus applies verbatim.
\end{proof}

\subsubsection*{Time and Space Analysis}

\begin{proof}[Proof of \cref{thm:metropolis} (time and space)]
We bound only the time, as the number of queries never exceeds it.
By \cref{fact:ub-landmarks}\eqref{itm:ub-landmarks-size} with exponent $c+1$ and the correctness proof, with probability at least $1-2n^{-(c+1)}\ge1-n^{-c}$, all answers are correct and $|B|=\widetilde O(S)$ throughout, so $B$ and $D$ take $\widetilde O(S)$ words.
A call \UBLocate{$x$} stores $O(1)$ further words, makes $O(r^2\log n)$ steps, and examines at most $r$ neighbors in \cref{ln:ub-metro-locate-heavy-x} or \cref{ln:ub-metro-locate-heavy-y}, each taking $O(1)$ queries and $\widetilde O(1)$ time.
As an update makes at most three such calls and takes $\widetilde O(1)$ further time, the $i$-th update takes $\widetilde O(r^2)=\widetilde O(1+\min\{n,i\}^2/S^2)$ time, and $\widetilde O(S)$ words suffice.
\end{proof}

\subsection{Proof of \texorpdfstring{\cref{thm:smallspace}}{Theorem \getrefnumber{thm:smallspace}}}

As in \cref{sec:ub-large}, we give an algorithm for the list model that maintains a set $B$ of landmarks, initially $B_0=\{s,t\}$, and a union-find structure $D$ on $B$; the notation $B_k$ and the term landmark-free have the same meaning as there.

Choose sufficiently large constants $\kappa\ge\lambda>0$, and define, during update $i$,
\[
    m\coloneqq\lceil\kappa \cdot \log(2n)\rceil,
    \qquad
    N_x\coloneqq\left\lceil
        \kappa \lambda \cdot (\min\{n,i\}/S)d(x)\log^2(2n)
    \right\rceil.
\]

\SetKwFunction{UBWalk}{Walk}

\begin{algorithm}[htb]
\caption{Helper functions}
\label{alg:ub-helpers}
\UBFn{\UBWalk{$x,y,T$}}{
    \lIf{$d(x) = 0$\nllabel{ln:ub-walk-isolated}}{\Return $\mathsf{LandmarkFree}$}
    $(z,Z)\leftarrow(x,1)$\nllabel{ln:ub-walk-init}\;
    \While{true}{
        \lIf{$z=y$\nllabel{ln:ub-walk-target}}{\Return $\mathsf{Connected}$}
        \lIf{$z\in B$\nllabel{ln:ub-walk-root}}{\Return $z$}
        \lIf(\tcp*[f]{$N_x=\lceil\kappa\lambda(\min\{n,i\}/S)d(x)\log^2(2n)\rceil$}){$Z=N_x$\nllabel{ln:ub-walk-limit}}{\Return $\mathsf{LandmarkFree}$}
        \lIf{\Bernoulli{$1/(T+1)$} $ = 1$\nllabel{ln:ub-walk-kill}}{\Return $(\mathsf{Killed},Z)$}
        $z\leftarrow$ \Uniform{$N(z)$}; $Z\leftarrow Z+\mathbf{1}[z=x]$\nllabel{ln:ub-walk-step}\;
    }
}
\BlankLine
\UBFn{\UBAddRoot{$x$}}{
    \lIf{$x\in B$\nllabel{ln:ub-addroot-existing}}{\Return}
    $a\leftarrow$ \UBWalk{$x,\bot,\infty$}\nllabel{ln:ub-addroot-walk}\;
    $\mathtt{MakeSet}(x)$; $B\leftarrow B\cup\{x\}$\nllabel{ln:ub-addroot-insert}\;
    \lIf{$a\in[n]$\nllabel{ln:ub-addroot-union}}{$\mathtt{Union}(x,a)$}
}
\BlankLine
\UBFn{\UBRootEdge{$u,v$}}{
    \UBAddRoot{$u$}; \UBAddRoot{$v$}; $\mathtt{Union}(u,v)$\nllabel{ln:ub-rootedge}\tcp*[r]{as in \cref{alg:ub-large-helpers}}
}
\end{algorithm}

\begin{algorithm}[!htb]
\caption{\texttt{Update}$(i,u,v)$}
\label{alg:ub-update}
\lIf{$i\le S$\nllabel{ln:ub-update-initial}}{\UBAddRoot{\Uniform{$[n]$}}; \UBRootEdge{$u,v$}; \Return}
\lIf{\Bernoulli{$S/i$} $ = 1$\nllabel{ln:ub-update-sample}}{\UBRootEdge{$u,v$}; \Return}
$c_x\leftarrow x$ if $x\in B$, else $\bot$, for $x\in\{u,v\}$\nllabel{ln:ub-update-init}\;
\lIf{$c_u\ne\bot$ and $c_v\ne\bot$\nllabel{ln:ub-update-initial-union}}{$\mathtt{Union}(c_u,c_v)$; \Return}
\For{$T=1,2,4,\ldots$\nllabel{ln:ub-update-scales}}{
    \ForEach{$(x,y)\in\{(u,v),(v,u)\}$ with $c_x=\bot$\nllabel{ln:ub-update-endpoints}}{
        \For(\tcp*[f]{$m=\lceil\kappa\log(2n)\rceil$}){$j=1,\ldots,m$\nllabel{ln:ub-update-trials}}{
            $a\leftarrow$ \UBWalk{$x,y,T$}\nllabel{ln:ub-update-walk}\;
            \lIf{$a\in\{\mathsf{Connected},\mathsf{LandmarkFree}\}$\nllabel{ln:ub-update-stop}}{\Return}
            \lIf{$a\in[n]$\nllabel{ln:ub-update-root}}{$c_x\leftarrow a$}
            \lIf{$c_u\ne\bot$ and $c_v\ne\bot$\nllabel{ln:ub-update-union}}{$\mathtt{Union}(c_u,c_v)$; \Return}
        }
    }
    \If{for some $x$ with $c_x=\bot$, all $m$ trials from $x$ at this scale were light kills\nllabel{ln:ub-update-promotion}}{
        \UBRootEdge{$u,v$}; \Return\nllabel{ln:ub-update-promoted-edge}\;
    }
}
\end{algorithm}

On update $i$, we call \UBUpdate{$i,u_i,v_i$}, using the functions of \cref{alg:ub-helpers,alg:ub-update}, and then report whether $\mathtt{Find}(s)=\mathtt{Find}(t)$, as in \cref{sec:ub-large}. More specifically, our functions do the following:
\begin{description}
    \item[\UBWalk{$x,y,T$}:] runs a random walk from $x$ in
    $G_{i-1}$, where $Z$ counts its visits to $x$, including the initial
    one. The walk stops as soon as it reaches the target $y$ (returning
    $\mathsf{Connected}$) or a landmark (returning that landmark), or when $Z$
    reaches $N_x$ (returning $\mathsf{LandmarkFree}$, which declares that
    $C_{i-1}(x)\cap B=\emptyset$); it also returns $\mathsf{LandmarkFree}$ at
    once if $d(x)=0$. Moreover, before each step, it is \emph{killed} with
    probability $1/(T+1)$, returning $(\mathsf{Killed},Z)$. For $i>S$,
    such a kill is \emph{light} if $Z\le Sd(x)(T+1)/(8i)$.
    A \emph{basic walk} is \UBWalk{$x,\bot,\infty$}: as $\bot$ is not a
    vertex and $1/(\infty+1)=0$, it has neither a target nor is it ever killed.
    \item[\UBAddRoot{$x$}:] is as in \cref{alg:ub-large-helpers}, except that it runs the basic walk instead of \UBLocate{$x$}.
    \item[\UBUpdate{$i,u,v$}:] If $i\le S$ or the update is \emph{sampled}, it proceeds as in \cref{alg:ub-large-update}.
    Otherwise, it runs a \emph{search}, which looks for a landmark in each of
    $C_{i-1}(u)$ and $C_{i-1}(v)$ in order to merge their classes without
    adding new landmarks. For each endpoint $x$, $c_x$ records a landmark known to
    lie in $C_{i-1}(x)$, or $\bot$ if none is known yet; initially,
    $c_x=x$ if $x\in B$, and $c_x=\bot$ otherwise. As soon as
    $c_u\ne\bot$ and $c_v\ne\bot$, the search calls
    $\mathtt{Union}(c_u,c_v)$ and returns.
    Until then, it proceeds through the \emph{scales} $T=1,2,4,\ldots$ in
    increasing order. At scale $T$, for each endpoint $x$ with $c_x=\bot$, it runs $m$
    \emph{trials} \UBWalk{$x,y,T$}, one at a time, where $y$ is the other
    endpoint. A trial that returns a landmark $a$ sets $c_x$ to $a$.
    A trial that returns $\mathsf{Connected}$ or $\mathsf{LandmarkFree}$ ends
    the update, since then $u$ and $v$ were already connected, or
    $C_{i-1}(x)$ contains no landmark, respectively; either way, no two landmark
    classes need to be merged.
    If, after the trials at scale $T$, some endpoint $x$ with $c_x=\bot$
    had all $m$ of its trials end in light kills, the update is
    \emph{promoted}: it calls \UBRootEdge{$u,v$} and returns. Otherwise,
    the search moves on to scale $2T$.
\end{description}

We now show that this algorithm satisfies \cref{thm:smallspace}, which we restate for convenience.

\begin{theorem}[\cref{thm:smallspace}]
Let $c>0$ be a constant, and let $n\ge2$ and $1\le S\le n$ be integers.
In the list model, there is a randomized algorithm with space $\widetilde O(S)$ for \STConn in the incremental setting such that, for every update sequence, with probability at least $1-n^{-c}$, all of its answers are correct and, for every $k\ge1$, both the total time and the total number of queries of its first $k$ updates are $\widetilde O(k(1+\min\{n,k\}/S))$.
\end{theorem}

Simulating each list query by a row scan of the adjacency matrix, with $O(n)$ queries and $O(1)$ words, gives the matrix-model bound of \cref{thm:smallspace}.

\subsubsection*{Correctness Analysis}
Correctness relies on the following lemma, proven in
\cref{sec:ub-search-proof}.

\begin{restatable}{lemma}{UBLandmarkFreeLemma}
\label{lem:ub-landmarkfree}
For every $c>0$, there is $K=K(c)$ such that, for
$\kappa\ge\lambda\ge K$, with probability at least $1-n^{-c}$, every
$\mathsf{LandmarkFree}$ declaration is correct, i.e., every call
\UBWalk{$x,y,T$} during an update $i$ returns $\mathsf{LandmarkFree}$ only if
$C_{i-1}(x)\cap B=\emptyset$.
\end{restatable}

\begin{proof}[Proof of \cref{thm:smallspace} (correctness)]
We choose $\kappa\ge\lambda$ large enough to apply \cref{lem:ub-landmarkfree} with exponent $c+2$, and let $\mathcal E_1$ be its event.
On $\mathcal E_1$, we show by induction on $k$ that, after update $k$, the classes of $D$ coincide with the components of $G_k$ on $B_k$, i.e., $\mathtt{Find}(x)=\mathtt{Find}(y)$ if and only if $C_k(x)=C_k(y)$ for all $x,y\in B_k$.
As $s,t\in B_k$, this makes every answer correct.
For $k=0$, it holds as $G_0$ is empty and each initial landmark has its own class.

Now assume it for $k=i-1$.
During update $i$, all random walks run in $G_{i-1}$, so a walk from $x$ returns only landmarks of $C_{i-1}(x)$ and, on $\mathcal E_1$, returns $\mathsf{LandmarkFree}$ only if $C_{i-1}(x)\cap B=\emptyset$ (\cref{lem:ub-landmarkfree}).
Hence \UBAddRoot{$x$} joins $x$ to the class of a landmark of its component, or opens a new class if there is none, which preserves the claim for $G_{i-1}$ and the current $B$.
Besides these calls, only the step that ends the update changes $B$ or $D$, and it is one of the following.
\begin{enumerate}[nosep,topsep=3pt,label=(\roman*)]
    \item $\mathtt{Union}(a,b)$ for landmarks $a\in C_{i-1}(u_i)$ and $b\in C_{i-1}(v_i)$, namely $u_i,v_i$ in \UBRootEdge{$u_i,v_i$} or $c_u,c_v$ in the search: this merges exactly the classes of the two components joined by $u_iv_i$, and does nothing if they coincide.
    \item A trial returns $\mathsf{Connected}$: then $C_{i-1}(u_i)=C_{i-1}(v_i)$, so $G_i$ and $G_{i-1}$ have the same components.
    \item A trial from $x$ returns $\mathsf{LandmarkFree}$: on $\mathcal E_1$, $C_{i-1}(x)$ contains no landmark, so two landmarks share a component in $G_i$ if and only if they do in $G_{i-1}$.
\end{enumerate}
In each case, the claim holds for $k=i$.
\end{proof}

\subsubsection*{Time and Space Analysis}
We bound only the time, as the number of queries never exceeds it.
To this end, let $\mathcal R$ be the set of updates $i$ that call
\UBRootEdge{$u_i,v_i$}, i.e., the updates $i\le S$ and the sampled and
promoted ones. Moreover, write
\[
    \mathcal I\coloneqq\{i:C_{i-1}(u_i)=C_{i-1}(v_i)\},
    \qquad R_i\coloneqq R_{G_{i-1}}(u_i,v_i) \ \text{for} \ i\in\mathcal I.
\]
Thus $\mathcal I$ consists of the updates whose endpoints were already
connected before the insertion, and $R_i$ is their effective resistance
in the \emph{old} graph, i.e., $G_{i-1}$.
Let $\mathsf U_i$ be the time cost of \UBUpdate{$i,u_i,v_i$} until it returns
or first calls \UBRootEdge{$u_i,v_i$}, excluding the latter's cost; the
initial call to \UBAddRoot{\Uniform{$[n]$}}, when $i\le S$, is included in $\mathsf U_i$.
Let $\mathsf A_i$ be the cost of that call to \UBRootEdge{$u_i,v_i$},
and set $\mathsf A_i=0$ when $i\notin\mathcal R$.

Our complexity analysis relies on the following four lemmas. The first one,
proven in \cref{sec:ub-search-proof} (the same where \cref{lem:ub-landmarkfree} is proven), bounds the cost of walks and
searches.

\begin{restatable}{lemma}{UBSearchLemma}
\label{lem:ub-search}
For every $c>0$, there is $K=K(c)$ such that, for
$\kappa\ge\lambda\ge K$, with probability at least $1-n^{-c}$, the following hold
simultaneously for every update $i>S$:
\begin{enumerate}[nosep,topsep=3pt]
    \item \label[ubsearchpart]{itm:ub-search-basic}
    Every basic walk \UBWalk{$x,\bot,\infty$} returns within
    $\widetilde O(\min\{n,i\}i/S^2)$ operations. If $C_{i-1}(x)\cap B=\emptyset$,
    this cost is additionally bounded by
    $\widetilde O(1+(\min\{n,i\}/S)\operatorname{vol}(C_{i-1}(x)))$.
    \item \label[ubsearchpart]{itm:ub-search-update}
    $\mathsf{U}_i = \widetilde O(\min\{n,i\}i/S^2)$,
    using only $O(\log(n\kappa))$ scales $T\le\widetilde O(\min\{n,i\}i/S^2)$.
    For each $x\in\{u_i,v_i\}$ with $C_{i-1}(x)\cap B_{i-1}=\emptyset$,
    we additionally have
    $\mathsf{U}_i = \widetilde O(1+(\min\{n,i\}/S)\operatorname{vol}(C_{i-1}(x)))$.
\end{enumerate}
\end{restatable}

The second one, proven in \cref{sec:ub-root-budget-proof}, shows
that few updates call \texttt{LandmarkEdge}.

\begin{restatable}{lemma}{UBRootBudgetLemma}
\label{lem:ub-root-budget}
For every $c>0$, there is $K=K(c)$ such that, for
$\kappa\ge\lambda\ge K$, with probability at least $1-n^{-c}$, simultaneously for
every $1\le k\le M$, we have $\sum_{i\in\mathcal R : i \le k}\max\{1,i/S\} =\widetilde O(k)$ and $|B_k|=\widetilde O(S)$.
\end{restatable}

The third one, proven in \cref{sec:ub-internal-cost-proof}, bounds
the cost of an update whose endpoints are already connected in terms of
their effective resistance.

\begin{restatable}{lemma}{UBInternalCostLemma}
\label{lem:ub-internal-cost}
For every $c>0$, there is $K=K(c)$ such that, for $\kappa\ge\lambda\ge K$, with probability at least $1-n^{-c}$, simultaneously for every $i>S$ with $i \in \mathcal{I}$, we have $\mathsf{U}_i = \widetilde O(1+iR_i/S)$.
\end{restatable}

The fourth one, proven in \cref{sec:ub-cumulative-resistance-proof},
bounds the sum of these resistances.

\begin{restatable}{lemma}{UBResistanceLemma}
\label{lem:ub-cumulative-resistance}
With probability one, we have $\sum_{i\in\mathcal I : i \leq k} R_i=O(\min\{n,k\}\log^2  n)$ simultaneously for every $1\le k\le M$.
\end{restatable}

\begin{proof}[Proof of \cref{thm:smallspace} (complexity)]
We also choose $\kappa\ge\lambda$ large enough to apply
\cref{lem:ub-search,lem:ub-root-budget,lem:ub-internal-cost} with exponent
$c+2$, and let $\mathcal E\subseteq\mathcal E_1$ be the joint event of
\cref{lem:ub-landmarkfree,lem:ub-search,lem:ub-root-budget,lem:ub-internal-cost}.
Then $\Pr[\mathcal E]\ge1-4n^{-(c+2)}\ge1-n^{-c}$, and all answers are
correct on $\mathcal E$, so it remains to bound the cost and the space on
$\mathcal E$.

Fix $k\in[M]$ and let $\mathcal M\coloneqq[k]\setminus(\mathcal R\cup\mathcal I)$.
Since $\mathsf A_i=0$ for $i\notin\mathcal R$, the first $k$ updates cost
\[
    \sum_{i\le k}(\mathsf U_i+\mathsf A_i)
    =\sum_{i\in[k]\cap\mathcal R}(\mathsf U_i+\mathsf A_i)
    +\sum_{i\in[k]\cap(\mathcal I\setminus\mathcal R)}\mathsf U_i
    +\sum_{i\in\mathcal M}\mathsf U_i,
\]
plus $\widetilde O(k)$ for reporting the answers (\cref{lem:union-find}).
It suffices to show that each sum is $\widetilde O(k(1+\min\{n,k\}/S))$.
Every update $i\le S$ calls \texttt{LandmarkEdge}, so $i>S$ in the last two
sums.

\emph{Updates in $\mathcal R$.}
We show $\mathsf U_i+\mathsf A_i=\widetilde O((1+\min\{n,i\}/S)\max\{1,i/S\})$ for
$i\in\mathcal R$, which suffices by \cref{lem:ub-root-budget}, as $\min\{n,i\}\le\min\{n,k\}$.
If $i\le S$, all earlier updates called
\texttt{LandmarkEdge} too, so their endpoints are landmarks and every vertex
outside $B_{i-1}$ is isolated in $G_{i-1}$. \texttt{AddLandmark} walks only
from non-landmarks, so every walk of update $i$ returns at once, and
$\mathsf U_i+\mathsf A_i=\widetilde O(1)$. If $i>S$,
\cref{lem:ub-search}\eqref{itm:ub-search-update} gives
$\mathsf U_i=\widetilde O(\min\{n,i\}i/S^2)$, and
\cref{lem:ub-search}\eqref{itm:ub-search-basic} bounds
each of the at most two basic walks of \texttt{LandmarkEdge} by the same
amount.

\emph{Updates in $\mathcal I\setminus\mathcal R$.}
By \cref{lem:ub-internal-cost,lem:ub-cumulative-resistance}, the second
sum is
\[
    \widetilde O\Bigl(\sum_{i\in[k]\cap\mathcal I}(1+iR_i/S)\Bigr)
    =\widetilde O\Bigl(k+(k/S)\sum_{i\in[k]\cap\mathcal I}R_i\Bigr)
    =\widetilde O(k(1+\min\{n,k\}/S)).
\]

\emph{Updates in $\mathcal M$.}
Each $i\in\mathcal M$ merges the distinct components $C_{i-1}(u_i)$ and
$C_{i-1}(v_i)$; we charge $\mathsf U_i$ to the one with fewer vertices,
$D_i$ (ties broken arbitrarily).
Thus every $z\in D_i$ has $|C_i(z)|\ge2|C_{i-1}(z)|$. Since component
sizes never decrease and are at most $n$, each vertex lies in at most
$\log_2n$ of the sets $D_i$. By \cref{lem:ub-search}\eqref{itm:ub-search-update},
$\mathsf U_i=\widetilde O(\min\{n,i\}i/S^2)$ if $D_i \cap B_{i-1} \neq \emptyset$, and
$\mathsf U_i=\widetilde O(1+(\min\{n,i\}/S)\operatorname{vol}_{G_{i-1}}(D_i))$
otherwise. As $i\le k$, $\min\{n,i\}\le\min\{n,k\}$, $G_{i-1}\subseteq G_k$, and $B_{i-1}\subseteq B_k$,
either way
\[
    \mathsf U_i=\widetilde O\Bigl(1+(\min\{n,k\}/S)\sum_{z\in D_i}d_{G_k}(z)
    +(\min\{n,k\}k/S^2)\,|D_i\cap B_k|\Bigr).
\]
Summing over $i\in\mathcal M$ counts each vertex at most $\log_2n$ times.
As $\sum_zd_{G_k}(z)=2k$, the third sum is thus
$\widetilde O(k+k\min\{n,k\}/S+(\min\{n,k\}k/S^2)|B_k|)$, which is $\widetilde O(k(1+\min\{n,k\}/S))$ by
$|B_k|=\widetilde O(S)$ (\cref{lem:ub-root-budget}). Hence, on
$\mathcal E$, the first $k$ updates cost $\widetilde O(k(1+\min\{n,k\}/S))$,
simultaneously for all $k\in[M]$.

\emph{Space.}
On $\mathcal E$, we have $|B|\le|B_k|=\widetilde O(S)$ during update $k$
(\cref{lem:ub-root-budget}), so the landmarks and $D$ occupy
$\widetilde O(S)$ words (\cref{lem:union-find}). Walks run one at a time,
and all other variables, such as $z$, $Z\le N_x$, $c_u$, $c_v$, and
$T=\widetilde O(\min\{n,k\}k/S^2)$ (\cref{lem:ub-search}\eqref{itm:ub-search-update}),
take $O(1)$ words.
\end{proof}

\subsubsection{A foreword on the random walks' history}

The proofs of \cref{lem:ub-landmarkfree,lem:ub-search,lem:ub-root-budget,lem:ub-internal-cost}
analyze one walk at a time, or one \emph{batch}, i.e., the $m$ trials
from an endpoint at one scale. We define the \emph{history} $\mathcal H$ of a walk
or batch to be the outcome of all random choices made before it starts.
As the insertions are fixed in advance, $\mathcal H$ determines the
current update $i$, the graph $G_{i-1}$, the landmark set $B$, and the
arguments $x,y,T$, while the walk's own random choices are independent
of $\mathcal H$. Hence, given $\mathcal H$, a walk is a simple random
walk from $x$ in the fixed graph $G_{i-1}$ with fixed stopping rules,
and the trials of a batch are independent such walks, as long as they
run. 

\subsubsection{Random walks and landmark-free declarations}
\label{sec:ub-search-proof}

Here, we prove \cref{lem:ub-landmarkfree,lem:ub-search}. Both rely on
\cref{prop:ub-sampling} that proves the simple fact that sampled landmarks make walks hit them quickly
and landmark-free components small. 
To this end, fix a constant $\alpha$ with
$2\ln\kappa\le \alpha\le\sqrt\kappa$, which each proof below chooses
sufficiently large in terms of its exponent $c$; this is possible as
$\kappa\ge K$ is large. Moreover, for a walk
$(X_t)_{t\ge0}$ in $G_{i-1}$, put $\mathcal V_t\coloneqq\{X_0,\ldots,X_t\}$
and $\sigma_i\coloneqq\inf\{t:\operatorname{vol}(\mathcal V_t)\ge4\alpha i\log(2n)/S\}$.
The \emph{landmark samples} are the uniform vertices drawn by updates $i\le S$ and the sampling coins of updates $i>S$; they are independent of the update sequence.

\begin{proposition}
\label{prop:ub-sampling}
With probability at least $1-O(n^{3-\alpha})$ over the landmark samples,
the following hold for all $i>S$, the current landmark set $B\supseteq B_{i-1}$, and components
$C$ of $G_{i-1}$, where $r\coloneqq\lceil 2\alpha\min\{n,i\}\log(2n)/S\rceil$:
\begin{enumerate}[label=(\alph*),ref=\alph*,nosep,topsep=3pt]
    \item \label[ubsamplingpart]{itm:ub-sampling-walk}
    $\Pr_z[\sigma_i<\tau(B)]\le(2n)^{-\alpha}$ for $z\in[n]$.
    \item \label[ubsamplingpart]{itm:ub-sampling-distance}
    If $C\cap B\ne\emptyset$, then
    $R_{G_{i-1}}(z,B)\le\operatorname{dist}_{G_{i-1}}(z,B)\le r$
    for $z\in C\setminus B$.
    \item \label[ubsamplingpart]{itm:ub-sampling-rootfree}
    If $C\cap B=\emptyset$, then $|C|<r$ and $\operatorname{vol}(C)<4\alpha i\log(2n)/S$.
\end{enumerate}
\end{proposition}
\begin{proof}
Throughout, put $q_i\coloneqq4\alpha i\log(2n)/S$, so that $\sigma_i=\inf\{t:\operatorname{vol}(\mathcal V_t)\ge q_i\}$.
For $i>S$, let $L_i$ consist of $s$, $t$, the uniform landmarks, and the endpoints of those $e_j$, $j<i$, with $j\le S$ or update $j$ sampled; then $L_i\subseteq B_{i-1}\subseteq B$, and, as $L_i$ is the set $B_{i-1}$ of \cref{alg:ub-large-update} for the same landmark samples, \cref{fact:ub-landmarks}\eqref{itm:ub-landmarks-hit} applies to it.

We first prove \cref{itm:ub-sampling-walk}. For $z\in[n]$, let $p_{i,z}\coloneqq\Pr_z[\sigma_i<\tau(L_i)\mid L_i]$ for a walk independent of the landmark samples.
For every fixed trajectory of the walk with $\sigma_i<\infty$, we have $\operatorname{vol}(\mathcal V_{\sigma_i})\ge q_i$, so $L_i$ misses $\mathcal V_{\sigma_i}$ with probability at most $e^{-Sq_i/(2i)}=(2n)^{-2\alpha}$ by \cref{fact:ub-landmarks}\eqref{itm:ub-landmarks-hit}.
Hence $\Ex[p_{i,z}]\le(2n)^{-2\alpha}$, and Markov's inequality gives $\Pr[p_{i,z}>(2n)^{-\alpha}]\le(2n)^{-\alpha}$. By a
union bound over the $O(n^3)$ pairs $(i,z)$, all $p_{i,z}\le(2n)^{-\alpha}$,
except with probability $O(n^{3-\alpha})$. On this event,
\cref{itm:ub-sampling-walk} follows as $\tau(B)\le\tau(L_i)$.

For \cref{itm:ub-sampling-distance,itm:ub-sampling-rootfree}, let $W$ be the first $r$ vertices of a breadth-first search from $z$ in $G_{i-1}$, for $i>S$ and $z\in[n]$ with $|C_{i-1}(z)|\ge r$; then, as $r\ge2\alpha\log(2n)\ge2$, no vertex of $W$ is isolated in $G_{i-1}$, so $\Pr[L_i\cap W=\emptyset]\le e^{-Sr/(2\min\{n,i\})}\le(2n)^{-\alpha}$ by \cref{fact:ub-landmarks}\eqref{itm:ub-landmarks-hit}.
By a union bound over the $O(n^3)$ pairs $(i,z)$, all such $W$ satisfy
$L_i\cap W\ne\emptyset$, except with probability $O(n^{3-\alpha})$.
On this event, let $z\in C\setminus B$. If $|C|\ge r$, the prefix $W$ from
$z$ satisfies $W\cap B\supseteq W\cap L_i\ne\emptyset$, so
$\operatorname{dist}(z,B)<r$; this also holds if $|C|<r$ and
$C\cap B\ne\emptyset$. By \cref{fact:resistance}\eqref{itm:resistance-rayleigh},
$R_{G_{i-1}}(z,B)\le\operatorname{dist}(z,B)$. This proves
\cref{itm:ub-sampling-distance}, and shows $|C|<r$ if $C\cap B=\emptyset$;
then a walk from $z$ never leaves $C$ and visits all of $C$ with
probability one, so $\operatorname{vol}(C)\ge q_i$ would give
$\Pr_z[\sigma_i<\tau(B)]=1$, contradicting \cref{itm:ub-sampling-walk}. This proves
\cref{itm:ub-sampling-rootfree}.
\end{proof}

We move on to the proofs of \cref{lem:ub-search} and \cref{lem:ub-landmarkfree}, respectively.
In both proofs, we fix landmark samples satisfying \cref{prop:ub-sampling}
and include all of them, even later ones, in every history.

\UBSearchLemma*
\begin{proof}
We bound each walk and batch given its history and take a union bound.

\proofparagraph{\Cref{itm:ub-search-basic}.}
Consider a basic walk from $x$ during update $i>S$, let $\mathsf L_x$ be
its number of steps, and abbreviate $C\coloneqq C_{i-1}(x)$,
$r\coloneqq\lceil 2\alpha\min\{n,i\}\log(2n)/S\rceil$, and $q_i\coloneqq4\alpha i\log(2n)/S$. We
show that, given its history, with probability at least $1-2(2n)^{-\alpha}$,
\begin{equation}
\label{eq:ub-basic-walk-cost}
    \mathsf L_x=\widetilde O(\min\{n,i\}i/S^2),
    \qquad\text{and}\qquad
    \mathsf L_x=\widetilde O\bigl(1+(\min\{n,i\}/S)\operatorname{vol}(C)\bigr)
    \ \text{ if $C\cap B=\emptyset$.}
\end{equation}
If $C\cap B\ne\emptyset$, the walk stops in at most $\tau(B)$ steps, so it suffices to
show $\tau(B)=\widetilde O(\min\{n,i\}i/S^2)$ with probability at least $1-2(2n)^{-\alpha}$. For
$v\in C\setminus B$, let $Y_v\coloneqq|\{t<\tau(B):X_t=v\}|$ be the number
of visits to $v$ before $\tau(B)$. These visits occur during
$[\tau(v),\tau(B))$, so, by the strong Markov property at $\tau(v)$, $Y_v$ is
distributed as $V_v(B)$ for a fresh walk from $v$ if $\tau(v)<\tau(B)$, and
$Y_v=0$ otherwise. By \cref{lem:random-walk-facts} and
\cref{prop:ub-sampling}\eqref{itm:ub-sampling-distance}, $V_v(B)$ is
geometric with mean $d(v)R_{G_{i-1}}(v,B)\le rd(v)$, so
\[
    \Pr[Y_v>2\lceil \alpha\log(2n)\rceil rd(v)]\le e^{-2\alpha\log(2n)}\le(2n)^{-2\alpha}.
\]
Moreover,
$\tau(B)\le\sigma_i$ with probability at least $1-(2n)^{-\alpha}$ by
\cref{prop:ub-sampling}\eqref{itm:ub-sampling-walk}, and then the vertices
$v$ with $Y_v>0$ have total volume below $q_i$. Hence, except with
probability $(2n)^{-\alpha}+n(2n)^{-2\alpha}\le2(2n)^{-\alpha}$,
\[
    \tau(B)
    =\sum_{v\in C\setminus B}Y_v
    \le2\lceil \alpha\log(2n)\rceil r\sum_{v:\,Y_v>0}d(v)<2\lceil \alpha\log(2n)\rceil rq_i
    =\widetilde O(\min\{n,i\}i/S^2).
\]

If $C\cap B=\emptyset$, then $\mathsf L_x=0$ if $d(x)=0$; otherwise, the
walk stops at its $N_x$-th visit to $x$, and, at every time, given the
walk so far, its expected remaining duration is at most
\begin{align*}
    \max_{z\in C}\Ex_z[\tau(x)]+N_x\Ex_x[\tau^+(x)]
    &\le\operatorname{vol}(C)\max_{z\in C}R_{G_{i-1}}(z,x)+N_x\operatorname{vol}(C)/d(x)
    &&\text{(\cref{lem:random-walk-facts})}\\
    &\le r\operatorname{vol}(C)+N_x\operatorname{vol}(C)/d(x)
    &&\text{(\cref{fact:resistance}\eqref{itm:resistance-rayleigh}, \cref{prop:ub-sampling}\eqref{itm:ub-sampling-rootfree})}\\
    &\le\operatorname{vol}(C)\bigl(r+\kappa\lambda(\min\{n,i\}/S)\log^2(2n)+1\bigr)
    &&\text{(definition of $N_x$)}.
\end{align*}
By Markov's inequality, the walk thus stops within each block of twice as
many steps with probability at least $1/2$ if it has not stopped before,
so after $\lceil \alpha\log(2n)\rceil$ blocks, i.e.,
$\widetilde O\bigl((\min\{n,i\}/S)\operatorname{vol}(C)\bigr)$ steps, it has stopped
except with probability $2^{-\alpha\log(2n)}=(2n)^{-\alpha}$. As
$\operatorname{vol}(C)<q_i$ by
\cref{prop:ub-sampling}\eqref{itm:ub-sampling-rootfree}, this also gives
the first bound in \cref{eq:ub-basic-walk-cost}.

\proofparagraph{\Cref{itm:ub-search-update}.}
Fix $i>S$. A sampled update has $\mathsf U_i=O(1)$, so consider an update
that is not sampled, and so executes the search in
lines~\ref{ln:ub-update-init}--\ref{ln:ub-update-promoted-edge} of
\cref{alg:ub-update}. As it adds no landmarks before
line~\ref{ln:ub-update-promoted-edge}, $B=B_{i-1}$ during its trials. A trial
\UBWalk{$x,y,T$} makes the same steps as a basic walk from $x$, to which
\cref{eq:ub-basic-walk-cost} applies, until it reaches $y$ or is killed,
and its killing coins are independent of its steps. Let $L_*$ be the
bound in \cref{eq:ub-basic-walk-cost} that applies to $x$, i.e.,
$L_*=\widetilde O(1+(\min\{n,i\}/S)\operatorname{vol}(C_{i-1}(x)))$ if
$C_{i-1}(x)\cap B=\emptyset$, and $L_*=\widetilde O(\min\{n,i\}i/S^2)$ otherwise.
For $T\ge2L_*$, the trial passes the
killing tests of its first $L_*$ steps with probability
$(1-1/(T+1))^{L_*}\ge1/2$, so it returns a landmark, $\mathsf{Connected}$, or
$\mathsf{LandmarkFree}$ with probability at least $(1-2(2n)^{-\alpha})/2\ge1/4$.

Call the batch from $x$ at the smallest scale $T\ge2L_*$ \emph{critical};
it \emph{fails} if all its $m$ trials are killed, which, given its
history, has probability at most $(3/4)^m$. Otherwise, $c_x$ is set or
the update ends, the latter if $C_{i-1}(x)\cap B=\emptyset$, as then no
trial from $x$ returns a landmark. So, unless a critical batch fails, the
search ends by the smallest scale $T\ge2L_*$ for each endpoint $x$ with
$c_x=\bot$; promotions only end it earlier. Killing ends each trial at
scale $T$ within $\alpha(T+1)\log(2n)$ steps with probability
at least $1-(2n)^{-\alpha}$. So each scale $T$ costs $\widetilde O(T)$, and summing up
to scale $\max\{1,4L_*\}$ gives $\mathsf U_i=\widetilde O(1+L_*)$, proving
\cref{itm:ub-search-update}, as $L_*=\widetilde O(\min\{n,i\}i/S^2)$ in either case.

\proofparagraph{Union bound.}
A walk fails if it is a basic walk exceeding its bound in
\cref{eq:ub-basic-walk-cost}, or is a trial longer than $\alpha(T+1)\log(2n)$ steps; given its history, this has
probability $O((2n)^{-\alpha})$. Each update has at most two critical
batches. Until the first failure, each update uses only the
$O(\log(n\kappa))$ scales $T\le\max\{1,4L_*\}=\widetilde O(\min\{n,i\}i/S^2)$, as
$\lambda\le\kappa$ and $\alpha\le\sqrt\kappa$. So the
$M\le\binom n2$ updates run $O(n^2\kappa\log(n)\log(n\kappa))$ trials and
$O(n^2)$ basic walks. As $(2n)^{-\alpha}\le(2n)^{-\alpha/2}\kappa^{-1}$ by
$\alpha\ge2\ln\kappa$, a union bound with \cref{prop:ub-sampling} shows that
some failure occurs with probability $O(n^{3-\alpha/2}+n^2(3/4)^m)\le n^{-c}$
for sufficiently large $\alpha$ and $\kappa$. Otherwise, both parts hold for every update $i>S$.
\end{proof}

\UBLandmarkFreeLemma*
\begin{proof}
Note that a $\mathsf{LandmarkFree}$ declaration can be wrong only if $C_{i-1}(x)\cap B\ne\emptyset$, $x$ is not a landmark, $d(x)>0$ and $i>S$. In this case, a walk declares $\mathsf{LandmarkFree}$ only if $V_x(B)\ge N_x$. By
\cref{lem:random-walk-facts} and
\cref{prop:ub-sampling}\eqref{itm:ub-sampling-distance}, $V_x(B)$
is geometric with mean at most $d(x)r$ for $r\coloneqq\lceil 2\alpha\min\{n,i\}\log(2n)/S\rceil$;
as $\kappa\lambda\ge3\kappa\ge 2\alpha^2+\alpha+1$ for $\lambda\ge3$ and
$\alpha\le\sqrt\kappa$, and as $\min\{n,i\}\ge S$,
\[
    \Pr[\UBWalk{$x,y,T$}=\mathsf{LandmarkFree}\mid\mathcal H]
    \le\Pr\nolimits_x[V_x(B)\ge N_x]
    \le\bigl(1-1/(d(x)r)\bigr)^{N_x-1}
    \le(2n)^{-\alpha}.
\]

By \cref{lem:ub-search}\eqref{itm:ub-search-update} with exponent $c+1$,
except with probability $n^{-(c+1)}$, each update uses only
$O(\log(n\kappa))$ scales, so all walks are among the first
$J=O(n^2\kappa\log(n)\log(n\kappa))$ ones, as each update runs at most
three basic walks. A union bound over these $J$ walks and
\cref{prop:ub-sampling} shows that some declaration is incorrect with
probability at most $n^{-(c+1)}+O(n^{3-\alpha})+J(2n)^{-\alpha}\le n^{-c}$
for sufficiently large $\alpha$ and $K$.
\end{proof}

\subsubsection{Landmark budget}
\label{sec:ub-root-budget-proof}

Here, we prove \cref{lem:ub-root-budget}, for which we first need a
potential for landmark sets of general graphs.

For a graph $G = (V,E)$, a set $B \subseteq V$, and $T\ge1$, the
\emph{auxiliary walk} is a simple random walk in $G$ that stops at $B$
and is killed with probability $1/(T+1)$ before each step; let $p_B(v)$
be its killing probability from $v$, so $p_B(b)=0$ for each $b \in B$. In other words, an auxiliary walk from $x$ is a trial \UBWalk{$x,y,T$} of
\cref{alg:ub-helpers} without the target $y$ and without the visit limit
$N_x$: it ends either at a landmark or by a kill, and $p_B(x)$ is the
probability of the latter. 

We define the potential
$\Phi_T(G,B)\coloneqq\sum_vd_G(v)p_B(v)$ as a measure of how often walks of
about $T$ steps miss $B$. For $x\notin B$ with $d_G(x)>0$, let $Z_x$
count the visits to $x$ of the auxiliary walk from $x$, including the
initial visit and the one at which it is killed, and let
$\mu_x\coloneqq\Ex_x[Z_x]$.

\begin{lemma}
\label{lem:ub-root-potential}
For every graph $G$, set $B$, and $T\ge1$, the following hold:
\begin{enumerate}[label=(\alph*),ref=\alph*,nosep,topsep=3pt]
    \item \label[ubpotentialpart]{itm:ub-potential-edge}
    $\Phi_T(G+uv,B)-\Phi_T(G,B)\le2$ for $uv\notin E(G)$.
    \item \label[ubpotentialpart]{itm:ub-potential-root}
    $\Phi_T(G,B)-\Phi_T(G,B\cup\{x\})
    =d_G(x)(T+1)p_B(x)^2/\mu_x$ for $x\notin B$ with $d_G(x)>0$.
    \item \label[ubpotentialpart]{itm:ub-potential-monotone}
    $\Phi_T(G,B')\le\Phi_T(G,B)$ for $B'\supseteq B$.
\end{enumerate}
\end{lemma}
\begin{proof}
For \cref{itm:ub-potential-edge}, put
$J_G(f)\coloneqq\mathcal E_G(f)+(1/T)\sum_vd_G(v)(1-f(v))^2$ for
$f:V\to\mathbb R$ with $f(b)=0$ for all $b\in B$. For $v\notin B$, the
auxiliary walk from $v$ is killed at once with probability $1/(T+1)$, and
otherwise moves to a uniform neighbor $w$ and continues as an auxiliary
walk from $w$. Hence
\[
    p_B(v)=\frac{1}{T+1}+\frac{T}{T+1}\cdot\frac{1}{d_G(v)}\sum_{w\in N_G(v)}p_B(w),
    \ \text{from which it follows that}\
    (\mathcal L_Gp_B)(v)=\frac{d_G(v)\bigl(1-p_B(v)\bigr)}{T},
\]
which says that the gradient of the convex function $J_G$ on these
functions vanishes at $p_B$. Since
$\mathcal E_G(p_B)=\sum_{v\notin B}p_B(v)(\mathcal L_Gp_B)(v)$ we have
\begin{align*}
    \min\nolimits_{f:\,f(b)=0\ \forall b\in B}J_G(f)=J_G(p_B)
    &=\frac1T\sum_{v\notin B}d_G(v)p_B(v)\bigl(1-p_B(v)\bigr)
    +\frac1T\sum_vd_G(v)\bigl(1-p_B(v)\bigr)^2\\
    &=\frac1T\sum_vd_G(v)\bigl(1-p_B(v)\bigr)
    =\frac{2|E(G)|-\Phi_T(G,B)}{T}.
\end{align*}
For every $f$, we have
$\mathcal E_{G+uv}(f)=\mathcal E_G(f)+\bigl(f(u)-f(v)\bigr)^2\ge\mathcal E_G(f)$
and $d_{G+uv}(w)\ge d_G(w)$ for all $w$, so adding $uv$ cannot lower the
minimum but raises $|E(G)|$ by one; hence $\Phi_T$ grows by at most~$2$.

For \cref{itm:ub-potential-root}, let $g(v,y)$, for $v,y\notin B$, be
the expected number of visits to $y$ by the auxiliary walk from $v$, so
$g(x,x)=\mu_x$. By the strong Markov property, the walk from $v$ visits
$x$ with probability $g(v,x)/\mu_x$ and is then killed with probability
$p_B(x)$; making $x$ a landmark removes exactly these kills, so
$p_B(v)-p_{B\cup\{x\}}(v)=g(v,x)p_B(x)/\mu_x$. By reversibility,
$d_G(v)g(v,x)=d_G(x)g(x,v)$
(see also \cite[Proposition~2.1]{lyonsperes2021}). Each visit
outside $B$ makes one killing test, so
$\sum_{v\notin B}g(x,v)=(T+1)p_B(x)$. Thus
\[
    \Phi_T(G,B)-\Phi_T(G,B\cup\{x\})
    =\frac{d_G(x)p_B(x)}{\mu_x}\sum_{v\notin B}g(x,v)
    =\frac{d_G(x)(T+1)p_B(x)^2}{\mu_x}.
\]

For \cref{itm:ub-potential-monotone}, couple the auxiliary walks for
$B$ and $B'$ through the same steps and killing coins: the walk for $B'$
stops no later, so it is killed no more often, i.e., $p_{B'}\le p_B$
pointwise.
\end{proof}

\UBRootBudgetLemma*
\begin{proof}
Call $W_i\coloneqq\max\{1,i/S\}$ the \emph{weight} of update $i$, so that
the first bound asks for the total weight of the updates in $\mathcal R$
among the first $k$. These are of three types:
\begin{enumerate*}[label=(\roman*),ref=\roman*]
    \item\label[ubbudgettype]{itm:ub-budget-initial} the updates $i\le S$;
    \item\label[ubbudgettype]{itm:ub-budget-sampled} the updates sampled
    with probability $S/i$; and
    \item\label[ubbudgettype]{itm:ub-budget-promoted} the promoted updates.
\end{enumerate*}
We bound the weight of each type separately, and then derive the second
bound from the first.

\proofparagraph{Weight of \cref{itm:ub-budget-initial}.}
These updates have weight $1$, so those among the first $k$ weigh at most
$k$ in total.

\proofparagraph{Weight of \cref{itm:ub-budget-sampled}.}
Let $X$ count all sampled updates $i>S$. It is a sum of independent
indicators with mean $\sum_{S<i\le M}S/i=O(S\log(2n))$.
By a Chernoff bound, with probability at least $1 - n^{-(c+3)}$, we have
$X=O_c(S\log(2n))=\widetilde O(S)$. On this event, simultaneously for
every $k>S$, the sampled updates $i\le k$ have total weight at most
$(k/S)X=\widetilde O(k)$; for $k\le S$, there are none.

\proofparagraph{Weight of \cref{itm:ub-budget-promoted}.}
We use the potentials $\Phi_T$ of \cref{lem:ub-root-potential}, one per
scale $T$.
Let $\mathcal P_T$ be the updates promoted at scale $T$, and fix a batch
from $x$ at scale $T$ of an update $i>S$, with history $\mathcal H$. As
the search adds no landmarks, its trials run in $G\coloneqq G_{i-1}$ with
landmarks $B=B_{i-1}$, and $x\notin B$ as $c_x=\bot$. Assume $d_G(x)>0$, as
otherwise trials return $\mathsf{LandmarkFree}$. As $W_i=i/S$, a kill is
light if $Z\le h_x\coloneqq d_G(x)(T+1)/(8W_i)$; let
$\beta_x\coloneqq\Pr_x[\mathsf{Killed},\,Z_x\le h_x]$ for the auxiliary
walk in $G$ with landmarks $B$ at scale $T$. A trial is this auxiliary walk
with extra stops at $y$ and at $Z=N_x$, and its counter $Z$ counts
visits to $x$ as $Z_x$ does, so its light kills are also ones of the
coupled auxiliary walk. As the trials are independent given
$\mathcal H$, all $m$ end in light kills with probability at most
$\beta_x^m\le2^{-m}$ if $\beta_x<1/2$. On the event of
\cref{lem:ub-search} with exponent $c+3$, there are $O(\log(n\kappa))$
scales per search by \cref{lem:ub-search}\eqref{itm:ub-search-update},
hence $O(n^2\log(n\kappa))$ batches. By a union bound, except with
probability $n^{-(c+3)}+O(2^{-m}n^2\log(n\kappa))$, this event holds and
every $x$ that triggers a promotion has $\beta_x\ge1/2$.

For such an $x$, $p_B(x)\ge\beta_x\ge1/2$. By the strong Markov
property at visits to $x$, $Z_x\sim\Geo(1/\mu_x)$, so
$1/2\le\beta_x\le1-(1-1/\mu_x)^{\lfloor h_x\rfloor}\le h_x/\mu_x$, i.e.,
$\mu_x\le2h_x$. The promotion calls \texttt{LandmarkEdge}, so
$B_i\supseteq B\cup\{x\}$. Hence
\cref{lem:ub-root-potential}\eqref{itm:ub-potential-monotone}
and~\eqref{itm:ub-potential-root} give
\[
    \Phi_T(G,B)-\Phi_T(G,B_i)
    \ge\frac{d_G(x)(T+1)p_B(x)^2}{\mu_x}
    \ge\frac{d_G(x)(T+1)}{8h_x}=W_i.
\]
Now, each update first adds landmarks, which never increases $\Phi_T$ by
\cref{lem:ub-root-potential}\eqref{itm:ub-potential-monotone}, and then
inserts $e_i$, raising it by at most $2$ by
\cref{lem:ub-root-potential}\eqref{itm:ub-potential-edge}. As
$\Phi_T(G_0,B_0)=0\le\Phi_T(G_k,B_k)$, telescoping gives
$\sum_{i\in\mathcal P_T\cap[k]}W_i\le2k$. Summing over the
$O(\log(n\kappa))$ scales, the promoted updates $i\le k$ weigh
$\widetilde O(k)$. Adding the weights of the three types proves the lemma's
first bound.

\proofparagraph{Number of landmarks.}
Put $Q_0\coloneqq0$ and $Q_k\coloneqq\sum_{i\in\mathcal R\cap[k]}W_i$, so
that $Q_i-Q_{i-1}$ is $W_i$ if $i\in\mathcal R$ and $0$ otherwise. Since
$1\le SW_i/i$ for every $i$, summation by parts and the first bound
$Q_i=\widetilde O(i)$ give
\begin{align*}
    |\mathcal R|
    &\le\sum_{i\in\mathcal R}\frac{SW_i}{i}
    =S\sum_{i=1}^M\frac{Q_i-Q_{i-1}}i
    =\frac{SQ_M}{M}+S\sum_{i=1}^{M-1}Q_i\Bigl(\frac1i-\frac1{i+1}\Bigr)\\
    &=\frac{SQ_M}{M}+S\sum_{i=1}^{M-1}\frac{Q_i}{i(i+1)}
    =\widetilde O(S)+\widetilde O(S)\sum_{i=1}^{M-1}\frac1{i+1}
    =\widetilde O(S).
\end{align*}
Besides $s,t$, uniform samples add at most $S$ landmarks and each \texttt{LandmarkEdge}
at most two, so $|B_k|\le2+S+2|\mathcal R|=\widetilde O(S)$. The failure
probabilities sum to at most $n^{-c}$ for large $\kappa$.
\end{proof}

\subsubsection{Internal-update cost}
\label{sec:ub-internal-cost-proof}

Here, we prove \cref{lem:ub-internal-cost}.

\UBInternalCostLemma*
\begin{proof}
Fix $i>S$ with $i\in\mathcal I$. A sampled update has $\mathsf U_i=O(1)$,
so consider an update that is not sampled, i.e., the search in
lines~\ref{ln:ub-update-init}--\ref{ln:ub-update-promoted-edge} of
\cref{alg:ub-update}. Choose $\alpha$ sufficiently large in terms of
$c,\kappa$, and let $T_*$ be the smallest scale with
$T_*\ge \alpha(1+iR_i/S)\log(2n)$. Then $T_*=\widetilde O(1+iR_i/S)$ depends
only on $i$, and $\log_2T_*=O(\log n)$, as $i\le\binom n2$ and $R_i\le n-1$.

The search ends by scale $T_*$ unless a trial at that scale is a
nonlight kill. Indeed, if it does not, some endpoint $x$ still has
$c_x=\bot$ after scale $T_*$, so none of its $m$ trials at this scale
returned a landmark, $\mathsf{Connected}$, or $\mathsf{LandmarkFree}$. All were
thus kills, and one was nonlight, as the update was not promoted.

We claim that such kills are unlikely. Indeed, given its history, a trial \UBWalk{$x,y,T_*$}
from either endpoint $x$ follows a simple random walk in $G_{i-1}$
until, at the latest, it reaches $y$. So a nonlight kill requires $V_x(\{y\})>h_x\coloneqq Sd(x)(T_*+1)/(8i)$. As
$y\in C_{i-1}(x)$, \cref{lem:random-walk-facts} gives
$V_x(\{y\})\sim\Geo(1/\mu)$ with mean $\mu\coloneqq d(x)R_i\ge1$, and
$h_x/\mu\ge(\alpha/8)\log(2n)$ by the choice of $T_*$. Hence
\[
    \Pr[\text{nonlight kill}\mid\mathcal H]
    \le\Pr\nolimits_x[V_x(\{y\})>h_x]
    =(1-1/\mu)^{\lfloor h_x\rfloor}
    \le e^{1-h_x/\mu}
    \le e(2n)^{-\alpha/8}.
\]
A trial at scale $T$ is killed before each step with
probability $1/(T+1)$, so, given its history, it takes more than
$\alpha(T+1)\log(2n)$ steps with probability at most $(2n)^{-\alpha}$. A union
bound over the at most $2m$ trials per scale shows that, except with
probability
$2m(1+\log_2T_*)(2n)^{-\alpha}+2m\,e(2n)^{-\alpha/8}\le n^{-(c+3)}$, no trial at
scale $T_*$ is a nonlight kill and every trial at scale $T\le T_*$ takes
$\widetilde O(T)$ steps; then the search ends by scale $T_*$, and
$\mathsf U_i=\widetilde O(1+2+\dots+T_*)=\widetilde O(1+iR_i/S)$. A union
bound over the $M\le\binom n2$ updates proves the lemma.
\end{proof}

\subsubsection{Cumulative resistance}
\label{sec:ub-cumulative-resistance-proof}

Here, we prove \cref{lem:ub-cumulative-resistance}.

\UBResistanceLemma*
\begin{proof}
As $\mathcal I\cap[k]$ and the $R_i$ with $i\le k$ depend only on $e_1,\ldots,e_k$, we may assume $k=M$.
Fix a connected component $C$ of $G_M$ with $h=|C|\ge2$, and let
$\mathcal I_C\coloneqq\{i\in\mathcal I:u_i,v_i\in C\}$.
It suffices to prove that $\sum_{i\in\mathcal I_C}R_i=O(h\log^2h)$.
Indeed, every $i\in\mathcal I$ lies in $\mathcal I_C$ for exactly one such
$C$, as $u_i\ne v_i$ are connected in $G_{i-1}\subseteq G_M$.
As every vertex of such a $C$ is an endpoint of some $e_j$, we have $\sum_C|C|\le\min\{n,2M\}$, so summing over $C$ then gives $\sum_{i\in\mathcal I}R_i\le\sum_CO(|C|\log^2|C|)=O(\min\{n,M\}\log^2n)$.

To bound said quantity $\sum_{i\in\mathcal I_C}R_i$, we set
$L\coloneqq16\ln(2h)$ and we write:
\[
    \sum_{i\in\mathcal I_C}R_i
    =\sum_{i\in\mathcal I_C}\Bigl(\min\{R_i,L\}+\int_L^{\infty}\mathbf 1[R_i\ge a]\,da\Bigr)
    \le2L\cdot\underbrace{\sum_{i\in\mathcal I_C}\log(1+R_i)}_{(\dagger)}
    +\int_L^{\max\{L,h\}}\underbrace{\bigl|\{i\in\mathcal I_C:R_i\ge a\}\bigr|}_{(\ddagger)}\,da,
\]
where we use $\min\{t,L\}\le2L\log(1+t)$ for $t\ge0$, and that
term~$(\ddagger)$ vanishes for $a>h-1$, as $R_i\le\operatorname{dist}_{G_{i-1}}(u_i,v_i)\le h-1$
by \cref{fact:resistance}\eqref{itm:resistance-rayleigh}.

Thus, our focus is to bound terms~$(\dagger)$ and~$(\ddagger)$. Both bounds use the
set $F$ of edges $e_j$ with $u_j,v_j\in C$ and $j\notin\mathcal I$. Each
such $e_j$ joins two components of $G_{j-1}$, while the edges $e_j$ with
$j\in\mathcal I$ join none. Hence $F$ is acyclic and connects $C$, i.e.,
$F$ is a spanning tree of $C$ with $h-1$ edges.

\begin{subclaim}
\label{clm:UBResistanceClaim}
Term~$(\dagger)$ satisfies $\sum_{i\in\mathcal I_C}\log(1+R_i)\le h\log h$.
\end{subclaim}
\begin{subproof}
Let $\operatorname{Tree}(H)$ be the number of spanning trees of a graph
$H$. For a connected graph $H$ and an edge $e=uv\notin E(H)$,
\cite[Section~3, before Lemma~3.1]{lipattersonyizhang2018} gives
\begin{equation}
\label{eq:ub-tree-insertion}
    \operatorname{Tree}(H+e)=\operatorname{Tree}(H)\bigl(1+R_H(u,v)\bigr).
\end{equation}
We apply \cref{eq:ub-tree-insertion} along a reordered insertion sequence
on $C$: first the tree $F$, then the edges $e_i$, $i\in\mathcal I_C$, in
increasing order of $i$. The graph just before $e_i$ is
$H_i\coloneqq G_{i-1}[C]\cup F$, as every edge of $G_M[C]$ not of the
form $e_j$, $j\in\mathcal I_C$, lies in $F$. Each $H_i$ is connected, as
it contains $F$, and the sequence goes from $F$, with
$\operatorname{Tree}(F)=1$, to $G_M[C]$, with
$\operatorname{Tree}(G_M[C])\le h^{h-2}$ by Cayley's formula. Taking
logarithms in \cref{eq:ub-tree-insertion} and telescoping gives
\[
    \sum_{i\in\mathcal I_C}\log\bigl(1+R_{H_i}(u_i,v_i)\bigr)
    =\log\operatorname{Tree}(G_M[C])-\log\operatorname{Tree}(F)
    \le(h-2)\log h.
\]
It remains to show $R_{H_i}(u_i,v_i)=R_i$ for $i\in\mathcal I_C$. No edge
of $G_{i-1}$ has exactly one endpoint in $C$, as $C$ is a union of
components of $G_{i-1}$, so
\cref{fact:resistance}\eqref{itm:resistance-restrict} gives
$R_i=R_{G_{i-1}[C]}(u_i,v_i)$. Next, $H_i$ arises from $G_{i-1}[C]$ by
inserting the edges $e_j\in F$ with $j>i$ in increasing order of $j$.
Each such $e_j$ has its endpoints in different components of the current
graph, as the latter is a subgraph of $G_{j-1}$, in which $e_j$ joins two
components. Hence \cref{fact:resistance}\eqref{itm:resistance-bridge},
applied once per insertion, gives
$R_{H_i}(u_i,v_i)=R_{G_{i-1}[C]}(u_i,v_i)=R_i$.
\end{subproof}

\begin{subclaim}
\label{clm:UBResistanceClaimTwo}
For every $a\ge L$, term~$(\ddagger)$ satisfies
$|\{i\in\mathcal I_C:R_i\ge a\}|\le8h\ln(2h)/a$.
\end{subclaim}
\begin{subproof}
Let $J$ be the graph on $C$ with edge set
$F\cup\{e_i:i\in\mathcal I_C,\ R_i\ge a\}$. As $|F|=h-1$ and
$e_i\notin F$ for $i\in\mathcal I_C$, we have to show
$|E(J)|\le h-1+8h\ln(2h)/a$.

Every cycle $Q$ of $J$ has length at least $a+1$. Indeed, let $e_j$ be
the edge of $Q$ with the largest index. Then $Q-e_j$ is a $u_j$--$v_j$
path in $G_{j-1}$. So $j\in\mathcal I$, which rules out $e_j\in F$, and
$a\le R_j\le\operatorname{dist}_{G_{j-1}}(u_j,v_j)\le|Q|-1$ by
\cref{fact:resistance}\eqref{itm:resistance-rayleigh}.

Set $r\coloneqq\lfloor a/3\rfloor$ and
$\delta\coloneqq8\ln(2h)/a\le1/2$, and say that an edge \emph{leaves} a
vertex set $U$ if exactly one of its endpoints lies in $U$. Let $J'$ be a
nonempty induced
subgraph of $J$, let $z$ be a vertex of $J'$, and let $U_\rho$ be the set
of vertices at distance at most $\rho$ from $z$ in $J'$. Then $J'[U_r]$
is a tree: it is connected, and an edge of $J'[U_r]$ outside the
breadth-first search tree from $z$ would close a cycle of length at most
$2r+1\le a$. Hence, for $\rho<r$, distinct edges leaving $U_\rho$ end at
distinct vertices of $U_{\rho+1}\setminus U_\rho$. If every $U_\rho$ with
$\rho<r$ had more than $\delta|U_\rho|$ leaving edges, then
$ h\ge|U_r|>(1+\delta)^r\ge e^{\delta r/2}\ge e^{\delta a/8}=2h$,
using $\ln(1+\delta)\ge\delta/2$ and $r\ge a/4$, the latter as
$a\ge L\ge12$; a contradiction. So some $U\coloneqq U_\rho$ with $\rho<r$
induces a tree, with $|U|-1$ edges, and has at most $\delta|U|$ leaving
edges.

Starting from $J'=J$, repeatedly remove such a set $U$ from $J'$ until no
vertex is left. The removed sets partition $C$, and every edge of $J$
lies inside, or leaves, the first removed set containing one of its
endpoints. Hence $|E(J)|\le\sum_U\bigl(|U|-1\bigr)+\delta\sum_U|U|
\le h-1+\delta h=h-1+\frac{8h\ln(2h)}a$.\qedhere
\end{subproof}

By \cref{clm:UBResistanceClaim,clm:UBResistanceClaimTwo}, we conclude
\[
    \sum_{i\in\mathcal I_C}R_i
    \le2Lh\log h+\int_L^{\max\{L,h\}}\frac{8h\ln(2h)}a\,da
    \le32h\ln(2h)\log h+8h\ln(2h)\ln h
    =O(h\log^2h).\qedhere
\]
\end{proof}

\newcommand{\ftDarts}{\vec E}
\newcommand{\ftSucc}{\operatorname{succ}}
\newcommand{\ftPrec}{\operatorname{prec}}
\newcommand{\ftTour}{\operatorname{t}}
\newcommand{\ftGap}{\operatorname{gap}}
\newcommand{\ftLm}{B}
\newcommand{\ftSub}[2]{[#1]_{#2}}
\newcommand{\ftFree}{\mathsf{LandmarkFree}}
\newcommand{\ftConn}{\mathsf{Connected}}
\newcommand{\ftInv}{\textup{(Inv)}}
\newcommand{\ftInvW}{\textup{(Inv$'$)}}
\newcommand{\ftOp}[1]{\mathtt{#1}}
\SetKwFunction{FTNavigate}{Navigate}
\SetKwFunction{FTRepair}{Repair}
\SetKwFunction{FTUpdate}{ForestUpdate}

\subsection{Fully Dynamic Algorithm on Forests}
\label{sec:forests}

In this subsection, we prove \cref{thm:forests}, so we consider a fully dynamic update sequence $(e_i=u_iv_i)_{i=1}^M$ such that every $G_k$, the graph after the first $k$ updates, is a forest, i.e., every insertion $e_i$ joins two distinct components of $G_{i-1}$ and every deletion $e_i$ removes an edge of $G_{i-1}$; in the list model, we also fix $\Delta\ge1$ such that every $G_k$ has maximum degree at most $\Delta$.

\paragraph{Circular Lists.}
For a set $U$, let $\mathcal L(U)$ be the set of \emph{lists} $L=(z_1,\ldots,z_h)$ of $h\ge0$ distinct elements of $U$, and $\mathcal C(U)$ the set of \emph{circular lists}, i.e., of nonempty lists up to rotation, where $(z_1,\ldots,z_h)=(z_2,\ldots,z_h,z_1)$.
For $L\in\mathcal L(U)\cup\mathcal C(U)$, we write $|L|\coloneqq h$ and also use $L$ for the set $\{z_1,\ldots,z_h\}$.
We denote with $\circ$ the concatenation, where $z\in U$ stands for $(z)$, with $\mathsf{close}(L)\in\mathcal C(U)$ the circular list of a nonempty $L\in\mathcal L(U)$, and, for $z_j\in L$, with $\mathsf{split}(L,z_j)$ the lists $(z_1,\ldots,z_{j-1})$ and $(z_j,\ldots,z_h)$ if $L\in\mathcal L(U)$, and the list $(z_j,\ldots,z_h)\circ(z_1,\ldots,z_{j-1})$ if $L\in\mathcal C(U)$, which is well defined as the $z_i$ are distinct.
For $L\in\mathcal C(U)$, we define the \emph{successor} $\ftSucc_L(z_j)\coloneqq z_{j+1}$ and the \emph{predecessor} $\ftPrec_L(z_j)\coloneqq z_{j-1}$ of $z_j$, where $z_{h+1}\coloneqq z_1$ and $z_0\coloneqq z_h$.

\paragraph{Euler Tours.}
For a graph $G=(V,E)$ with $V\subseteq[n]$, let $\ftDarts(G)\coloneqq\{(w,x)\mid wx\in E\}$ be the set of its \emph{oriented edges}, and, for $x\in V$ with $d_G(x)\ge1$, we also write $N_G(x)$ for the circular list of the neighbors of $x$ in increasing order; a forest has $|\ftDarts(G)|=2|E|\le2n-2$ oriented edges.
An \emph{Euler tour} of a tree $T$ with at least one edge is a $Z=(z_1,\ldots,z_h)\in\mathcal C(\ftDarts(T))$ with $Z=\ftDarts(T)$ such that $z_i=(w,x)$ and $z_{i+1}=(w',x')$ satisfy $x=w'$ for every $i\in[h]$.
We call it \emph{canonical} if, for every $(w,x)\in Z$,
\begin{equation}\label{eq:forests-succ}
    \ftSucc_Z((w,x))=(x,\ftSucc_{N_T(x)}(w)),\qquad\text{and hence}\qquad\ftPrec_Z((w,x))=(\ftPrec_{N_T(w)}(x),w).
\end{equation}
Note that the canonical Euler tour $\ftTour(T)$ of $T$ is unique.
As the $N_T(x)$ are sorted by vertex, $\ftTour(T)$ does not depend on the order of the adjacency lists in the list model.

For a forest $G$, let $\mathcal T(G)\coloneqq\{G[C]\mid C\text{ is a component of $G$ with }|C|\ge2\}$ be the set of its trees with at least one edge, and, for $z\in\ftDarts(G)$, let $T_G(z)$ be the $T\in\mathcal T(G)$ with $z\in\ftDarts(T)$, and $\ftSucc_G(z)\coloneqq\ftSucc_{\ftTour(T_G(z))}(z)$.

\begin{remark}
\label{rem:forests-navigation}
For an edge $wx$ of a forest $G$, the first and the last neighbor of $x$ in the order $w+1,\ldots,n,1,\ldots,w-1$, or $w$ if there is none, are $\ftSucc_{N_G(x)}(w)$ and $\ftPrec_{N_G(x)}(w)$.
Hence, in the matrix model, if the oracle describes $G$ up to one known pair $e$, then scanning the row of $x$ and flipping the entry of $e$ yields both with at most $n$ queries in time $O(n)$ with $O(1)$ words, and, in the list model, scanning the list of $x$ instead of the row yields both in time $O(d_G(x))=O(\Delta)$ with $O(1)$ words, as it suffices to see every neighbor of $x$ once, in any order, so, by \cref{eq:forests-succ}, we can navigate the $\ftTour(T)$ for $T\in\mathcal T(G)$ without storing them.
\end{remark}

The following lemma shows that an update modifies the Euler tours only locally.

\begin{lemma}\label{lem:forests-splice}
Let $G$ be a forest, $T\in\mathcal T(G)$, $uv$ an edge of $T$, and $G'\coloneqq G-uv$.
For $x\in\{u,v\}$ and $y \in \{u,v\} \setminus \{x\}$, let $T_x\coloneqq G'[C_{G'}(x)]$, $a_x\coloneqq(\ftPrec_{N_T(x)}(y),x)$, $b_x\coloneqq(x,\ftSucc_{N_T(x)}(y))$, and $L_x\coloneqq\mathsf{split}(\ftTour(T_x),b_x)$ if $T_x\in\mathcal T(G')$, and $L_x\coloneqq()$ otherwise.
Then $\mathcal T(G')\setminus\{T_u,T_v\}=\mathcal T(G)\setminus\{T\}$ and
\[
    \ftTour(T)=\mathsf{close}(L_u\circ(u,v)\circ L_v\circ(v,u)).
\]
Moreover, $L_x\circ(x,y)$ starts with $b_x$, and $(y,x)\circ L_x$ ends with $a_x$; in particular, the first elements of the nonempty $L_x$ form $\{b_u,b_v\}\cap\ftDarts(G')$.
\end{lemma}

\begin{proof}
Deleting $uv$ splits $T$ into $T_u$ and $T_v$ and keeps all other trees of $G$, so $\mathcal T(G')\setminus\{T_u,T_v\}=\mathcal T(G)\setminus\{T\}$.
For $x\in\{u,v\}$, we have $a_x=(y'',x)$ and $b_x=(x,y')$ for the predecessor $y''$ and the successor $y'$ of $y$ in $N_T(x)$.
If $T_x\in\mathcal T(G')$, then $N_T(x)$ arises from $N_{T_x}(x)$ by inserting $y$, so $a_x,b_x\in\ftTour(T_x)$ and $y'$ follows $y''$ in $N_{T_x}(x)$; as $N_T(z)=N_{T_x}(z)$ for the other vertices $z$ of $T_x$, \cref{eq:forests-succ} for $\ftTour(T_x)$ shows that $L_x$, which starts with $b_x$, ends with $\ftPrec_{\ftTour(T_x)}(b_x)=a_x$ and that, in $L_x$, every $(w,z)\neq a_x$ is followed by $(z,\ftSucc_{N_T(z)}(w))$.
Otherwise, $L_x=()$ and $(a_x,b_x)=((y,x),(x,y))$.
In both cases, $L_x\circ(x,y)$ starts with $b_x$ and $(y,x)\circ L_x$ ends with $a_x$, so, in $Z\coloneqq\mathsf{close}(L_u\circ(u,v)\circ L_v\circ(v,u))$, $(y,x)$ is followed by $b_x$ and $a_x$ by $(x,y)$.
Hence every $(w,z)\in Z$ is followed by $(z,\ftSucc_{N_T(z)}(w))$, and, as $Z=\ftDarts(T)$, $Z$ is the canonical Euler tour of $T$.
\end{proof}

For update $i$, we apply \cref{lem:forests-splice} with $uv\coloneqq e_i$ to the forest among $G_{i-1}$ and $G_i$ that contains $e_i$ and its tree $T$ containing $e_i$, so that $G'$ is the other forest, and let $X$ be the set of first elements of the nonempty lists among $L_u$, $L_v$, and, for a deletion, $(u,v)$ and $(v,u)$, i.e., $X=\{b_u,b_v\}\cap\ftDarts(G_{i-1})$ for an insertion and $X=\{b_u,(u,v),b_v,(v,u)\}$ for a deletion.
While we process update $i$, the oracle still describes $G_{i-1}$, which differs from $G_i$ only in $e_i$, so, by \cref{rem:forests-navigation}, we can compute $b_u$ and $b_v$ and evaluate $\ftSucc_{G_{i-1}}$ and $\ftSucc_{G_i}$ in time $O(n)$, or $O(\Delta)$ in the list model, with $O(1)$ words.

\paragraph{Sparse Representation.}
Set $\ell\coloneqq\lceil2n/S\rceil$, so that $2\le\ell\le3n/S$, and call a tree $T$ \emph{short} if $|\ftTour(T)|<\ell$, and \emph{long} otherwise.
We keep in memory only a set $\ftLm$ of oriented edges, the \emph{landmarks}, partitioned into a set $\mathcal S\subseteq\mathcal L(\ftLm)\cup\mathcal C(\ftLm)$ of nonempty lists and circular lists, and, for $\lambda\in\ftLm$, we let $S_\lambda$ be the $L\in\mathcal S$ with $\lambda\in L$.
For a list or circular list $Z=(z_1,\ldots,z_h)$, let $\ftSub{Z}{\ftLm}\coloneqq(z_{j_1},\ldots,z_{j_k})$ for the indices $j_1<\cdots<j_k$ with $z_{j_1},\ldots,z_{j_k}\in\ftLm$, which is circular if $Z$ is and $k\ge1$; thus $\ftSub{L\circ L'}{\ftLm}=\ftSub{L}{\ftLm}\circ\ftSub{L'}{\ftLm}$, and $\ftSub{\mathsf{close}(L)}{\ftLm}=\mathsf{close}(\ftSub{L}{\ftLm})$ if $L\cap\ftLm\neq\emptyset$.
For a forest $G$ and $\lambda\in\ftLm\cap\ftDarts(G)$, the \emph{gap} after $\lambda$ is $\ftGap_G(\lambda)\coloneqq\min\{j\ge1\mid\ftSucc_G^j(\lambda)\in\ftLm\}$, where $\ftSucc_G^j$ is the $j$-fold iterate of $\ftSucc_G$, so that $\sum_{\lambda\in\ftTour(T)\cap\ftLm}\ftGap_G(\lambda)=|\ftTour(T)|$ for every $T\in\mathcal T(G)$ with $\ftTour(T)\cap\ftLm\neq\emptyset$.
We maintain the following invariant for the current forest $G$:
\begin{enumerate}[nosep,topsep=3pt]
    \item[\ftInv] For every $T\in\mathcal T(G)$, $\ftTour(T)\cap\ftLm\neq\emptyset$ if and only if $|\ftTour(T)|\ge\ell$; $\ell\le\ftGap_G(\lambda)\le2\ell$ for all $\lambda\in\ftLm$; and $\mathcal S=\{\ftSub{\ftTour(T)}{\ftLm}\mid T\in\mathcal T(G),\ \ftTour(T)\cap\ftLm\neq\emptyset\}$.
\end{enumerate}
We access $\mathcal S$ only through the following interface, which we implement in \cref{sec:forests-lists}, where $\lambda,\lambda'\in\ftLm$ and $z\notin\ftLm$:
\begin{itemize}[nosep,topsep=3pt]
    \item $z\in\ftLm$ and $\ftOp{Root}(\lambda)$: test membership, and return an identifier of $S_\lambda$, so that $\ftOp{Root}(\lambda)=\ftOp{Root}(\lambda')$ if and only if $S_\lambda=S_{\lambda'}$.
    \item $\ftOp{Prev}(\lambda)$ and $\ftOp{Cut}(\lambda)$: return $\ftPrec_{S_\lambda}(\lambda)$ if $S_\lambda\in\mathcal C(\ftLm)$, and replace $S_\lambda$ in $\mathcal S$ by the nonempty lists of $\mathsf{split}(S_\lambda,\lambda)$.
    \item $\ftOp{Create}(z)$, $\ftOp{Insert}(z,\lambda)$, and $\ftOp{Delete}(\lambda)$: add $\mathsf{close}(z)$ to $\mathcal S$, insert $z$ into $S_\lambda$ directly before $\lambda$, and remove $\lambda$ from $S_\lambda$, from $\mathcal S$ if $|S_\lambda|=1$, and from the set $W$ below.
    \item $\ftOp{Close}(L_1,\ldots,L_r)$ for $r=O(1)$ arguments, each a list in $\mathcal S\cap\mathcal L(\ftLm)$, given by any of its elements, or some $z\notin\ftLm$: replace them in $\mathcal S$ by $\mathsf{close}(L_1\circ\cdots\circ L_r)$.
\end{itemize}
Our algorithm maintains \ftInv{} for $G_i$ after update $i$.
While processing update $i$, it additionally stores a set $W$ of landmarks, which contains every landmark whose gap the update creates or shortens, and its functions use $\ftSucc_G$ for the \emph{current forest} $G$, which is $G_{i-1}$ until \FTUpdate{} sets $G\leftarrow G_i$.

\begin{algorithm}[htb]
\caption{Helper functions on forests}
\label{alg:forests-helpers}
\UBFn{\FTNavigate{$z,\bar z$}}{
    $(z',j)\leftarrow(z,0)$\tcp*[r]{the target $\bar z$ is optional}
    \lRepeat{$z'\in\ftLm\cup\{z,\bar z\}$}{$z'\leftarrow\ftSucc_G(z')$; $j\leftarrow j+1$}
    \lIf{$z'=\bar z$}{\Return $\ftConn$}
    \lIf{$z'\in\ftLm$}{\Return $(z',j)$}
    \Return $\ftFree$\;
}
\BlankLine
\UBFn{\FTRepair{}}{
    \While{$W\ne\emptyset$}{
        $\lambda\leftarrow$ any landmark in $W$; $W\leftarrow W\setminus\{\lambda\}$; $(\lambda',j)\leftarrow$ \FTNavigate{$\lambda$}\;
        \lWhile{$j<\ell$ and $\lambda'\ne\lambda$}{$\ftOp{Delete}(\lambda')$; $(\lambda',j)\leftarrow$ \FTNavigate{$\lambda$}}
        \lIf{$j<\ell$}{$\ftOp{Delete}(\lambda)$}
        \lIf{$j>2\ell$}{$\ftOp{Insert}(\ftSucc_G^\ell(\lambda),\lambda')$}
    }
}
\end{algorithm}

\begin{algorithm}[htb]
\caption{\texttt{ForestUpdate}$(i,u,v)$}
\label{alg:forests-update}
$b_x\leftarrow(x,\ftSucc_{N_T(x)}(y))$ for $x\in\{u,v\}$ and $y \in \{u,v\} \setminus \{x\}$; $W\leftarrow\emptyset$\;
\ForEach(\tcp*[f]{$T$ and $X$ as after \cref{lem:forests-splice}}){$z\in X\setminus\ftLm$}{
    $r\leftarrow$ \FTNavigate{$z$}\;
    \lIf{$r=\ftFree$}{$\ftOp{Create}(z)$}
    \lElse{$\ftOp{Insert}(z,\lambda)$ for $r=(\lambda,j)$}
    $W\leftarrow W\cup\{z,\ftOp{Prev}(z)\}$\;
}
$\ftOp{Cut}(z)$ for every $z\in X$\tcp*[r]{splices the circular lists as in \cref{lem:forests-splice}}
\lIf{$e_i$ is inserted}{$\ftOp{Close}(\ftSub{L_u}{\ftLm},(u,v),\ftSub{L_v}{\ftLm},(v,u))$; $W\leftarrow W\cup\{(u,v),(v,u)\}$}
\lElse{$\ftOp{Delete}((u,v))$; $\ftOp{Delete}((v,u))$; $\ftOp{Close}(\ftSub{L_x}{\ftLm})$ for every $x\in\{u,v\}$ with $L_x$ nonempty}
$G\leftarrow G_i$; \FTRepair{}\tcp*[r]{restores \ftInv{} for $G_i$}
\lIf{$s$ or $t$ has no neighbor in $G_i$}{\Return $\mathbf 1[s=t]$}
$r\leftarrow$ \FTNavigate{$(s,s'),(t,t')$}; $r'\leftarrow$ \FTNavigate{$(t,t')$}\tcp*[r]{$s',t'$: neighbors of $s,t$ in $G_i$}
\Return $\mathbf 1[r=\ftConn$, or $r=(\lambda,j)$ and $r'=(\lambda',j')$ with $\ftOp{Root}(\lambda)=\ftOp{Root}(\lambda')]$\;
\end{algorithm}

On update $i$, we call \FTUpdate{$i,u_i,v_i$}, using the functions of \cref{alg:forests-helpers,alg:forests-update}.
More specifically, our functions do the following:
\begin{description}
    \item[\FTNavigate{$z,\bar z$}:] walks along $\ftTour(T_G(z))$ from $z$ until it reaches the target $\bar z$, if given, returning $\ftConn$, a landmark $\lambda$, returning $\lambda$ and the number $j$ of steps, or $z$ again, returning $\ftFree$; thus, if $z\in\ftLm$ and no target is given, then $j$ is the gap after $z$.
    \item[\FTRepair{}:] processes the landmarks in $W$ one at a time: while the gap after $\lambda$ is less than $\ell$, it deletes the next landmark, which adds its gap to that of $\lambda$, unless $\lambda$ is the only landmark on $\ftTour(T_G(\lambda))$, in which case it deletes $\lambda$; if the gap then exceeds $2\ell$, it inserts the oriented edge $\ell$ steps after $\lambda$ as a landmark.
    \item[\FTUpdate{$i,u,v$}:] computes $b_u$ and $b_v$ from the rows or lists of $u$ and $v$ in $G_{i-1}$ and makes every $z\in X$ a landmark, by inserting it before the first landmark after it, or by creating $\mathsf{close}(z)$ if $\ftTour(T_G(z))$ has no landmark; this shortens or creates only the gaps after $z$ and after its predecessor, which join $W$.
    It then splits the stored circular lists at these landmarks and joins and closes the pieces as in \cref{lem:forests-splice}, where $\ftSub{L_x}{\ftLm}$ is the piece starting with $b_x$, or empty if $b_x=(x,y)$.
    Finally, it sets $G\leftarrow G_i$, restores \ftInv{} for $G_i$, finds neighbors $s'$ of $s$ and $t'$ of $t$ in $G_i$ by scanning their rows or lists in $G_{i-1}$ and accounting for $e_i$, and reports whether the walk from $(s,s')$ reaches $(t,t')$, or a landmark in the same stored circular list as the walk from $(t,t')$.
\end{description}

We now show that this algorithm satisfies \cref{thm:forests}, which we restate for convenience.

\forestsUB*

\subsubsection*{Correctness Analysis}
Correctness relies on \cref{lem:forests-splice} and on the following lemma.
As \FTRepair{} runs on intermediate states in which the gaps are not yet balanced, we state it under the following weaker condition on the current forest $G$:
\begin{enumerate}[nosep,topsep=3pt]
    \item[\ftInvW] For every $T\in\mathcal T(G)$, $\ftTour(T)\cap\ftLm=\emptyset$ implies $|\ftTour(T)|<\ell$; $\ftGap_G(\lambda)\le2\ell$ for all $\lambda\in\ftLm$; and $\mathcal S=\{\ftSub{\ftTour(T)}{\ftLm}\mid T\in\mathcal T(G),\ \ftTour(T)\cap\ftLm\neq\emptyset\}$.
\end{enumerate}
Note that \ftInv{} implies \ftInvW, and that, under \ftInvW, \FTNavigate{$z$} takes at most $2\ell$ steps if $\ftTour(T_G(z))\cap\ftLm\neq\emptyset$, and fewer than $\ell$ steps otherwise, with one evaluation of $\ftSucc_G$ and one membership test per step.

\begin{lemma}
\label{lem:forests-repair}
Let $G=G_i$, suppose that \ftInvW{} holds for $G_i$, and let $W$ contain every landmark with gap less than $\ell$.
Then \FTRepair{} restores \ftInv{} for $G_i$ and, if $W$ initially has $m$ elements, makes $O(m\ell)$ evaluations of $\ftSucc_G$ and membership tests and $O(m)$ further operations of the interface.
\end{lemma}

\begin{proof}
It is straightforward to check that every outer iteration, for some $\lambda$ taken from $W$, maintains the assumptions of the lemma: the inner loop merges the gap after $\lambda$ with the following gaps, each at most $2\ell$, until it is at least $\ell$, and thus less than $3\ell$, or deletes $\lambda$ if $\lambda$ is the only landmark on $\ftTour(T_G(\lambda))$, which then is short, and the insertion splits a gap $j>2\ell$ into $\ell$ and $j-\ell\in(\ell,2\ell)$.
Hence, once $W=\emptyset$, all gaps lie in $[\ell,2\ell]$, and \ftInv{} follows from \ftInvW.
Every deleted $\lambda'$ except the last one of each outer iteration has gap less than $\ell$ and thus lies in $W$, which never grows, so there are at most $m$ outer and $2m$ inner iterations, each with $O(\ell)$ evaluations of $\ftSucc_G$ and membership tests and $O(1)$ further operations.
\end{proof}

\begin{proof}[Proof of \cref{thm:forests} (correctness)]
We show by induction on $k$ that \ftInv{} holds for $G_k$ after update $k$, which is clear for $k=0$.
For update $i$, with $u\coloneqq u_i$ and $v\coloneqq v_i$, the loop over $X$ makes every $z\in X$ a landmark, which keeps \ftInvW{} for $G_{i-1}$ and changes only the gaps after $z$ and after its predecessor, which join $W$.
Then, by \cref{lem:forests-splice} and as $\ftSub{\cdot}{\ftLm}$ commutes with $\circ$ and $\mathsf{close}$, splitting the elements of $\mathcal S$ at $X$ and closing the pieces replaces the $\ftSub{\ftTour(T')}{\ftLm}$ for $T'\in\mathcal T(G_{i-1})\setminus\mathcal T(G_i)$ by those for $T'\in\mathcal T(G_i)\setminus\mathcal T(G_{i-1})$; clearly, all other gaps are preserved, and the new landmarks $(u,v)$ and $(v,u)$ of an insertion have gap $1$ and lie in $W$.
Thus, \ftInvW{} holds for $G_i$ and $W$ contains every landmark with gap less than $\ell$, so \FTRepair{} restores \ftInv{} for $G_i$ by \cref{lem:forests-repair}.

Finally, $s$ and $t$ are connected in $G_i$ if and only if $s=t$, or both have neighbors $s'$ and $t'$ in $G_i$ and $T_{G_i}((s,s'))=T_{G_i}((t,t'))$.
If both have neighbors, the walk from $(s,s')$ stays in $\ftTour(T_{G_i}((s,s')))$, so it returns $\ftConn$ only if the two trees coincide and $\ftFree$ only if they differ; otherwise, it returns a landmark $\lambda$, and, by \ftInv, the trees coincide if and only if the walk from $(t,t')$ returns a landmark $\lambda'$ with $\ftOp{Root}(\lambda)=\ftOp{Root}(\lambda')$.
\end{proof}

\subsubsection*{Time and Space Analysis}

\begin{proof}[Proof of \cref{thm:forests} (time and space)]
Fix an update $i$.
Before it, \ftInv{} holds for $G_{i-1}$, so, for every long $T\in\mathcal T(G_{i-1})$, the gaps on $\ftTour(T)$ sum to $|\ftTour(T)|$ and are at least $\ell$, and there are at most $(2n-2)/\ell<S$ landmarks.
As the update creates only $O(1)$ further landmarks, every operation of the interface takes time $O(\log(S+1))$ by \cref{lem:forests-lists}.
For each of the at most four $z\in X$, it calls \FTNavigate at most once, which makes at most $2\ell$ evaluations of $\ftSucc_G$ and membership tests by \ftInvW{} for $G_{i-1}$, and performs $O(1)$ further operations.
Splicing takes $O(1)$ operations, and afterwards $|W|\le8$, so \FTRepair{} makes $O(\ell)$ evaluations of $\ftSucc_G$ and membership tests and $O(1)$ further operations by \cref{lem:forests-repair}.
Finally, answering the query scans two rows or lists and calls $\ftOp{Root}$ twice and \FTNavigate twice, with at most $2\ell$ steps each by \ftInv{} for $G_i$.
In total, the update makes $O(\ell)$ evaluations of $\ftSucc_{G_{i-1}}$ and $\ftSucc_{G_i}$ and row or list scans of $G_{i-1}$, each in time $O(n)$, or $O(\Delta)$ in the list model, by \cref{rem:forests-navigation}, and performs $O(\ell)$ operations of the interface, each in time $O(\log(S+1))$.
As $\ell\le3n/S$ and $S\le n$, its worst-case time is $O(\ell(n+\log(S+1)))=O(n^2/S)$ in the matrix model, and $O(\ell(\Delta+\log(S+1)))=O((n/S)(\Delta+\log S))$ in the list model, as $\Delta\ge1$.
Its space is $O(S)$, as at most $S+O(1)$ landmarks are stored at any time, which take $O(S)$ words by \cref{lem:forests-lists}, and all other variables take $O(1)$ words, as $|X|\le4$ and $|W|\le8$ at all times, and so does each scan of a row or list by \cref{rem:forests-navigation}.
\end{proof}

\subsubsection{Implementing the circular lists}
\label{sec:forests-lists}

\begin{lemma}\label{lem:forests-lists}
For at most $k\ge1$ landmarks at any time, the interface of the sparse representation can be implemented with $O(k)$ words such that every operation takes $O(\log(k+1))$ worst-case time.
\end{lemma}
\begin{proof}
We store every list in $\mathcal S$ as a red-black tree with parent pointers whose in-order is the list, and every circular list in $\mathcal S$ as such a tree of one of its rotations, marked as circular; two balanced search trees store $W$ and map every landmark to its node, and are updated when a landmark is created or deleted.
All trees take $O(k)$ words, have height $O(\log(k+1))$, and support inserting a node before another, deleting a node, and joining and splitting trees in $O(\log(k+1))$ worst-case time \cite[Ch.~4]{Tarjan83}.
Thus, every operation takes $O(\log(k+1))$ time: $z\in\ftLm$ is a lookup, $\ftOp{Root}(\lambda)$ the root above $\lambda$, $\ftOp{Prev}(\lambda)$ its in-order predecessor, or else the last node, $\ftOp{Create}(z)$ creates a circular one-node tree, $\ftOp{Insert}$ and $\ftOp{Delete}$ insert or delete a node, $\ftOp{Cut}(\lambda)$ splits before $\lambda$ and, if circular, joins the suffix with the prefix and removes the mark, and $\ftOp{Close}$ creates nodes for its new landmarks, joins its nonempty arguments in order, and marks the result as circular.
\end{proof}

\appendix

\section{Further Motivation}\label{sec:motivation}

Our model describes a graph that is owned by one system and monitored by
another. The graph is already stored as part of an application's data and
is maintained by its owner, while a separate monitor with limited private
memory must keep an answer up to date as the graph changes. The monitor is
notified of each update and can inspect the graph, but it may not annotate
the graph's representation or build an unrestricted auxiliary index: the
graph is read-only \emph{to the monitor}, even though its owner keeps
modifying it. Neo4j's change-data-capture interface, for instance, exposes
the creation, modification, and deletion of nodes and relationships to
other applications \cite{Neo4jCDC}, although such a feed does not by
itself provide the consistent snapshots and update ordering that our model
assumes (Section~\ref{sec:model}). The relevant resources are the monitor's
additional memory and the work it performs to maintain its answer after
each update. The following two illustrative scenarios both lead to dynamic
undirected $s$--$t$ connectivity; we do not claim that particular deployed
systems implement our exact oracle.

\paragraph{Monitoring network connectivity.}
A network-management system maintains the topology of a network whose links
are bidirectional, so that the topology is an undirected graph; link
failures and recoveries delete and insert edges, and building or restoring
the network gives a natural insertion-only phase. A separate monitoring
task must track whether two designated sites $s$ and $t$ are connected. It
can inspect the topology through the management system's interface, but it
has a limited memory allocation and no permission to modify the topology.
Maintaining $s$--$t$ connectivity after every update is exactly its task at
the level of the graph; routing policies, link directions, capacities, and
latencies would call for richer problems.

\paragraph{Auditing record linkage.}
A data-integration pipeline gradually accepts pairwise links asserting that
two records refer to the same entity. Since such assertions are symmetric,
we can model the records as vertices and the accepted links as undirected
edges; the connected components are then exactly the equivalence classes
implied by the accepted links. An independent auditor watches two records
$s$ and $t$ that are known to describe different entities, and must raise
an alarm at the first insertion that joins them through a chain of accepted
links. The alarm then stays set, mirroring the monotonicity of incremental
connectivity; if links can later be withdrawn, the problem becomes fully
dynamic. A disjoint-set structure would maintain the entire partition with
$O(n)$ additional words, but an auditor responsible for a single binary
alarm may have far less memory available, while the accepted links remain
in the database, where they can be reread.

\paragraph{Sources of the memory restriction.}
Restrictions on private memory may come from a memory allocation for an
independent analysis, from access permissions, or from the absence of a
suitable server-side index. When several analyses ask connectivity
questions about the same graph, a single shared index could serve them all;
our setting is most relevant when no such index is available, or when the
analyses work with different views of the graph. For a filtered view, our
model applies when the view's edge updates are supplied explicitly; a
single change to the source data that alters many edges of the view is not
automatically a single update.

\paragraph{Systems context.}
Systems research also supports separating a large read-only graph from a
smaller writable memory: the Read-Only Semi-External model \cite{ROSE21}
and the Sage graph engine \cite{Sage20} keep the graph in memory that
their algorithms only read, and give them writable memory proportional to
the number of vertices (see Appendix~\ref{sec:related}). Our question
concerns maintaining an answer as an externally owned graph changes, with
workspace budgets that include the regime far below the number of vertices.

\section{Related Work}\label{sec:related}

\paragraph{Static time--space tradeoffs.}
The randomized line of work begins with the logarithmic-space random walk
of Aleliunas et al.~\cite{AKLLR79}. Broder, Karlin, Raghavan, and Upfal
\cite{BKRU94} introduced a landmark approach that stores connectivity
information about sampled vertices and connects them using short walks; up
to logarithmic factors and the admissible space ranges, their tradeoff has
time--space product $m^2$. Barnes and Feige \cite{BF96} improved the
product to $m^{3/2}n^{1/2}$, and Feige \cite{Feige97} obtained $mn$
throughout the space range, together with finer degree-sensitive bounds.
Feige also gave fast logarithmic-space randomized algorithms for testing
whether the entire graph is connected \cite{Feige96}. Nonaka, Ono,
Sadakane, and Yamashita \cite{NOSY10} showed that a suitable Metropolis
walk has $O(n^2)$ hitting time and $O(n^2\log n)$ cover time, independently
of the number of edges. Building on these walks, Kosowski
\cite{Kosowski13} obtained the static tradeoff
\[
        T=\softO\!\left(m+\frac{n^2}{W}\right)
\]
with a workspace of $\softO(W)$ words; the additive $m$ term is part of
the bound and matters for dense graphs. Kosowski's degree-splitting
construction relies on efficient access to both neighbor ports and reverse
ports, whereas our list model supplies only degrees and indexed
neighbors.\footnote{In Kosowski's representation, given adjacent $u,v$,
the position of $u$ in $v$'s adjacency list can be computed in
polylogarithmic time. This operation is not supplied by
$\operatorname{Degree}$ and $\operatorname{Neighbor}$ alone, so his static
bound cannot simply be invoked verbatim for our weaker oracle.}
Our worst-case algorithm (Theorem~\ref{thm:metropolis}) also searches with
Metropolis walks. All of these algorithms are static: their walks and
landmark information refer to a single fixed graph. The central challenge
behind our algorithms is to maintain such information in small space while
the graph grows.

\paragraph{Deterministic space-efficient algorithms.}
A complementary line of work minimizes deterministic space. Savitch's
recursion uses $O(\log^2 n)$ bits for directed and undirected reachability,
in quasipolynomial time \cite{Savitch70}. Barnes and Ruzzo \cite{BR91}
obtained deterministic polynomial-time algorithms with sublinear space, and
Nisan \cite{Nisan94} gave a polynomial-time, $O(\log^2 n)$-space algorithm
through $\mathsf{RL}\subseteq\mathsf{SC}$. The space bound was improved to
$O(\log^{3/2}n)$ by Nisan, Szemer\'edi, and Wigderson \cite{NSW92} and to
$O(\log^{4/3}n)$ by Armoni et al.~\cite{ATWZ00}. Trifonov
\cite{Trifonov08} independently obtained $O(\log n\log\log n)$ space,
while Reingold achieved the optimal $O(\log n)$ bound \cite{Reingold08}.
Since undirected connectivity is complete for symmetric logarithmic space
\cite{LP82}, Reingold's theorem also shows that $\mathsf{SL}=\mathsf{L}$.
It settles the space complexity of undirected connectivity, but it does not
supply the fast intermediate-space tradeoffs we seek. Whether randomness
helps in logarithmic space more generally, that is, whether
$\mathsf{RL}=\mathsf{L}$, has been a central open problem in complexity
theory. OpenAI has recently claimed a proof that
$\mathsf{L}=\mathsf{RL}=\mathsf{BPL}$ \cite{OAI26}; at the time of
writing, this claim has not been independently verified. Like Reingold's
theorem, such a result concerns polynomial-time computation in logarithmic
space and does not by itself address the time--space tradeoffs studied
here. At
larger space budgets, Elmasry, Hagerup, and Kammer \cite{EHK15} give basic
graph algorithms on read-only input, including breadth-first search and
connected components, using $O(n)$ bits and linear time.

\paragraph{Compact connectivity indexes.}
Ben-David et al.~\cite{BDFGGMSh18} construct connectivity oracles for
bounded-degree graphs using an implicit decomposition, in a model that
charges more for writes than for reads. They store information about only
$O(n/k)$ centers and answer a component-label query in $O(k)$ expected work
using $O(k\log n)$ temporary words. This is a close precedent for
recovering connectivity information from a small stored summary together
with continued access to the graph. Their input, however, is static, and
they separate persistent storage from temporary symmetric memory, whereas
we charge all writable memory to a single bound.

\paragraph{Lower bounds for restricted machines.}
Cook and Rackoff \cite{CR80} introduced the jumping-automaton framework for
maze threadability. Beame et al.~\cite{BBRRT96,BBRRT99} established
time--space tradeoffs for graph traversal by graph automata and walking
automata, and Edmonds \cite{Edmonds98} proved further tradeoffs for
undirected $s$--$t$ connectivity on graph automata. Related directed results
include those of Barnes and Edmonds \cite{BE98} and of Edmonds, Poon, and
Achlioptas \cite{EPA99}. These models restrict how a machine navigates the
graph and remembers vertices, so their lower bounds are not general
word-RAM lower bounds and do not directly imply ours.

\paragraph{Dynamic connectivity.}
The conventional dynamic problem supports connectivity queries for
arbitrary pairs while maintaining a data structure under edge updates.
Early fully dynamic techniques include Frederickson's work on minimum
spanning trees \cite{Frederickson85}. Henzinger and King \cite{HK99}
obtained randomized polylogarithmic amortized time per operation, and Holm,
de Lichtenberg, and Thorup \cite{HLT01} obtained deterministic
$O(\log^2 n)$ amortized update time. Thorup \cite{Thorup00} and
Wulff-Nilsen \cite{Wulff13} further developed the time and space bounds,
the latter achieving $O(\log^2 n/\log\log n)$ amortized update time, and
Huang et al.~\cite{HHKPT23} achieved $O(\log n(\log\log n)^2)$ amortized
expected update time. For worst-case guarantees, Kapron, King, and Mountjoy
\cite{KKM13} gave a randomized data structure with polylogarithmic
worst-case update time. Gibb et al.~\cite{GKKT15} improved the update
bound and gave an $O(n\log^2 n)$-word implementation; their
\emph{sublinear space} refers to savings relative to storing all edges of a
dense graph, not to $o(n)$ words. Meierhans and Probst Gutenberg
\cite{MG26} obtained expected polylogarithmic worst-case update time
against adaptive updates, and Meierhans, Probst Gutenberg, and Yeh
\cite{MGY26} subsequently obtained polylogarithmic worst-case update
bounds with high probability against adaptive updates. These data
structures achieve strong time and adversary guarantees, but they retain far
more vertex-level state than the small-workspace regime studied here.

\paragraph{Lower bounds for dynamic connectivity.}
Fredman and Saks \cite{FS89} introduced the chronogram method and
established foundational cell-probe lower bounds for dynamic data
structures, including bounds for disjoint-set problems and a lower bound
of $\Omega(\log n/\log\log n)$ for maintaining prefix parities. Miltersen,
Subramanian, Vitter, and Tamassia \cite{MSVT94} and Henzinger and Fredman
\cite{HF98} independently observed a simple reduction from prefix parity
to dynamic connectivity, which yields the same lower bound for dynamic
connectivity. P{\u a}tra{\c s}cu and Demaine \cite{PD06} proved logarithmic amortized
lower bounds and update--query tradeoffs for dynamic connectivity, even on
collections of paths. P{\u a}tra{\c s}cu and Thorup \cite{PT11} showed
that for incremental connectivity, worst-case insertion time
$o(\log n/\log\log n)$ forces worst-case query time $n^{1-o(1)}$; for fully
dynamic connectivity, they gave the analogous query lower bound when the
amortized update time is $o(\log n)$, even allowing Las Vegas
randomization.

Going beyond logarithmic lower bounds has required new techniques. For
dynamic weighted range counting, a problem whose queries return many bits,
Larsen \cite{Lar12} proved a lower bound of
$\Omega((\log n/\log\log n)^2)$.
For problems with Boolean answers, Larsen, Weinstein, and Yu \cite{LWY20}
proved the first super-logarithmic lower bound, $\widetilde\Omega(\log^{3/2}n)$,
and Ko \cite{KoBool26} recently obtained an
$\Omega((\log n/\log\log n)^2)$ lower bound for such a problem. Larsen and
Yu \cite{LY26} obtained the first super-logarithmic lower bound for a
natural graph problem, incremental reachability in directed acyclic graphs,
and Ko \cite{Ko26} has very recently improved it: every zero-error
cell-probe data structure for this problem with worst-case update time
$t_u$ and expected query time $t_q$ satisfies
$\max(t_u,t_q)=\Omega((\log n/\log\log n)^2)$, and the same bound holds for
incremental undirected shortest paths and subgraph connectivity.
Polynomial cell-probe lower bounds remain out of reach. P{\u a}tra{\c s}cu
\cite{Pat10} introduced the multiphase problem and conjectured a
polynomial lower bound for it, which would imply polynomial lower bounds
for dynamic reachability and subgraph connectivity, among other problems.
Polynomial lower bounds for the multiphase problem are known only for
restricted classes of data structures, namely non-adaptive
\cite{BL15} and semi-adaptive \cite{KW20,DL20} ones; related results
include threshold lower bounds, in which very fast queries force
polynomial update time \cite{CGL15}, and lower bounds for a 1.5-round
version of P{\u a}tra{\c s}cu's communication game \cite{CEEP16}.

All of these results charge accesses to a maintained data structure and
study the balance between update and query costs. Our lower bounds instead
charge probes or time with respect to the current input and constrain the
entire writable workspace; moreover, the terminal pair is fixed and an
answer is required after every update. In particular, our $\Omega(n^2/B)$
lower bound is polynomial because the workspace is restricted, and it has
no bearing on the open questions about cell-probe lower bounds above.

\paragraph{Streaming and sketching.}
Graph streaming also separates the input from a small working memory, but
under a different access restriction: a stream does not provide random
access to the current graph, whereas our algorithms can inspect it
repeatedly. The semi-streaming model of Feigenbaum et al.~\cite{FKMSZ05}
allows $\softO(n)$ space, and the linear sketches of Ahn, Guha, and
McGregor \cite{AGM12a,AGM12b} recover spanning forests from
insertion--deletion streams. Streaming lower bounds include Sun and
Woodruff's $\Omega(n\log n)$-bit bound for global connectivity in insertion
streams \cite{SW15} and Nelson and Yu's $\Omega(n\log^3 n)$-bit bound for
spanning forest computation in dynamic streams \cite{NY19}, and Yu
\cite{Yu21} proved a tight lower bound for distributed connectivity
sketches. These bounds concern different outputs or communication
restrictions, and should not be read as identical lower bounds for
fixed-pair connectivity.
Huang and Peng \cite{HP19} obtained $o(n)$-space dynamic-stream algorithms
for approximate graph statistics and property-testing tasks, rather than
for exact fixed-pair connectivity in arbitrary graphs. Demetrescu,
Finocchi, and Ribichini \cite{DFR09} studied space--pass tradeoffs in a
streaming model that can write intermediate streams. Such passes and
writable streams offer another way to trade memory for repeated access, but
they do not coincide with our read-only oracle for the current graph.

\paragraph{Sensitivity oracles.}
Sensitivity oracles preprocess a graph so that connectivity queries can be
answered after a batch of failures or changes \cite{PT07,DP10,LLPSW26}.
Their preprocessing, stored index, and bounded-change interface differ from
our task of maintaining a fixed-terminal answer throughout an online update
sequence with small total workspace.

\paragraph{Directed and quantum connectivity.}
Directed reachability has a different space complexity: besides Savitch's
recursion, Barnes et al.~\cite{BBRS98} gave a polynomial-time algorithm
using $n/2^{\Theta(\sqrt{\log n})}$ space. Dynamic directed reachability
admits algebraic update--query tradeoffs via dynamic matrix inversion
\cite{Sankowski04,BNSS19}; these are not sublinear-workspace results for
undirected connectivity. In the quantum model, Apers et al.~\cite{AJPW23}
obtained $\softO(n)$ time and $O(\log n)$ space for undirected connectivity
with list-type graph access, improving on the known classical small-space
algorithms, and Jeffery and Pass \cite{JP26} gave a quantum time--space
tradeoff for directed $s$--$t$ connectivity. These results use resources and
oracle operations different from those of the classical dynamic algorithms
considered here.

\paragraph{Read-only inputs with auxiliary memory.}
The separation between an externally stored input and auxiliary writable
information is well established. In systems, the Read-Only Semi-External
(ROSE) model of Blelloch et al.~\cite{ROSE21} stores the edges of a graph
in read-only external memory, accessed in blocks, and requires the vertices
to fit in internal memory; its motivations include sharing one graph among
concurrent analyses, keeping a compressed representation without
algorithm-induced rewrites, and avoiding expensive writes to external
memory. In the Parallel Semi-Asymmetric Model of Dhulipala et
al.~\cite{Sage20}, the graph resides in nonvolatile memory, where writes
are expensive, and mutable memory is proportional to the number of
vertices; their Sage engine never writes to the graph's memory, and their
experiments give practical evidence for this design. These models have
access and cost measures different from ours, and they provide writable
memory proportional to the number of vertices, whereas we consider an input
that changes over time and workspace budgets that include the regime far
below this one; our model should not be identified with either. In systematic data
structures, the input
is stored verbatim and queries use a smaller auxiliary index; G\'al and
Miltersen \cite{GM07} proved probe--redundancy tradeoffs in this setting.
The recent query-with-sketch framework of Garg et al.~\cite{GHLPPY26} is
particularly relevant: it explicitly permits the auxiliary information to
change between queries and proves amortized probe lower bounds. There,
however, the input is static and accesses to the auxiliary information are
not charged, so the framework does not itself formulate our evolving-graph
workspace problem. Dynamic descriptive complexity likewise distinguishes
changing input relations from maintained auxiliary relations, beginning
with DynFO \cite{PI97}; reachability can be maintained in this framework
\cite{DKMSZ18}, where the central resource is the expressiveness of
first-order updates rather than sequential word-RAM time under a workspace
budget. Dynamic time--space questions have also been studied for algebraic
problems \cite{RT97}, and dynamic algorithms with external oracle access
appear in geometric problems such as union-volume estimation
\cite{BBCW26}. Our model applies this separation to exact undirected
fixed-pair connectivity under an online update sequence, with persistent
and temporary memory charged to the same bound.

\section*{Acknowledgements}

\noindent\begin{minipage}[t]{\dimexpr\linewidth-5.5cm\relax}%
\vspace{0pt}%
\indent This research was supported by the European Research Council (ERC) under the European Union's Horizon 2020 research and innovation programme (Grant agreement No.\ 101019564), the Austrian Science Fund (FWF) under grant DOI 10.55776/Z422, by the National Science Foundation under NSF Grant CCF-2330048, the Binational Science Foundation under BSF Grant 2024233, by a Siebel Fellowship and by a Simons Investigator Award.

\end{minipage}
\begin{minipage}[t]{5cm}%
\vspace{0pt}%
\includegraphics[width=5cm]{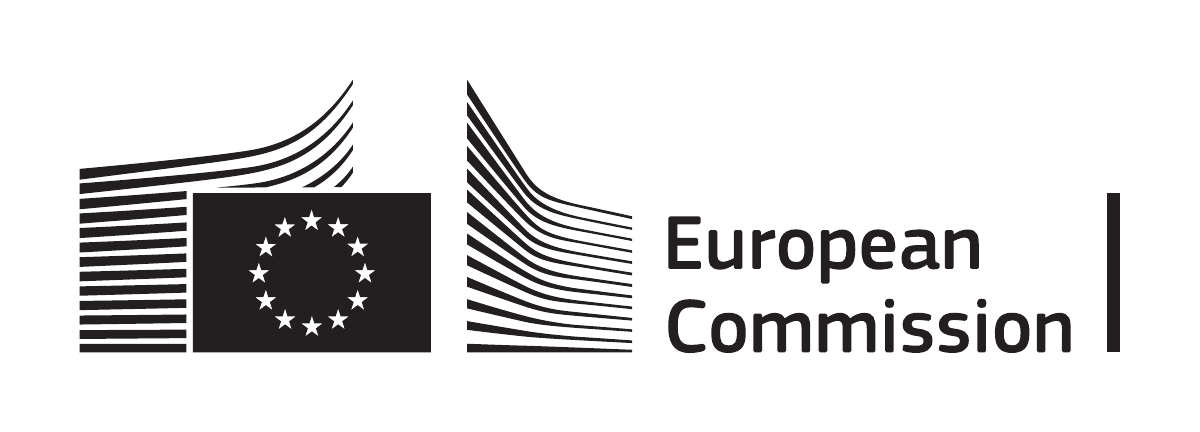}%
\end{minipage}

\smallskip
\noindent For open access purposes, the authors have applied a CC BY public copyright license to any author-accepted manuscript version arising from this submission. Views and opinions expressed are however those of the author(s) only and do not necessarily reflect those of the European Union or the European Research Council Executive Agency. Neither the European Union nor the granting authority can be held responsible for them.

\section*{AI Acknowledgment}

The authors independently designed the new model in the dynamic setting. For the lower bounds, the initial ideas came from the authors, and the AI tools Astra (OpenAI) and Claude Opus (Anthropic) were used to work out the technical details. For the upper bounds, the ideas for \cref{thm:largespace,thm:forests} came from the authors, while the proofs of \cref{thm:metropolis,thm:smallspace} were obtained with AI tools, in interaction with the authors.
AI tools were also used to generate figures. 

\bibliographystyle{alpha}
\bibliography{references}
\end{document}